\documentclass[12pt]{article}

\usepackage{graphics,psfrag}
\usepackage{amsthm,amssymb,epsfig,amsmath,euscript,array,cite}
\usepackage{color}
\usepackage{graphicx}
\usepackage{amscd}
\usepackage{slashed}
\usepackage{hyperref}
\newcommand{\be}{\begin{equation}}
\newcommand{\ee}{\end{equation}}
\def\bea{\begin{eqnarray}}
\def\eea{\end{eqnarray}}

\def\nn{\nonumber}

\newcommand{\beq}{\begin{equation}}
\newcommand{\eeq}{\end{equation}}
\newcommand{\ben}{\begin{eqnarray}}
\newcommand{\een}{\end{eqnarray}}
\newcommand{\bes}{\begin{subequations}}
\newcommand{\ees}{\end{subequations}}
\newcommand{\blg}{\begin{align}}
\newcommand{\elg}{\end{align}}

\def\one{\mbox{1 \kern-.59em {\rm l}}}
\def\dphi1{{\dot\phi_1}}
\def\dphi2{{\dot\phi_2}}
\def\dphi3{{\dot\phi_3}}
\def\dphi{{\dot\phi}}

\def\={\, =\, }
\def\a{\alpha}      
\def\b{\beta}       
    
\def\d{\delta}    
\def\e{\epsilon}

\def\l{\lambda} \def\L{\Lambda}
 
\def\o{\omega}

\def\s{\sigma}

\makeatletter \@addtoreset{equation}{section} \makeatother

\begin{document}

\begin{titlepage}
\hfill MCTP-14-28\\


\begin{center}

{\Large \bf
Type IIB Supergravity Solutions with AdS${}_5$ }\\[3mm]
{\Large \bf From  Abelian and Non-Abelian T Dualities}
\vspace{14pt}

{\bf \small Niall T. Macpherson$^{a,d}$, Carlos N\'u\~nez$^a$\footnote{Also at CP3 Origins. Odense, SDU.},
 Leopoldo A. Pando Zayas$^b$}\\[2mm]
{\bf \small Vincent G. J. Rodgers$^c$ and Catherine A. Whiting$^c$ }

\vspace{.2cm}
{\it ${}^a$ School of Physical Sciences, Swansea University,\\
 Singleton Park, Swansea, SA2 8PP, UK}\\[2mm]
{\it ${}^b$ Michigan Center for Theoretical Physics,
 Randall Laboratory of Physics,\\
 University of Michigan,
 Ann Arbor, MI 48109, USA}\\[2mm]
{\it ${}^c$ Department of Physics and Astronomy,
  The University of Iowa,\\
 Iowa City, IA 52242, USA}\\[2mm]
{\it ${}^d$ Dipartimento di Fisica,
\it    Universit$\grave{a}$ di Milano-Bicocca,\\
  I-20126 Milano, Italy}

\vspace{14pt}

\end{center}
\begin{abstract}
We present a large class of new backgrounds that are
solutions of type IIB supergravity with a warped AdS${}_5$ factor,
non-trivial axion-dilaton, $B$-field and three-form Ramond-Ramond
flux but  yet have no five-form flux.  We obtain these solutions and
many of their variations by judiciously applying  non-Abelian
and Abelian T-dualities, as well as coordinate shifts to
AdS${}_5\times X_5$  IIB supergravity solutions with
$X_5=S^5, T^{1,1}, Y^{p,q}$.
We address a number of issues pertaining to charge quantization in the context of non-Abelian T-duality.
We comment on some properties of the expected dual super
conformal field theories by studying their CFT
central charge holographically. We also use the structure of
the supergravity Page charges, central charges and some probe branes to infer aspects of
the dual super conformal field theories.
\end{abstract}

\end{titlepage}

\setcounter{page}{1} \renewcommand{\thefootnote}{\arabic{footnote}}
\setcounter{footnote}{0}
\newpage

\tableofcontents
\section{Introduction}
In its most precise formulation, the AdS/CFT correspondence conjectures an equivalence between string theory in AdS${}_5\times S^5$ with $N$ units of Ramond-Ramond five-form flux and ${\cal N}=4$ supersymmetric Yang Mills with $SU(N)$ gauge  group \cite{Maldacena:1997re,Witten:1998qj,Gubser:1998bc,Aharony:1999ti}. There are many versions of the correspondence which extend to string theory on other manifolds and, respectively, other field theories. One particularly intuitive entry in the AdS/CFT dictionary is how conformal invariance in the  corresponding field theory  is translated into an isometry of the metric in AdS${}_5\times X_5$:  a rescaling of the radial direction in the   AdS${}_5$ component of the metric corresponds to change in the energy scale of the field theory. This gravitational isometry gives a geometrical perspective to the idea that in conformal field theories, re-scaling of the coordinates and the energy scale together leave the theory invariant.

The central role that the AdS${}_5$ component of space-time plays  is that its  $SO(4,2)$ isometry will dictate that a (super)conformal field theory will be its dual. This   has prompted the search for solutions in IIB superstrings and M-theory that contain AdS${}_5$ as a space-time  factor. Some fairly systematic attacks have been launched in such search. For example, in \cite{Gauntlett:2004zh}, a  search in the context of M-theory yielded many interesting new backgrounds. Similar methods, based on a combination of supersymmetry and other endemic symmetries, were  extended to type IIB \cite{Gauntlett:2005ww}, an interesting subcase relevant  to this manuscript was presented in \cite{Colgain:2011hb}.
As in the references above,
the search for gravity solutions in general has historically
been based on symmetries \cite{Stephani:624239}.
There is although a relatively new conceptual strategy, distinct from merely exploiting symmetries to identify solutions.
It is based on the
use of symmetry transformations to generate solutions.
Indeed, solution-generating techniques have already been successfully
applied for  black hole and intersecting brane solutions
in string theory; for a review see \cite{Peet:2000hn}.
This particular strategy has also been  used in the context of the AdS/CFT correspondence, where a $U(1)\times U(1)$ isometry
was exploited to generate  the  Lunin-Maldacena
backgrounds \cite{Lunin:2005jy}.

A generalization of T-duality, called non-Abelian T-Duality (NATD), was suggested some time ago \cite{Ossa:1992vc,Fridling:1983ha,Fradkin:1984ai}.   Many investigations of this possibly new symmetry followed that focused on transformations of chiral $\sigma$ models \cite{Giveon:1993ai,Gasperini:1993nz,Alvarez:1994np,Elitzur:1994ri,Sfetsos:1994vz,Curtright:1994be,Alvarez:1995uc} ; for a
review see \cite{Alvarez:1994dn,Giveon:1994fu}.  However the question of how this symmetry would manifest itself in the Ramond-Ramond sector remained unanswered.    Recently, there has been a revival of NATD and in particular, the crucial
extension to the Ramond-Ramond sector has
been proposed \cite{Sfetsos:2010uq, Lozano:2011kb}.
This resurrected symmetry has already  been used
to generate solutions from various  seed backgrounds
in the context of the AdS/CFT correspondence
\cite{Itsios:2013wd, Caceres:2014uoa,
Lozano:2012au, Itsios:2012zv, Macpherson:2013lja, Gevorgyan:2013xka,
Lozano:2013oma, Gaillard:2013vsa, Elander:2013jqa, Zacarias:2014wta,
Pradhan:2014zqa, Lozano:2014ata}.
Several investigations  about the
interplay of NATD and physical properties
of holographically interesting backgrounds were
discussed in  \cite{Itsios:2012zv, Itsios:2013wd, Barranco:2013fza,
Gaillard:2013vsa, Lozano:2013oma, Caceres:2014uoa, Lozano:2014ata}.

The main goal of this paper is to further use T-duality and NATD  to  construct supergravity backgrounds that contain an AdS${}_5$ factor in the metric.  Furthermore,
we will also investigate some of the salient properties of the
dual super-conformal field theories. Our strategy for the construction
of such backgrounds relies on starting with seed backgrounds of
the form AdS${}_5\times X_5$  with $X_5=S^5, T^{1,1}, Y^{p,q}$
and applying a set of Abelian and non-Abelian T-dualities on
the $X_5$ factor. The dual SCFT to AdS${}_5\times X_5$ is
well understood and we use this information to deduce some of
the properties of the dual SCFT for the cases of T-dualized backgrounds.

The paper is organized as follows.
In Section \ref{Sec:NATD-IIB} we start with revisiting NATD for $AdS_5\times X_5$ where $X_5=S^5, T^{1,1}, Y^{p,q}$.
This analysis has already been presented in the
literature \cite{Itsios:2012zv, Itsios:2013wd, Sfetsos:2014tza}.
We pay, however, special attention to normalization factors,
including powers of $\alpha'$  as this will  be important in
subsequently  interpreting the gravity results from the field theory
point of view.
Furthermore we lift these solutions to eleven dimensions in Section \ref{Sec:mtheory} in order to probe their structure in M-theory.
Whilst the present  work can be understood strictly in the context of
constructing supergravity backgrounds; our ultimate motivation
is to investigate the dual field theories that arise through the  AdS/CFT correspondence.  In  Section
\ref{Sec:Page}, we examine the structure of
Page charges in supergravity and its implications on
the field theory, we also compute the central charge of
the dual field theories.

In Section \ref{Sec:NATD-T}, we present one of the main results
of this work.
There, we apply another T-duality to
the backgrounds discussed in Section \ref{Sec:NATD-IIB} which leads to interesting backgrounds for  IIB.
In the work of Lunin and Maldacena
\cite{Lunin:2005jy}, they
generated a plethora of interesting solutions by  performing a T-duality, followed by a shift in one of the coordinates, and then followed  by yet another T-duality (TsT)  on gravity theories
with  $U(1)\times U(1)$ isometry.  Motivated by this procedure,  we consider an
NATD-s-T transformation as our backgrounds have $SU(2)\times U(1)$ isometry.
Thus, in Section \ref{Sec:NATD-T},
we also present  a sample of the one-parameter family of solutions that
contain an $AdS_5$ factor.  It should be mentioned that some of these backgrounds
have  singularities
in the Ricci scalar.

In Section \ref{Sec:Susy}, we show that the solutions we
constructed using NATD followed by
a T-duality (but without the introduction of
a free parameter)
are supersymmetric and also  explicitly show their G-structure.
We discuss the interpretation of the dual field theories in Section \ref{Sec:FT} and conclude in Section \ref{Sec:Conclusions}.

For the sake of coherence as well as a sense of  completeness, we relegate a number of notational and technical issues to
the appendices. Appendix \ref{App:Rules}, for example,  describes our prescriptions for Abelian and Non-Abelian
T-dualities. In the appendices, we also present various
new one-parameter family of solutions obtained by considering
different actions of Abelian and Non-Abelian T-dualities on the AdS${}_5\times S^5$ background with various shifts of $U(1)$ isometries.
In applying NATD to backgrounds with $SU(2)$
isometry, there is a three-dimensional space of
parameters that can be introduced via gauge fixing.
Although formally NATD allows for the introduction of three parameters,
we  explicitly classify a sample of   twenty of the most obvious   possibilities and determine whether their volume form can lead to  consistent non-degenerate backgrounds. We further discuss the gauge ambiguity that might arise in NATD and show how the previously mentioned solutions are parameterized in Appendix \ref{App:GFix}.

\section{Non-Abelian T-duality for IIB Freund-Rubin
backgrounds}\label{Sec:NATD-IIB}

In this section we review the results of applying non-Abelian
T-dualities to the  $AdS_5\times S^5$, $ AdS_5\times T^{1,1}$ and
$AdS_5\times Y^{p,q}$ backgrounds.
It is common throughout the literature to work in units where $\alpha'=1$.
But in order to make clear some aspects of the
field theory dual
to the backgrounds we will  obtain,
we find it useful to focus on the normalization
and the factors of $\alpha'$. Therefore, in this section we present
the results with the appropriate factors of $L$
and $\alpha'$ restored for the benefit of the
reader.  However, we do not pay attention to factors of $g_s$, thus we
set $g_s=1$ throughout the paper.
In  Appendix \ref{App:Rules}, we review the B{\"u}scher rules
for Abelian T-duality and their extension to Non-Abelian T-duality,
with the proper factors of $\alpha'$.
Below, we summarize the results for the NATD applied to the backgrounds
of the form $AdS_5\times X^5$ mentioned above.
\subsection{$AdS_5\times S^5$}

We start with our conventions for the metric on $AdS_5\times S^5$,
\be
ds^2=4ds^2(AdS_5)+ds^2(S^5),
\ee
where
\be
\label{Ads5linenew}
ds^2(AdS_5)=(\frac{R^2(dx_{1,3}^2)}{L^2}+\frac{L^2 dR^2}{R^2}),\quad
ds^2(S^5)=L^2(4(d\alpha^2+\sin^2\alpha d\theta^2)+\cos^2\alpha ds^2(S^3)).
\ee
The metric on $S^3$ is defined as
\be
ds^2(S^3)= d\beta^2+d\phi^2+d\psi^2+2\cos\beta d\psi d\phi.
\ee
Here, and throughout this paper, we take $0\leq \beta\leq \pi,\ 0\leq \phi\leq2\pi,\ 0\leq \psi\leq 4\pi$. The attentive reader will notice that we have introduced
non-standard factors of $4$. This
was driven by demanding that the the $S^3$ needed
for NATD has a simple form (this just means that
its radius is $R_{S^3}=2$). The above geometry is
supported by the self-dual RR 5-form flux,
\be
F_5=\frac{4}{L}(d\text{Vol}(AdS_5)-d\text{Vol}(S^5)).
\ee
We will pay particular attention to normalizations,
as they will be relevant for our discussion of
properties of the dual field theory. In particular,
for the RR 5-form flux we take
\be
\frac{1}{(4\pi \alpha')^2}\int_{S^5}F_5=N,
\ee
which leads to the result $L^4= \frac{1}{4}\pi  N \alpha'^2 $.  Note that, using the normalization above, this is consistent with the usual result, $R^4=4\pi N \alpha'^2$.
\subsubsection{NATD of $AdS_5\times S^5$}
\label{NATD AdS5xS5}

We present now the results of a NATD transformation on the $S^3$
displayed Eq.(\ref{Ads5linenew}). These were
originally presented
in \cite{Sfetsos:2010uq}.
The gauge fixing we use is $(v_1,v_2,v_3)
\to (\rho,\chi,\xi)$. That is, the Lagrange multipliers introduced in NATD (see appendix  \ref{App:Rules} for details)
are written in spherical polar coordinates, where
$v_1=\rho \cos\xi \sin\chi,\;\; v_2=\rho \sin\chi \sin\xi,\;\;
v_3=\rho \cos\chi$. The range of the angles are $0<\chi<\pi$
and $0<\xi<2\pi$.
We will discuss general gauge fixing procedures and present the results
of alternate gauge fixings in Appendix \ref{App:GFix}.
We have included the correct factors of $\alpha'$,
appearing from the duality transformation, to
emphasize that the dual coordinates $(\rho, \chi, \xi)$
remain dimensionless,
\begin{align}
&\hat{ds}^2=4ds^2(AdS_5)+4L^2(d\alpha^2+\sin^2\alpha d\theta^2)
+\frac{\alpha '^2 d\rho^2 }{L^2\cos^2\alpha }
+ \frac{\alpha '^2 L^2 \rho ^2 \cos ^2\alpha
   \left(\mathit{d}\xi ^2 \sin ^2\chi +\mathit{d}\chi
   ^2\right)}{\alpha '^2 \rho ^2+L^4 \cos ^4\alpha },\nn \\&
   \hat{B}=\frac{\alpha '^3 \rho ^3 \sin \chi  \mathit{d}\xi \wedge
   \mathit{d}\chi }{\alpha'^2 \rho ^2+L^4 \cos ^4\alpha }, \quad
   e^{-2\hat{\Phi}}=L^2\cos^2\alpha\left(\frac{L^4 \cos ^4\alpha +\alpha'^2  \rho^2 }{\alpha '^3}\right).
\label{NATDads5s5}\end{align}
Notice that the dilaton has a singularity at $\alpha=\pi/2$.
Indeed, this is a curvature singularity as can be seen from the
10D string frame Ricci Scalar,
\bea
\mathcal{\hat{R}}=\frac{3 \sec ^2\alpha +\frac{4 \left(-7 \rho ^2 \alpha '^2+3 L^4
   \cos 2 \alpha +3 L^4\right)}{\rho ^2 \alpha '^2+L^4 \cos
   ^4\alpha }+\frac{28 \rho ^4\alpha '^4}{\left(\rho ^2
   \alpha '^2+L^4 \cos ^4\alpha \right)^2}-6}{2 L^2}.
\eea
The singularity appears because we are dualising on a manifold
that shrinks to zero-size at $\alpha=\pi/2$
---see Eq.(\ref{Ads5linenew}).
The non-trivial dual RR fluxes are given by,
\begin{align}
&\hat{F}_2=\frac{8 L^4 \sin \alpha  \cos ^3\alpha  \mathit{d}\alpha \wedge
   \mathit{d}\theta }{\alpha '^{3/2}}, \nn \\&
   \hat{F}_4=\frac{8 L^4 \rho ^3 \alpha '^{3/2} \sin \alpha  \cos ^3\alpha  \sin
   \chi  \mathit{d}\alpha \wedge \mathit{d}\theta \wedge \mathit{d}\xi \wedge
   \mathit{d}\chi }{\rho ^2 \alpha '^2+L^4 \cos ^4\alpha },\nn \\&
   \hat{F}_6=\frac{2\sqrt{\alpha'}\rho d\rho\wedge d\text{Vol}(AdS_5)}{L},\nn \\&
   \hat{F}_8=\frac{2 L^3 \rho ^2  \alpha '^{3/2} \cos ^4\alpha  \sin \chi
   \mathit{d}\xi \wedge \mathit{d}\rho \wedge \mathit{d}\chi \wedge d\text{Vol}(AdS_5)}{\rho ^2
\alpha '^2+L^4 \cos ^4\alpha },
\end{align}
satisfying $\star \hat{F}_2=\hat{F}_8,\ \ \star\hat{F}_4=-\hat{F}_6$.  Unless stated otherwise the RR p-forms quoted in the text are those that appear in the equations of motions: $\hat{F}_4=dC_3 -C_1\wedge H_3$.

Let us now move to present the results for the minimally SUSY background
$AdS_5\times T^{1,1}$.
\subsection{$AdS_5\times T^{1,1}$}
In this section we discuss the background originally
presented by Klebanov and Witten \cite{Klebanov:1998hh}.
The field theory dual to this supergravity background has played
an important role in the understanding of the AdS/CFT correspondence
beyond the maximally supersymmetric context,
see \cite{Herzog:2001xk} for a review.

\noindent
The metric is given by,
\begin{eqnarray}
ds^2&=&ds^2(AdS_5)+L^2ds_{T^{1,1}}^2, \nonumber \\
ds_{T^{1,1}}^2&=&\lambda_1^2(\sigma_{\hat{1}}^2+
\sigma_{\hat{2}}^2)+\lambda_2^2(\sigma_{1}^2+\sigma_{2}^2)+
\lambda^2(\sigma_3+\text{cos}\theta_1d\phi_1)^2,
\label{metrict11xx}\end{eqnarray}
where $\lambda^2=\frac{1}{9},\ \lambda_1^2=\lambda_2^2=\frac{1}{6}$ and
\begin{eqnarray}
\sigma_{\hat{1}}&=&\text{sin}\theta_1d\phi_1, \qquad   \sigma_{\hat{2}}=d\theta_1,  \nonumber \\
\sigma_1&=&\text{cos}\psi\ \text{sin}\theta_2d\phi_2-\text{sin}\psi\ d\theta_2, \quad \sigma_2=\text{sin}\psi\ \text{sin}\theta_2\ d\phi_2+\text{cos}\psi\ d\theta_2, \nonumber \\
\sigma_3&=&d\psi+\text{cos}\theta_2\ d\phi_2.
\end{eqnarray}
The metric of $AdS_5$ was written in Eq.(\ref{Ads5linenew}).
The full background includes a self-dual RR five-form,
\be
F_5=\frac{4}{L}(d\text{Vol}(AdS_5)-d\text{Vol}(T^{1,1})).
\ee
We use the normalization
\be
\frac{1}{(4\pi \alpha')^2}\int_{T^{1,1}}F_5=N_{D3}.
\ee
The Einstein equations of motion then lead to
\be
L^4= \frac{27}{4}\pi  \alpha'^2 \,\,N_{D3}.
\ee
Let us study the action of NATD on one of the $SU(2)$ isometries
displayed by the background in Eq.(\ref{metrict11xx}).
\subsubsection{NATD of $AdS_5\times T^{1,1}$}\label{Sec:NATD-T11}
We now consider the NATD of the Klebanov-Witten background.
The result was originally presented in \cite{Itsios:2013wd}, \cite{Itsios:2012zv}. Unlike in those works, we will
choose a gauge where $(v_1,v_2,v_3)\to(\rho,\chi,\xi)$. As above,
the Lagrange multipliers will be
 written in spherical polar coordinates with
the angles varying as,
$0\le \chi \le \pi,\;\;0\le \xi\le 2\pi$.
We start by presenting the expressions for the NS fields,
\bea
 d\hat{s}^2&=&\frac{r^2}{L^2}dx_{1,3}^2+ \frac{L^2}{r^2}dr^2
 + L^2\lambda_1^2 (d\theta_1^2 +\sin^2\theta_1 d\phi_1^2)\nn \\
&~&+\frac{\alpha '^2 }{QL^2} \bigg[(\lambda _2^4 L^4 (\mathit{d}\rho  \cos \chi
   -\rho  \mathit{d}\chi  \sin \chi )^2+\lambda ^2 \lambda _2^2 L^4 \left(\rho
   ^2 \mathit{d}\chi ^2 \cos ^2\chi +\rho  \mathit{d}\rho  \mathit{d}\chi
   \sin 2 \chi \right. \nn \\
        &~& \left. +\sin ^2\chi  \left(\mathit{d}\rho ^2+\rho ^2
   \left(\mathit{d}\xi +\mathit{d}\phi _1 \cos
  \theta _1\right){}^2\right)\right)+ \rho^2 \alpha'{}^2\mathit{d}\rho ^2\bigg],\nn \\[2mm]
 QL^2  \hat{B}_2&=&\frac{1}{2} \rho ^2 \alpha '^3 \sin \chi  \left(\left(\lambda
   ^2-\lambda _2^2\right) \sin 2 \chi  \mathit{d}\xi \wedge \mathit{d}\rho +2
   \rho  N\mathit{d}\xi \wedge \mathit{d}\chi \right)\nn \\
   &~ & -\lambda ^2 \alpha ' \cos   \theta _1 \left(\cos \chi  \left(\rho ^2 \alpha '^2+\lambda
   _2^4 L^4\right)\mathit{d}\rho \wedge
   \mathit{d}\phi _1 -\lambda _2^4 L^4 \rho  \sin \chi  \mathit{d}\chi \wedge
   \mathit{d}\phi _1\right),\nn \\[2mm]
    e^{-2\hat{\Phi}}&=& \frac{QL^2}{\alpha '^3},
\label{3metric}
\eea
with,
\beq
\label{Eq:QN}
Q= \left(\lambda ^2 \lambda _2^4
   L^4+  \rho ^2 \alpha '^2 N
   \right),\quad N= ( \lambda ^2\cos ^2\chi
   +\lambda _2^2   \sin ^2\chi).
\eeq
Although the range of the coordinate $\rho$ has not been established yet,
we notice that if it were compact---
we will later argue that $0\le \rho\le \pi$--- the quantity $Q$
would then be bounded, leading to a completely smooth background.
The RR fields are given by
\bea
\hat{F}_2&=& \frac{4 \lambda  \lambda _1^2 \lambda _2^2 L^4 \sin \theta
   _1 \mathit{d}\theta _1\wedge \mathit{d}\phi
   _1}{\alpha '^{3/2}}, \nonumber\\[2mm]
 Q \hat{F}_4&=&2 \lambda  \lambda _1^2 \lambda _2^2  L^2 \rho ^2
  \alpha '^{3/2}\sin \chi \sin \theta _1\mathit{d}\theta _1\wedge \mathit{d}\phi\wedge
  \bigg[\left(\lambda ^2-\lambda _2^2\right) \sin 2 \chi  \mathit{d}\xi \wedge \mathit{d}\rho
\nn\\
        &~&+2 \rho  N \mathit{d}\xi \wedge \mathit{d}\chi
   _1\bigg],\nonumber\\[2mm]
 \hat{F}_6&=&\frac{4  \rho \sqrt{\alpha '}
  d\text{Vol($AdS_5$)}\wedge \mathit{d}\rho }{L} ,\\[2mm]
        QL \hat{F}_8&=&4  \lambda ^2 \lambda _2^4 \rho ^2  \alpha
   '^{3/2} \sin \chi
 d\text{Vol}(AdS_5)\wedge \mathit{d}\rho \wedge \mathit{d}\chi \wedge
 ( \cos \theta _1 \mathit{d}\phi _1+\mathit{d}\xi ).\nn
   \label{RRfields}
\eea
The coordinate $\xi$, plays the role of the
R-symmetry after  NATD.
Note, again, that we write the shifted $F_p$,
in particular, $\hat{F}_4=dC_3-C_1\wedge H_3$.
Let us now move into our last case study.


\subsection{$AdS_5\times Y^{p,q}$}
We will study here the action of NATD on the geometry
$AdS_5\times Y^{p,q}$.
We will follow the conventions of
\cite{Gauntlett:2004zh, Gauntlett:2004yd, Gauntlett:2004hh}.
We start by presenting the background,
\bea
 ds^2&=&ds^2(AdS_5)+L^2ds^2(Y^{p,q}),\nn \\[2mm]
ds^2(Y^{p,q})&=&\frac{1-y}{6}(\sigma_1^2+\sigma_2^2)+
\frac{1}{wy}\mathit{d}y^2+\frac{v}{9}\sigma_3^2+
w(\mathit{d}\alpha+f\sigma_3)^2,
\eea
where the $\sigma$'s were defined above, and
\bea
w=\frac{2(b-y^2)}{1-y},\quad v=\frac{b-3y^2+2y^3}{b-y^2},
\quad f=\frac{v-2y+y^2}{6(b-y^2)},
\eea
with the quantity $b$ given by,
\bea
b=\frac{1}{2}-\frac{(p^2-3q^2)}{4p^3}\sqrt{4p^2-3q^2}.
\eea
The ranges of $\alpha$ and $y$ are $0\leq \alpha\leq 2\pi l$,
$y_1\leq y\leq y_2$, where the numbers $l, y_1, y_2$ are,
\bea
&y_1=\frac{1}{4p}(2p-3q-\sqrt{4p^2-3q^2}),\\& y_2
=\frac{1}{4p}(2p+3q-\sqrt{4p^2-3q^2}),\nn \\&
l=\frac{q}{3q^2-2p^2+p\sqrt{4p^2-3q^2}}.
\eea
The self-dual RR flux is,
\be
F_5=\frac{4}{L}(d\text{Vol}(AdS_5)-d\text{Vol}(Y^{p,q})).
\ee
Again, we normalize the $F_5$
\be
\frac{1}{(4\pi \alpha')^2}\int_{S^5}F_5=N,
\ee
which leads to the relation between L and N,
$L^4=\frac{9\pi g_s N\alpha'^2}{2l(y_2^2-y_1^2+2(y_1-y_2))}$.
We now proceed to apply a NATD on the $SU(2)$ isometry parametrised
by $\sigma_i$'s.
\subsubsection{NATD of $AdS_5\times Y^{p,q}$}\label{NATD:ypq}
If we choose a gauge fixing such that we keep
all Lagrange multipliers $(v_1,\ v_2,\ v_3)$ and, as we did above,
we change coordinates to spherical polar coordinates
$(\rho,\ \chi,\ \xi)$; where
$v_1=\rho\sin\chi\cos\xi,\ v_2=\rho\sin\chi\sin\xi,\ v_3=\rho\cos\chi$.
We obtain,
\bea
d\hat{s}^2&=&ds^2(AdS_5)+\frac{L^2}{vw}\mathit{d}y^2+L^2k^2\mathit{d}\alpha^2+\frac{\alpha'^2}{\Upsilon}\bigg[6L^2\rho^2m\sin^2\chi(h\mathit{d}\alpha+\sqrt{g}\mathit{d}\xi)^2\nn \\
&~&+\left(\frac{36\alpha'^2}{L^2}\rho^2+L^2m^2\cos^2\chi\right)\mathit{d}\rho^2+L^2\rho m^2\sin\chi(\rho\sin\chi\mathit{d}\chi^2-2\cos\chi\mathit{d}\rho\mathit{d}\chi)\nn \\
&~&+6L^2gm(\sin\chi\mathit{d}\rho+\rho\cos\chi\mathit{d}\chi)^2\bigg],\nn \\[2mm]
\Upsilon\hat{B}&=&\alpha'\bigg[\sqrt{g}h\cos\chi(36\alpha'^2\rho^2+L^4m^2)\mathit{d}\alpha\wedge\mathit{d}\rho-L^4\sqrt{g}hm^2\rho\sin\chi\mathit{d}\alpha\wedge\mathit{d}\chi\nn \\
&~&+\alpha'^2\rho^2\sin\chi\big(6(6g-m)\cos\chi\sin\chi \mathit{d}\xi\wedge\mathit{d}\rho+6\rho(m\sin^2\chi+6g\cos^2\chi) \mathit{d}\xi\wedge\mathit{d}\chi\big)\bigg],\nn \\[2mm]
e^{-2\hat{\Phi}}&=&\frac{L^2}{36\alpha'^3}\Upsilon .
\eea
Where $\Upsilon=g(36\alpha'^2\rho^2\cos^2\chi+L^4m^2)+
6\alpha'^2\rho^2m\sin^2\chi$, and
\be
\label{ghkmYpq}
g=\frac{v}{9}+wf^2,\quad h=\frac{wf}{\sqrt{g}},
\quad k=\sqrt{\frac{vw}{9g}},\quad m=1-y .
\ee
The RR fields will read,
\bea
 \hat{F}_2&=&\frac{2L^4m}{9\alpha'^{3/2}}
\mathit{d}\alpha\wedge\mathit{d}y, \nn \\[2mm]
3\Upsilon\hat{F}_4&=&4L^4\alpha'^{3/2}\rho^2 m
\sin\chi\mathit{d}y\wedge\mathit{d}\alpha
\wedge\mathit{d}\xi\wedge\bigg[\cos\chi\sin\chi(m-6g)
\mathit{d}\rho\nn \\&~&-\rho(6g\cos^2\chi+m\sin^2\chi)\mathit{d}\chi\bigg].
\eea

Had we chosen a gauge such that the
Lagrange multipliers $(v_1,\ v_2,\ v_3)$ are changed to cylindrical
polar coordinates
$(\rho,\ \xi,\ x)$ with, $v_1=\rho\sin\xi,\ v_2=\rho\cos\xi$, we would have
obtained the recent result of \cite{Sfetsos:2014tza}.
This completes the presentation of the three backgrounds and their NATD's.
Below we will work with these new IIA backgrounds.
Let us close the section with some general comments.
First of all (as it is obvious), we have checked that all
equations of motion (Einstein, Dilaton, Maxwell and Bianchi) are satisfied\footnote{This check is actually superfluous for the NATD of $AdS_5\times S^5$, as it is proved that all equations of motion and Bianchi identities are implied when one dualises any solution with an $SO(4)$ isometry on one of its $SU(2)$ subgroups in \cite{Itsios:2012dc}.}
The generated solutions are non-singular
in the cases of the NATD of $AdS_5\times T^{1,1}$
and $AdS_5\times Y^{p,q}$.

Finally, as  shown in \cite{Sfetsos:2010uq}, the lift to M-theory
of the solution described around
Eq.(\ref{NATDads5s5}) gives an eleven-dimensional background quite similar
to the Gaiotto-Maldacena geometries \cite{Gaiotto:2009gz}.
The M-theory lift of the geometries presented
in Section \ref{Sec:NATD-T11} is in the same way,
resemblant of the ${\cal N}=1$ version
of the Gaiotto $T_{N}$ theories (as shown in
\cite{Itsios:2012zv, Itsios:2013wd}). Both dual field
theories can be thought to be connected by an RG flow induced by
 a relevant operator.
It would be interesting to understand
if there is a way to connect these with the
solution presented in Section \ref{NATD:ypq}.
It would also be of interest to connect the Type IIA geometries
with the solutions discussed by \cite{ReidEdwards:2010qs} and
\cite{Aharony:2012tz}.

\section{M-theory lifts of NATD of Freund-Rubin solutions }
\label{Sec:mtheory}

In this section we consider the M-theory lift of the
solutions generated in the previous section.
Given a IIA background composed of a metric
in string frame $ds_{10}^2$  and with potentials $C_{(1)}, B_{(2)}$ and $C_{(3)}$ one constructs the corresponding M-theory solution composed of
$ds_{11}^2$ and $C_{(3)}^M$ as follows \cite{Cvetic:1999zs}:
\bea
ds_{11}^2&=& e^{-\frac{2}{3}\hat{\Phi}}ds_{10}^2 + e^{\frac{4}{3}\hat{\Phi}}(dy +C_{(1)})^2, \nonumber \\[2mm]
C^{M}_{(3)}&=& C^{IIA}_{(3)}+B_{(2)}\wedge dy, \qquad  {\rm or} \qquad
F^{M}_{(4)}=F^{IIA}_{(4)}+H_{(3)}\wedge dy, \nonumber
\eea
Note that $F_{(4)}^{IIA}$ differs from the value quoted in
previous sections precisely because it is the closed part
of $F_{(4)}^{IIA}=dC_3$. In the previous section
we wrote $\hat{F}_4=F_{(4)}^{IIA}-C_1\wedge H_3$ which is different from the one needed in this section.

One important aspect of any M-theory lifting is the fate of the M-theory circle as it geometrizes the coupling in the IIA frame.  As can be seen above, the radius is proportional to $e^{4\hat{\Phi}/3}$. Since we are dealing with dimensionful quantities, we must introduce the 11D $y$ coordinate with a length scale.  In this section we recall that the natural scale is the 11-d Planck length, $l_{P}$, which is related to the string theory scale as $l_P= g_s^{1/3} \sqrt{\alpha'}$.

The history of solutions in M-theory containing an $AdS_5$ factor dates back
to more than a decade ago starting with the work \cite{Maldacena:2000mw}.
Interpretations in the context of wrapped M5 branes were subsequently
systematically studied in \cite{Gauntlett:2006ux}.
More recently, attention to this type of solutions has surged
as potential gravity dual to Giaotto's theories
as discussed in \cite{Gaiotto:2009gz}. Other aspects of these
solutions including their origins as holographic RG flows has
recently been presented in \cite{Bah:2012dg}.
A more systematic approach to the construction of wrapped M5-branes
with an $AdS_5$ factor in M-theory has been presented recently in
\cite{Bah:2013qya}.
We should also mention the works
\cite{Cucu:2003bm} and \cite{Fayyazuddin:1999zu}--though the comparison
with our backgrounds is difficult.

We hope that some of the solutions we present in this section
might ultimately find a place in this bigger picture.

\subsection{M-theory lift of NATD of $AdS_5 \times S^5$}
Let us first consider the M-theory lift of the solution obtained by applying NATD to $AdS_5 \times S^5$.
The resulting M-theory background is \cite{Sfetsos:2010uq},
\bea
d\hat{s}^2_{11}&=&e^{-\frac{2}{3}\hat{\Phi}}ds^2(AdS_5)+e^{\frac{4}{3}\hat{\Phi}} (dy-2\frac{L^4\cos^4 \alpha}{\alpha'^{3/2}}d\theta)^2\nn \\
&~&+e^{-\frac{2}{3}\hat{\Phi}}\left(4L^2(d\alpha^2+\sin^2\alpha d\theta^2)
+\frac{\alpha '^2 d\rho^2 }{L^2\cos^2\alpha }
+ \frac{\alpha '^2 L^2 \rho ^2 \cos ^2\alpha
   \left(\mathit{d}\xi ^2 \sin ^2\chi +\mathit{d}\chi^2\right)}{\alpha '^2 \rho ^2+L^4 \cos ^4\alpha }\right)\nn\\[2mm]
B_2&=&\frac{\alpha '^3 \rho ^3 \sin \chi  \mathit{d}\xi \wedge \mathit{d}\chi }{\alpha'^2 \rho ^2+L^4 \cos ^4\alpha },  \quad
   e^{-2\hat{\Phi}}=L^2\cos^2\alpha\left(\frac{L^4 \cos ^4\alpha +\alpha'^2  \rho^2 }{\alpha '^3}\right), \nn\\[2mm]
F^{M}_{(4)}&=&F^{IIA}_{(4)}+H_{(3)}\wedge dy, \quad H_{(3)}= dB_2\nonumber \\[2mm]
F_{(4)}^{IIA}&=&\frac{2 L \alpha'{}^{3/2} \rho ^2 \sin\chi }{\left(\alpha'{}^2 \rho ^2+L^4 \cos^4 \alpha\right)^2}\left(4 L^3\alpha'{}^{3/2} \rho  \left(L+\alpha'^{1/2}\rho ^2\right) \cos^3 \alpha \sin \alpha \mathit{d}\alpha \wedge \mathit{d}\theta \wedge \mathit{d}\xi \wedge \mathit{d}\chi \right.\nn\\
&~&+\left. \left(\alpha'{}^{3/2}-L^3 \cos^4\alpha\right) \left(\alpha'{}^2 \rho ^2+3L^4 \cos^4\alpha \right) \mathit{d}\theta \wedge \mathit{d}\xi \wedge \mathit{d}\rho \wedge \mathit{d}\chi \right)
\label{MNATDads5s5}
\eea

The M-theory radius in this case,  $e^{4\hat{\Phi}/3}$ above, behaves like a trumpet starting out at non-vanishing size and blowing up when $\cos\alpha=0$. This singularity was already present in the IIA picture where it manifested itself as a curvature singularity.

\subsection{M-theory lift of NATD of $AdS_5 \times T^{1,1}$}
In this subsection we present the M-theory lift of the background obtained from applying NATD to $AdS_5\times T^{1,1}$; the result is
\bea
 \hat{ds}^2&=&e^{-\frac{2}{3}\hat{\Phi}}\bigg[\frac{r^2}{L^2}dx_{1,3}^2+ \frac{L^2}{r^2}dr^2
 + L^2\lambda_1^2 (d\theta_1^2 +\sin^2\theta_1 d\phi_1^2) \nn \\
 &~&+\frac{\alpha '^2 }{QL^2} \bigg(\lambda _2^4 L^4 (\mathit{d}\rho  \cos \chi
   -\rho  \mathit{d}\chi  \sin \chi )^2+\lambda ^2 \lambda _2^2 L^4 \big(\rho
   ^2 \mathit{d}\chi ^2 \cos ^2\chi +\rho  \mathit{d}\rho  \mathit{d}\chi
   \sin 2 \chi \nn \\
        &~&  +\sin ^2\chi  (\mathit{d}\rho ^2+\rho ^2
   (\mathit{d}\xi +\mathit{d}\phi _1 \cos
  \theta _1){}^2)\big)+ \rho ^2 \alpha'^2 \mathit{d}\rho ^2\bigg)\bigg]\nn \\
   &~&+e^{\frac{4}{3}\hat{\Phi}}\left(dy-\frac{4\lambda\lambda_1^2 \lambda_2^2 L^4}{\alpha'^{3/2}} \cos\theta_1 d\phi_1\right)^2 \nn, \\[2mm]
QL^2  \hat{B}_2&=&\frac{1}{2} \rho ^2 \alpha '^3 \sin \chi  \left(\left(\lambda
   ^2-\lambda _2^2\right) \sin 2 \chi  \mathit{d}\xi \wedge \mathit{d}\rho +2
   \rho  N\mathit{d}\xi \wedge \mathit{d}\chi \right)\nn \\
   &~& -\lambda ^2 \alpha ' \cos   \theta _1 \left(\cos \chi  \left(\rho ^2 \alpha '^2+\lambda
   _2^4 L^4\right)\mathit{d}\rho \wedge
   \mathit{d}\phi _1 -\lambda _2^4 L^4 \rho  \sin \chi  \mathit{d}\chi \wedge
   \mathit{d}\phi _1\right),\nn \\[2mm]
    e^{-2\hat{\Phi}}&=& \frac{QL^2}{\alpha '^3}.
                \eea
                The fluxes are:
\bea
  F^{M}_{(4)}&=&F^{IIA}_{(4)}+H_{(3)}\wedge dy, \quad H_{(3)}= dB_2\nonumber \\[2mm]
Q F_{(4)}^{IIA}&=&
4 L^4 \alpha'{}^{3/2} \lambda  \rho ^2 \cos\chi \sin\theta_1  \sin^2\chi \lambda _1^2 \lambda _2^2 \left(\lambda ^2-\lambda _2^2\right)  \mathit{d}\theta_1\wedge \mathit{d}\xi\wedge \mathit{d}\rho \wedge \mathit{d}\phi_1\nonumber \\
&~&-4L^4 \alpha'{}^{3/2} \lambda  \rho ^3 \sin\theta_1 \sin\chi \lambda _1^2 \lambda _2^2 \left(\lambda ^2 \cos^2\chi+\sin^2\chi \lambda _2^2\right)\mathit{d}\theta_1\wedge \mathit{d}\xi \wedge \mathit{d}\phi_1\wedge \mathit{d}\chi \nonumber \\
&~&-2 L^4 \alpha'{}^{3/2} \lambda^3 \rho ^2 \cos\theta_1 \sin\chi \lambda _1^2 \lambda _2^2\mathit{d}\xi \wedge \mathit{d}\rho \wedge \mathit{d}\phi_1\wedge \mathit{d}\chi
\bigg[-2 \alpha'{}^2 \lambda^2 \rho^2 \cos^2\chi  \nonumber \\
&~&+ \alpha'{}^2 \rho ^2 (3+\cos 2 \chi) \lambda _2^2+2 L^4 \lambda _2^4(\lambda ^2+2\lambda _2^2)\bigg]
\eea
Let us pay particular attention to the M-theory radius
\be
R_{11}=e^{\frac{2}{3}\hat{\Phi}} \sim \left(\lambda^2 \lambda_2^4+ \rho^2 \alpha'^2 (\lambda^2 \cos^2\chi + \lambda_2^2 \sin^2\chi\right)^{-1/3}.
\ee
One really remarkable aspect of this solution is the fact that the M-theory radius is bounded above and below. The means that the solutions is a completely smooth 11d supergravity background. It would be interesting to study this background in more detail and its field theory dual.

\subsection{M-theory lift of NATD of $AdS_5 \times Y^{p,q}$}
In this subsection we denote $\tilde{y}$ the M-theory circle to
avoid confusion with the $y$-coordinate  originally defined in $Y^{p,q}$.
The M-theory lifts reads \cite{Sfetsos:2014tza},
\bea
ds_{11}^2&=& e^{-\frac{2}{3}\hat{\Phi}}ds_{10}^2 + e^{\frac{4}{3}\hat{\Phi}}\left(d\tilde{y}-\frac{2L^4}{9\alpha'{}^{3/2}} y d\alpha\right)^2 , \nonumber \\[2mm]
ds^2_{10}&=&ds^2(AdS_5)+\frac{L^2}{vw}\mathit{d}y^2+\frac{1}{\Sigma}(\frac{\alpha'^2}{L^2}
(36x^2\alpha'^2+L^4m^2)\mathit{d}x^2+L^2(6\alpha'^2\rho^2h^2+\Sigma k^2)\mathit{d}\alpha^2\nn \\& &-12L^2\alpha'^2\rho^2\sqrt{g}hm\mathit{d}\alpha\mathit{d}\xi+\frac{6\alpha'^2}{L^2}
(6\alpha'^2\rho^2\mathit{d}\rho^2+L^4gm(\rho^2\mathit{d}\xi^2+\mathit{d}\rho^2)
+\frac{12\alpha'^2}{L^2}x\rho\mathit{d}x\mathit{d}\rho)),\nn \\[2mm]
B_2&=&\frac{\alpha'}{\Sigma}(6\alpha'^2\rho^2m\mathit{d}\xi\wedge\mathit{d}x
+\sqrt{g}h((36x^2\alpha'^2+L^4m^2)\mathit{d}\alpha
\wedge\mathit{d}x+36x\alpha'^2\rho\mathit{d}\alpha\wedge \mathit{d}\rho)\nn \\
&~&~~+36x\alpha'^2\rho g\mathit{d}\rho\wedge\mathit{d}\xi),\nn\\[2mm]
e^{-2\hat{\Phi}}&=&\frac{L^2}{36\alpha'^3}\Sigma
\eea
where $\Sigma=6\alpha'^2\rho^2m+g(36\alpha'^2x^2+L^4m^2)$. The 4-form field strength is given by
\bea
F^{M}_{(4)}&=&F^{IIA}_{(4)}+H_{(3)}\wedge dy, \quad H_{(3)}= dB_2\nonumber \\[2mm]
3 F^{IIA}_{(4)}&=&
-\frac{4 L^4 \alpha'{}^{3/2} \rho ^2 g m^2 \sin\chi\mathit{d}\alpha \wedge \mathit{d}\xi \wedge \mathit{d}\rho\wedge \mathit{d}\chi}{3 \left(g \left(36 \alpha'{}^2 \rho ^2 \cos^2\chi+L^4 m^2\right)+6 \alpha'{}^2 \rho ^2 m \sin^2\chi\right)^2}\left(9 \alpha'{}^2 \rho^2 (3+\cos 2\chi) m\right. \nn\\
&~&+\left.L^4m^3+3 gg\left(-36 \alpha'{}^2 \rho ^2 \cos^2\chi +L^4 m^2\right)\right)\nn\\
&~&+\frac{2 L^4 \alpha'{}^{3/2} \rho ^3 m \sin\chi\mathit{d}y\wedge \mathit{d}\alpha\wedge \mathit{d}\xi \wedge \mathit{d}\chi}
{\left(3 \left(g \left(36 \alpha'{}^2 \rho ^2 \cos^2\chi +L^4 m^2\right)+6 \alpha'{}^2 \rho^2 m \sin^2 \chi\right)^2\right)}
\left(432 \alpha'{}^2 \rho ^2 \cos^4\chi g^2 \right.\nn\\
&~&+\left. 12 \alpha'{}^2 \rho ^2 m^2 \sin^4\chi+g\left(L^4m^3 \sin^2\chi +36 \alpha'{}^2 \rho ^2 m \sin^2 2 \chi \right)+L^4 m^4 g'\sin^2\chi\right) \nn\\
&~&-\frac{2 L^4 \alpha'{}^{3/2} \rho ^2 \cos\chi m \sin^2\chi \mathit{d}y\wedge \mathit{d}\alpha \wedge \mathit{d}\xi \wedge \mathit{d}\rho }{3 \left(g \left(36 \alpha'{}^2 \rho ^2 \cos^2\chi+L^4 m^2\right)+6 \alpha'{}^2 \rho ^2 m \sin^2\chi \right)^2}\left(-432 \alpha'{}^2 \rho ^2 \cos^2\chi g^2\right. \nn\\
&~&+g m \left(36 \alpha'{}^2 \rho ^2 (1+2 \cos 2 \chi)+L^4 m^2\right) \left. +m^2 \left(12 \alpha'{}^2 \rho ^2 \sin^2\chi+g'\left(36 \alpha'{}^2 \rho ^2+L^4 m^2\right)\right)\right) \nn
\eea
One important aspect of this background is its M-theory radius, given roughly by $\Sigma^{-1/3}$; this quantity is well-behaved leading to a potential interpretation in the context of M-theory.
The M-theory backgrounds presented in this section have been previously presented in
\cite{Sfetsos:2010uq, Itsios:2013wd, Sfetsos:2014tza}.

\section{Page charges in gravity and field theory central charge }
\label{Sec:Page}
In this Section,
we will study a proposal to determine the range
of the $\rho$-coordinate after NATD.
Also, we will discuss two quantities
of physical importance, Page charges and central charge.
We will apply our results to the three backgrounds
in Section \ref{Sec:NATD-IIB}
\subsection{Global properties and Page charges}
Non-Abelian T-duality,
as well as regular Abelian T-duality,  is an
intrinsically local transformation. As mentioned,
we have checked explicitly in all the backgrounds presented in this
paper that the equations of motion are satisfied but we have no rule or intuition for determining the range of coordinates in the backgrounds.

In this subsection we discuss
a global issue that has haunted non-Abelian
T-duality for some time (see \cite{Alvarez:1993qi} for
early studies and \cite{Lozano:2013oma,Lozano:2014ata}
for recent discussions).
One of the difficulties with the interpretation
of the backgrounds is the lack of knowledge of the
range of the coordinates after  NATD.
The prescription we adopt  is {\it the same} as the one presented in
\cite{Lozano:2014ata}, but we are applying  it in a
background without singularities,
making the procedure more trustable \footnote{We thank
Yolanda Lozano for various discussions on this point.}. Hence this is another example of the prescription provided in \cite{Lozano:2014ata}.

We impose  bounds on the integral
via, $4\pi^2 \alpha' b_0=\int_{\Sigma_2} B_2 \in [0,1]$,
where $\Sigma_2$ is a suitably chosen two-manifold.
As we will discuss in our examples, this condition will imply
a bound on the range of the `radial'
coordinate $\rho$.  The  internal
space after the NATD will be compact.
When restricting the $B_2$-field to the manifold $\Sigma_2$ and computing
the quantity $b_0$, it will be periodically
identified as $b_0\sim b_0 +n$ when
we perform a large gauge transformation
\beq
B_2\to B_2 +n \pi \alpha' \Omega_2 .
\label{largegauge}
\eeq
Here $\Omega_2$ is a closed two-form non-vanishing asymptotically.
This condition will imply
that $\rho$ varies in $[n \pi, \pi(n+1)]$.
The range of the radial coordinate
is `quantised'.
Let us present various motivations for this condition.
\begin{itemize}
\item {String theory has the power to quantize certain symmetries, while
supergravity generically lacks such power.
The prototypical example is $SL(2,\mathbb{Z})$
versus $SL(2,\mathbb{R})$.
We are using this when imposing that $b_0\sim b_0+1$.}
\item{The condition $\frac{1}{4\pi^2\alpha'}
\int_{\Sigma_2}B_2\in (0,1)$ comes also
from the quantization of the string action,
$\exp\left(\frac{i}{4\pi^2\alpha'}\int_{\Sigma_2}B_2\right)$, as part of the string path integral.
This is similar to what happens
in quantum mechanics when coupling particles to
 a gauge field $A_\mu$. }
\item{ For the case of $AdS_5\times T^{1,1}$, in
the dual field theory, this condition is typically
related to a linear combination of gauge couplings.
Therefore, we are imposing that they remain well defined
under certain transformations of the rank of various gauge groups.}
\end{itemize}
We could, ultimately, disregard the string-theoretic motivations and accept the prescription as a way of completing the supergravity background. Note, and this is crucial, that we do not require a stringy object to form part of our background which remains strictly a supergravity one; we merely use string intuition to propose a way to impose global information on the local solution provided by NATD.

To gain further intuition into the
implications of this condition and to relate it to the Page charges of the
background, we can compare
it with a somewhat similar situation taking place in
the {\it cascade} of the Klebanov-Tseytlin-Strassler  system
\cite{Klebanov:2000nc,Klebanov:2000hb,Strassler:2005qs}\footnote{In the cases analyzed in this paper,
 we do not have a flow in energies, we are moving in the $\rho$ direction.
We are proposing that motions in the
$\rho$-coordinate correspond to Seiberg dualities between CFT's, all equivalent
to each other. We will return to some aspects of the field theory dual in section \ref{Sec:FT}.
}.
Let us recall some aspects of
the Klebanov-Tseytlin-Strassler
\cite{Klebanov:2000nc,Klebanov:2000hb} pair, a
quiver field theory with gauge group
$SU(kM)\times SU(kM+M)$ and its dual Type IIB solution.
In that background, one computes the Page charges for D3 and D5 branes and obtains,
\beq
Q_{P,D3}=0,\;\;\; Q_{P, D5}= M.
\label{yyx}\eeq
Restricting the $B_2$ field to a two-cycle and
performing a large gauge transformation of the $B_2$ field, one
obtains \cite{Benini:2007gx}
the effect equivalent to a
Seiberg duality when  flowing to the IR. The change in the Page charges is:
\beq
\Delta Q_{P, D5}=0, \;\;\;\; \Delta Q_{P, D3}=-M.
\label{xxy}\eeq
One of the roles of the Page charges in the context of the AdS/CFT
is to encode the information about Seiberg dualities.

In summary, we will impose that under large gauge
transformations of the $B_2$-field,
$b_0\sim b_0+1$. This will imply the quantisation of changes in
the $\rho$-coordinate. The Page charges will
change accordingly, hence suggesting a form of Seiberg duality
in our conformal field theories.
We will study this in each of the backgrounds presented in
Section \ref{Sec:NATD-IIB}. It is worth pointing out that
a sigma model approach to this problem was pursued
recently in \cite{Sfetsos:2014cea}, where it was shown for a particular
example that coordinates after NATD are indeed compact.

It is worthwhile to mention another viewpoint
one can adopt. We may think that translations in the $\rho$-coordinate
that increase $b_0\to b_0+n$ can be 'undone' by a large gauge transformation
of the $B_2$-field. This mixing between metric and B-field
points to a possible understanding in terms of non-geometric backgrounds
\cite{Berman:2013eva}.

\subsubsection{Page charges for NATD of $AdS_5\times S^5$}
In the following we will use,
\bea
2\kappa_{10}^2= (2\pi)^7 \alpha'^4, \;\;\; T_{Dp}=\frac{1}{(2\pi)^p
\alpha'^{\frac{p+1}{2}}}.\nonumber
\eea
For the NATD dual of $AdS_5\times S^5$ described in
Section \ref{NATD AdS5xS5},
we compute the Page charges using the definitions,
\bea
& &
Q_{P, D6}=\frac{1}{2\kappa_{10}^2T_{D6} }\int_{\Sigma_2}
(F_2-B_2 F_0)= N_{D6},\nonumber\\
& &
Q_{P, D4}=\frac{1}{2\kappa_{10}^2T_{D4}}\int_{\Sigma_4}(F_4- B_2\wedge F_2)=
N_{D4}=0.
\label{pagesxx}
\eea
Since $F_0=0$, the Page charge
involving quantization of the number of D6-branes, $N_{D6}$, amounts
to quantizing the $F_2$, leading to a constraint
from the SUGRA equations that determines the radius of the space
after NATD. Flux quantisation imposes, after NATD, the relation
\beq
L^4=\frac{1}{2}N_{D6}\alpha'^2 \Rightarrow Q_{D6}=N_{D6}.
\eeq
We then consider the  $S^2$ spanned by $\chi,\xi$ as this is where $B_2$ has legs. If we  further  restrict to $\alpha=\frac{\pi}{2}$ the NS two form reduces to \beq
B_2= \alpha' \rho Vol(S^2).
\eeq
If we examine the space spanned by $(\alpha,\chi,\xi)$ close to $\alpha= \pi/2$ we find that is conformally a singular cone with boundary $S^2$. This is reminiscent of what was found for the NATD of $AdS_4\times \mathbb{CP}^3$ in \cite{Lozano:2014ata} and shows that $S^2$ is indeed a cycle. Now  consider a large gauge transformation in $B_2$ of the form:
\beq
B_2\to B_2 + n\pi \alpha' \sin\chi d\xi\wedge d\chi,
\label{largexxx}\eeq
and we calculate the change in the Page charges to be,
\bea
 \Delta Q_{ D6}=0,\;\;\;
 \Delta Q_{ D4}=- n N_{D6},\nonumber
\eea
where we have used the relation $L^4=\frac{1}{2}N_{D6}\alpha'^2$ found above.
Hence, a large gauge transformation leaves untouched the
number of D6's charge, but changes the charge associated with D4-branes.
Note the striking similarity with the case of the KS-cascade, summarized around
Eqs. (\ref{yyx})-(\ref{xxy}), in the case $n=1$ for the gauge transformation
in Eq.(\ref{largexxx}).

Integrating $B_2$ with $\alpha=\frac{\pi}{2}$ gives,
\beq
b_0=\frac{1}{4\pi^2\alpha'}\int_{S^2}B_2=\frac{\rho}{\pi}.
\eeq
The periodic identification of $b_0$ would imply that the dual QFT
should be identified as we change  $\rho\sim \rho+\pi$.
Coming back to the ideas exposed above, we could associate
translations in the $\rho$-coordinate with an operation
similar to Seiberg duality--- in the conformal case for the example at hand.
It may seem strange the presence of dualities between CFTs
that change the CFT and some of its observables.
As a toy example we may think of ${\cal N}=1$ SQCD for $N_f=2N_c+\epsilon$.
the theory is not self dual, both electric and magnetic theories are conformal
in the IR.

We will study below similar structures in the case of $AdS_5\times T^{1,1}$
and $AdS_5\times Y^{p,q}$.

\subsubsection{Page charges for NATD of $AdS_5\times T^{1,1}$ }
Let us apply the previous analysis of Page charges and their changes to our background.
The Page charge for D6 and D4 branes and their respective changes
under a large gauge transformation in
the $B_2$-field, as indicated in Eq.(\ref{largegauge}).
For the NATD applied to the Klebanov-Witten background, we have,
\bea
& & F_2-F_0 B_2= \frac{4L^4 \lambda_1^4 \lambda}{g_s \alpha'^{3/2}}
\sin\theta_1d\theta_1\wedge d\phi_1, \quad  F_4=B_2\wedge F_2.
\eea
This result leads to the following
values for the Page charges,
\bea
Q_{P, D6}=N_{D6},\;\;\;
Q_{P, D4}=0.
\eea
Just like in the case of $S^5$, imposing the quantization of the
Page charge of D6-branes implies for the radius of the space
\beq
L^4=\frac{27}{2} N_{D6}\alpha'^2.
\eeq

We will choose a two-submanifold
\footnote{It would be interesting to have a criterium to select this
particular manifold, like the one discussed above Eq.(\ref{largexxx})--
see \cite{Lozano:2014ata}. }, $\Sigma_2=[\phi_1=2\pi-\xi, \chi]$,
for constant values
of $\rho$ and $\theta_1=0$. We evaluate  the $B_2$ field on the sub-manifold, i.e
\beq
B_2|_{\Sigma_2}= \alpha' \rho \sin\chi d\chi\wedge d\xi.
\eeq
Like above, we perform a large gauge transformation,
\beq
\Delta B_2= - n\pi \alpha' \sin\chi d\chi\wedge d\xi.
\eeq
Using this, we find that under large gauge transformations
of the $B_2$-field, the Page charges change according to,
\bea
\Delta Q_{P, D6}=0,\;\;\;\Delta Q_{P, D4}=- n N_{D6}.
\eea
Here again, large gauge transformations are linked
with an operation similar to  Seiberg duality
in the conformal field theory dual to our background.

Finally, imposing the identification of the quantity $b_0$, under the same gauge transformations, we find that
\beq
b_0=\frac{1}{4\pi^2\alpha'}\int_{\Sigma_2}B_2=\frac{\rho}{\pi},
\eeq
which again suggest that the field theory
description should change---with a Seiberg duality; for example---
every time we change the coordinate $\rho\to \rho+\pi$ in the dual
background. Here again, a motion in $\rho$ could be undone by a suitable
large gauge transformation, suggesting a non-geometric
interpretation of the background.
Let us briefly summarize the same calculations for
the case of the NATD of $AdS_5\times Y^{p,q}$.

\subsubsection{Page charges for NATD of $AdS_5\times Y^{p,q}$}
We will be quite brief here, as the results are very similar to those
discussed above.
In this case, the two cycle of interest
is $\Sigma_2=[y,\alpha]$ with $\alpha=2\pi-\xi$.
As noted in \cite{Sfetsos:2014tza}, we must also take $y=y_0$,
where $y_0$ is a solution of $h=\sqrt{g}$, with $h$ and $g$ functions
defined in Eq.(\ref{ghkmYpq}).

The quantization of the Page charge of D6 branes implies a relation,
\beq
L^4=\frac{9}{2l(y_2^2-y_1^2+2(y_1-y_2))}N_{D6}\alpha'^2
\Rightarrow Q_{D6}=N_{D6},\;\;
Q_{D4}=0.
\eeq
Just like above, applying a large gauge transformation, like the one of Eq.(\ref{largexxx}), implies that
\beq
\Delta Q_{D6}=0,\;\;\;\Delta Q_{D4}=-n N_{D6}.
\eeq
Similarly, periodic identification on $b_0$ implies that the $\rho$-coordinate
is divided in 'domains' and should be identified
$\rho\sim \rho+\pi$.

The structure we have identified in these three examples is quite similar. The Page charge of D6-branes is quantized. The Page charge of D4-branes is zero. It is possible to identify a two dimensional submanifold,
where the $B_2$ field takes a simple form. Large gauge transformations change the Page charge of D4 branes
in a multiple of the original D6 charge. This
suggests a form of duality between CFTs when moving in $\rho$.
Besides, imposing
the identification $b_0\sim b_0+1$, the characteristic of large gauge transformations also implies that the $\rho$-coordinate is also
identified\footnote{We are not saying that the background is periodic in
the coordinate $\rho$,
but that every time that we pass the position $\rho=n\pi$ in
the String Theory
we should change the CFT description.} $\rho\sim \rho+\pi$.
This also suggest that under changes in the $\rho$ direction,
the field theory undergoes a form of Seiberg-like transformation. The
bounds on the $\rho$ coordinate are quite welcomed.
Indeed, a KK reduction to five dimensions would lead to
a continuous spectrum of operators if $\rho$ had infinite range.

We will now move to the study of another observable, quite important in the understanding of the dual conformal field theory; the central charge. Relations with the Entanglement Entropy will also be discussed.
\subsection{On central charge and entanglement entropy}
The fate of some  physical observables under solution generating
techniques is an important question. Quantities such as temperature and
entropy have been shown to be frame-invariant under solution
generating techniques applied to black holes and black
branes \cite{Peet:2000hn}. In the context of the AdS/CFT, where geometric
backgrounds encode defining properties of the field theory dual,
the question of invariance of observables under frame changing
transformations becomes an important
one. In this section we will focus on the behavior of the central charge and
we will also comment on
the expressions defining the holographic entanglement entropy
under Abelian and Non-Abelian T-dualities.

The general prescription for the calculation of the field theory central charge was
introduced by Henningson and Skenderis in
\cite{Henningson:1998gx}. More directly related to our context are
\cite{Gauntlett:2005ww},\cite{Gabella:2009ni} and
the pedagogically lucid account of \cite{Ramallo:2013bua}.
Although completely consistent with the various presentations
mentioned above, in this manuscript we will follow, in particular, a slightly more general analysis due to \cite{Klebanov:2007ws}. Our goal is to compute simultaneously the central charge of the dual field theory and the entanglement entropy of a slab region.

Let us first summarize briefly the treatment of \cite{Klebanov:2007ws}.
These authors considered a generic metric in type II string theory dual to
a putative QFT in ($d+1$)-dimensions. In string frame, this  reads
\beq
ds^2= a dz_{1,d}^2 + ab dr^2 +g_{ij}d\theta^i d\theta^j.
\label{metricvvv}\eeq
 In general, there is a dilaton $\Phi$ as part of the
background. The functions $a,b$ are
typically functions of the radial coordinate $r$,
but this is not necessarily the case and it will not always
be the case for us.
The paper \cite{Klebanov:2007ws}
defines,
\be
\hat{H}= e^{-4\Phi}V_{int}^2 a^{d/2}, \qquad V_{int}=\int d\theta^i \sqrt{\det{g_{ij}}}.
\label{definiciones}
\ee
With these definitions the integral defining the EE (this
is the area of an eight-manifold)
that includes the internal space, the $d$ $z$-coordinates
and where $r$ is a function of one of the $z$-coordinates (that denotes
the separation between the entangled regions).

We introduce a sensible modification to the prescription of \cite{Klebanov:2007ws}.
Namely, it may be the case that the function $a$ does depend on the
internal coordinates $\vec{\theta}^i$; this possibility was not considered in \cite{Klebanov:2007ws}. In that case, we
define
\beq
\hat{V}_{int}=\int d\vec{\theta} \sqrt{e^{-4\Phi}\det[g_{int}] a^d},
\label{hatV}\eeq
so that the function $\hat{H}$ is in general given by,
\beq
\hat{H}= \hat{V}_{int}^2.
\eeq
Then, the  central charge for a QFT in $(d+1)$ spacetime dimensions is
defined to be
\cite{Klebanov:2007ws}:
\beq
\label{Eq:CC}
c= d^d \frac{b^{d/2} \hat{H}^{(2d+1)/2}}{G_N(\hat{H}')^d}.
\eeq
where $G_N=(l_p)^{D-1}=\alpha'^{\frac{D-1}{2}}$ for D space-time dimensions.  The $G_N$ factor is needed to cancel the length dimensions in $\hat{H}$.
We now apply these definitions to the calculation
of the functions $a,b,\hat{H}$
for the different backgrounds
discussed in Section \ref{Sec:NATD-IIB}.
Importantly, we will  require  that integrals over the internal $\rho$-coordinate
after NATD   is bounded between $[0,\pi]$.
We will present details of the calculations for the $S^5$ case,
and simply report the results  in the $T^{11}$ and $Y^{p,q}$ cases.

The point we want to make with these results is that in all cases, the central charge before and after the NATD, behaves as $N_c^2$, where  $N_c$ is the number of relevant branes before and after the duality (that is $N_c =N_{D3}$ before and $N_c=N_{D6}$ after the NATD). The numerical coefficients are changed by the duality. This we interpret as a change in the field theory dual to each background.
\subsubsection{Central charge for $AdS_5\times S^5$ and its NATD}
In this case, the functions referred to in Eq.(\ref{metricvvv})
characterizing the system are,
\beq
\alpha=\frac{4r^2}{L^2},\;\;\;b=\frac{L^4}{r^4},\;\;\; d=3.
\label{mismo}
\eeq
A straightforward calculation leads to
\beq
\hat{H}= (16L)^4 \pi^6 r^6,\;\;\;\;c=\frac{32\pi^3 L^8}{\alpha'^4}= 2\pi^5N_{D3}^2
\label{zaza1}\eeq
After the duality, we have the same
$a,b,d$ as in Eq.(\ref{mismo}). The quantity $\hat{V}_{int}$
can then be computed to be\beq
\hat{V}_{int}=\int \sqrt{e^{-4\Phi} \det (g)  a^3}= \frac{64}{3}\pi^5 L^2 r^3,
\eeq
where we remind the reader that the domain of $\rho$ is $[0,\pi]$.

Using this and the relation between $L$ and $N_{D6}$, we find that
after the NATD,
\beq
\hat{H}=\frac{1}{9}(8L)^4 \pi^{10} r^6,\;\;\;\;
c=\frac{8\pi^5 L^8}{3\alpha'^4}=\frac{2}{3} \pi^5 N_{D6}^2.
\label{zaza2}\eeq
We observe the usual gauge-theoretic dependence with the number
of degrees of freedom, but we also point out that the coefficient is
different, suggesting that the dual field theory has changed
by the effect of the NATD.
\subsubsection{Central charge for $AdS_5\times T^{1,1}$ and its NATD}
In this case, before and after the duality the functions and parameter $a,b,d$ are the same as those written in Eq.(\ref{mismo}). Before the duality, we find,
\beq
\sqrt{e^{-4\hat{\Phi}}\det( g) a^3}= L^2 r^3 \lambda\lambda_1^2\lambda_2^2 \sin\theta_1\sin\theta_2,
\eeq
after straightforward operations we obtain
\beq
c_{KW}=\frac{\pi^3L^8}{27\alpha'^4}=\frac{27}{8} \pi^5 N_{D3}^2.
\label{kwzzz}\eeq
After the NATD, we find
\beq
\sqrt{e^{-4\hat{\Phi}}\det (g )a^3}=L^2 r^3 \rho^2 \lambda\lambda_1^2\lambda_2^2 \sin\theta_1
\sin\chi.
\eeq
and performing integrals and straightforward algebra, we find
\beq
c_{\text{NATDKW}}=\frac{2L^8 \pi^5 \lambda \lambda_1^2\lambda_2^2}{3\alpha'^4}= \frac{9}{8}\pi^5 N_{D6}^2.
\label{kwnatdzzz}
\eeq
Again, emphasizing on the point that the coefficient differences
between Eqs.(\ref{kwzzz}) and (\ref{kwnatdzzz}) can be understood
as NATD changing the dual QFT .
\subsubsection{Central charge for M-theory lift of $AdS_5 \times S^5$}
For the case of the M-theory lift of $AdS_5 \times S^5$, we compute the Page charge of $F_4$ found in Eq.(\ref{MNATDads5s5}),
\be
Q_{P, M5}=\frac{1}{2\kappa_{11}^2T_{M5} }\int_{\Sigma_4}
F_4 = N_{M5}, \ee
where here, $\kappa_{11}=(2\pi)^4\alpha'^{9}$ and $T_{M5}=\frac{1}{(2\pi)^5\alpha'^3}$.
We consider the submanifold, $\Sigma_4$, with $\theta$ a constant and $\alpha$ is suitably chosen after integration.  (Note that $\alpha$ cannot be set to $\frac{\pi}{2}$ or the volume will vanish.)  We also exploit an ambiguity in the uplifting procedure in which we introduce the $y$ coordinate with a scaling factor of $(\frac{L^2}{\alpha'})^{\gamma}\sqrt{\alpha'}$ (instead of just $\sqrt{\alpha'}$, or $l_P$). Then, after imposing the charge quantization above, we find that
\be
L^4 = 2^{\frac{8}{\gamma}}(N_{M5})^{\frac{2}{\gamma}} \alpha'^2.
\ee
In order to compute the central charge one needs the functions
$a = e^{-\frac{2}{3}\hat{\Phi}} L^2 \frac{R^2}{L^4}$ and $b= \frac{L^4}{R^4}$.
The determinant factor of the internal metric is,
\bea
\sqrt{\det(g_{\text{int}})a^3}&=&32 L^{2(1+\gamma)}\alpha'^{\frac{1}{2}-\gamma} R^3\rho^2\cos^3\a \sin\a\sin\chi
\eea
Then one has
\be
 V_{int}=  \frac{128}{3}\pi^6R^3 L^{2(1+\gamma)}\alpha'^{\frac{1}{2}-\gamma}. \nn
\ee
where we have assumed the range of y to be $0\leq y\leq 2\pi$.  Using $G_{11}= \a'^{\frac{9}{2}}$, we compute the central charge to therefore be,
\be
c = \frac{16\pi^6}{3} (\frac{L^2}{\a'})^{4+\gamma}.
\ee
Now using the condition on L from above we find that
\be c = \frac{2^{8(1+\frac{2}{\gamma})}}{3}  \pi^6(N_{M5})^{1+\frac{4}{\gamma}}. \ee
For $\gamma = 4$, $c\sim (N_{M5})^2, $ while for $\gamma =2$, $c\sim (N_{M5})^3.  $ It is interesting that the $\gamma=4 $ scaling leaves the central charge invariant after the lift while $\gamma=2$ takes the system into a field theory that is similar to what is expected of the Gaiotto type theories.

\subsubsection{Brief comments on Entanglement Entropy.}
The Holographic Entanglement Entropy is a very interesting
observable. It can be calculated by solving a minimization problem for
an eight-manifold that hangs from radial infinity.
There are many analogies and important differences with the calculation
for Wilson loops \cite{Kol:2014nqa}. After the usual manipulations with
a Hamiltonian system we obtain two formulas for the Entanglement
Entropy and the separation between the two entangled regions.
They can be written in terms of $r_*$, the minimal radial position of the
hanging eight-manifold. They read,
\bea
& & L_{EE}(r_*)=2\sqrt{\hat{H}(r_*)}\int_{r_*}^\infty \frac{\sqrt{\beta(r)}}
{\sqrt{\hat{H}(r)-\hat{H}(r_*)}}dr,\\
& & \frac{2G_{10}}{V_3}S_{EE}(r_*)=\int_{r_*}^{r_{UV}}
\frac{\sqrt{\beta(r)} H(r)}{\sqrt{\hat{H}(r)-\hat{H}(r_*)}}.
\label{EELxxx}
\eea
There is, as in the case of Wilson loops, a substraction procedure.
This motivated the upper limit $r_{UV}$ in the integral
defining the Entanglement Entropy.

As observed above, in all of our examples, before and after
the NATD the functions $\alpha(r), \beta(r)$ are the same.
The changes occur in the internal volume $\hat{V}_{int}$ and consequently
in $\hat{H}$.

It is clear that the dependence of $S_{EE}$ on the separation $L_{EE}$
will be the same and driven by conformal invariance. The differences will be
in coefficients appearing in the function $\hat{H}(r)$, due to differences
in the volume of the internal manifold.
Compare for example the function $\hat{H}(r)$ for the case of
$AdS_5\times S^5$, before and after the NATD, as
calculated in Eqs.(\ref{zaza1})-(\ref{zaza2}).

\subsection{Quasi frame independence of the central charge volume form}
After having considered the examples most relevant to
this manuscript, we pose the question  pertaining to the invariance
of the functional form of
quantities such as the central charge and Entanglement Entropy
under Abelian and Non-Abelian T-dualities.

The question naturally arises in the context of solution-generating
techniques applied to backgrounds describing black holes
and intersecting branes. In fact,  in this section we borrow
heavily from an analysis of Horowitz and Welch
\cite{Horowitz:1993wt}. We consider the effects of performing
Abelian and Non-Abelian T-duality on the field theory central charge.
Let us start with the definition of central charge presented
in Eq.(\ref{Eq:CC}). The key functions to focus on are,
\be
H= \hat{V}_{int}^2, \qquad \hat{V}_{int}=\int d\vec{\theta} e^{-2\Phi} \det \sqrt{g_{int}}a^{3/2}.
\ee
It is easy to show that this combination is invariant under
Abelian T-duality. Namely, recall that under  Abelian T-duality,
the B\"uscher rule imply,
\be
\label{abelianDual}
\tilde{g}_{xx}=1/g_{xx}, \quad \tilde{\phi} =\phi -\frac{1}{2}\ln g_{xx}.
\ee
Therefore,
\be
e^{-2\tilde{\phi}}\sqrt{\tilde{g}_{xx}}= e^{-2\phi + \ln g_{xx}}\sqrt{\frac{1}{g_{xx}}}=e^{-2\phi} \sqrt{g_{xx}}. \label{centralchargevolumeform}
\ee
The above argument helps us establish that the
{\it central charge volume form} is invariant under Abelian T-duality.
It still leaves us with the daunting question of  what is the
range of integration. We need to access global information in the form
of range of coordinates to be able to conclusively establish
that the central charge of the field theory is frame independent.


In the NATD case a similar argument can be constructed albeit with more complicated expressions. Note, for example that the dilaton transforms as,
\begin{equation}
\hat{\Phi}=\Phi-\frac{1}{2}\text{ln}(\frac{\text{det}M}{\alpha'^3}).
\end{equation}

The complete form of the NATD transformation is given in
Appendix \ref{App:Rules}.
One can verify that the
central charge volume form, that is,
the un-integrated expression for $d\hat{V}_{int}$
is not invariant under NATD. But there is in all cases a very
interesting cancellation between the terms $e^{-4\Phi}$ and $\det[g_{int}]$.
Indeed, this was also observed in the case in which we flow away
from the fixed point in \cite{Itsios:2013wd}.
A general proof of this fact requires
certain identities of the seed B-field which was zero in our cases.
Let us move now to a different, more geometrical aspect of our study.

\section{New solutions in IIB via NATD and T-duality}\label{Sec:NATD-T}
In the next two sections we will switch our focus a bit.
Indeed, we will move into a more geometrical
part of our paper. We will present {\it new} solutions of Type
IIB Supergravity. These solutions as we anticipated in the Introduction,
will be {\it singular}. In some of the cases discussed below, they
will be SUSY preserving, in some other cases families of solutions
will be presented, but in all cases our new solutions will
present an $AdS_5$ factor and will avoid presently known
classifications
\cite{Gauntlett:2005ww}.

The guiding logic will be the following: We will start with the NATD of the
$AdS_5\times X^5$ backgrounds discussed in detail in
Section \ref{Sec:NATD-IIB}. It was shown that they are SUSY
preserving and in most cases
non-singular. We will then apply a T duality (or first shift
one of the coordinates with a parameter $\gamma$ and then apply a T-duality).
Our procedure is guided by the Lunin-Maldacena T-s-T transformations
\cite{Lunin:2005jy}. We have checked in Sections 5.1-5.3 and Section 5.6 below,
that the Einstein, Maxwell, dilaton and Bianchi equations are satisfied.

One point of interest for the solutions presented in the following
section will be to understand
the field theory meaning of this class of backgrounds,
where the five-form flux vanishes but that still
contain and AdS${}_5$ factor.
The natural tendency would be to interpret these backgrounds as
in the class of wrapped D5 branes but we will argue that
the answer is more subtle. We present other Type IIA solutions
in Appendix \ref{App:GFix} where we extend these type of backgrounds
by keeping some of the parameters involved in the procedure of NATD.

The Figure \ref{Fig:Map}
summarizes the idea of the procedure advocated above.
Let us present below our new backgrounds.
\begin{figure}
\begin{align*}
 \begin{CD}
 AdS_5 \times X_5 @>NATD>> AdS_5 \tilde{\times} Y_5 @>T-duality >> AdS_5
\tilde{\times} W_5 \\
   \\
  F_5 @>NATD>> (F_2, F_4) @>T-duality >> \left(\begin{tabular}{c} $F_3;F_1=0,F_5=0$ \\  $F_1,F_3;F_5=0$\end{tabular} \right) \\
  \\
    Q_{D3}=N @>NATD>> (Q_{D6}=N, Q_{D4}=0) @>T-duality >> \left(\begin{tabular}{c}$Q_{D7}=0,Q_{D5}=N,Q_{D3}=0$\\ $Q_{D7}=N,Q_{D5}=0,Q_{D3}=0$\end{tabular}\right)
\\
    \\
    \Delta Q_{D3}=0 @>NATD>> (\Delta Q_{D6}=0, \Delta Q_{D4}=nN) @>T-duality >> \left(\begin{tabular}{c}$\Delta Q_{D5}=0,\Delta Q_{D3}=0$\\ $\Delta Q_{D5}=nN,\Delta Q_{D3}=0$\end{tabular}\right)\\
 \end{CD}
\end{align*}
\caption{A schematic description of the supergravity solutions
discussed in this manuscript and the properties of RR fluxes that are
relevant for a field theory interpretation.
The expression between parenthesis
corresponds to the Page charges and their changes
under a large gauge transformation of the B-field.}\label{Fig:Map}
\end{figure}

\subsection{T-dual on the $\xi$-direction of the NATD of $AdS_5\times S^5$}
\label{TonxiS5}
In this section we present the results of performing an Abelian
T-duality on the $\xi$ angle of the background in
Section \ref{NATD AdS5xS5}, carrying through
explicitly the appropriate
powers of $\alpha'$ in the B{\"u}scher rules:
\begin{align}
\label{Eq:NATD-T-S5}
\tilde{\hat{ds}}^2=&ds^2(AdS_5)+4L^2(d\alpha^2+\sin^2\alpha d\theta^2)+\frac{\alpha'^2 d\rho^2}{L^2\cos^2\alpha}\nn \\&+\frac{\rho ^4 \alpha '^2 \mathit{d}\chi
   ^2-2 \rho ^3 \alpha '^2 \mathit{d}\xi
   \mathit{d}\chi  \csc \chi +\mathit{d}\xi ^2 \csc ^2\chi  \left(\rho ^2 \alpha '^2+L^4
   \cos ^4\alpha \right)}{L^2 \rho ^2\cos ^2\alpha  },\nn \\[2mm]
\tilde{\hat{B}}=&0,\quad
e^{-2\tilde{\hat{\Phi}}}=\frac{L^4\rho^2\cos^4\alpha\sin^2\chi}{\alpha'^2}.
\end{align}
As the dilaton indicates, there is a singularity at $\cos\alpha=0$. This is, indeed, a curvature singularity as can be seen from the 10d string frame Ricci scalar curvature:
\be
\tilde{\hat{\mathcal{R}}}=-\frac{4 L^2 \cos ^2\alpha  \sin ^2\chi }{\rho ^2\alpha
   '^2}-\frac{4 \sin ^2\chi  \left(1 +\sin ^2\chi\cos^2\alpha
   \right)}{L^2\cos^2\alpha}.
   \ee
   The RR Fluxes are
\begin{align}
\label{F3_equation}
&\tilde{\hat{F}}_3= -\frac{8 L^4 \sin \alpha  \cos ^3\alpha  \mathit{d}\alpha \wedge
   \mathit{d}\theta \wedge \mathit{d}\xi }{\alpha '}, \nn \\&
\tilde{\hat{F}}_7=-\frac{2 \rho  \alpha ' ( \mathit{d}\xi \wedge
   \mathit{d}\rho +\rho  \sin \chi  \mathit{d}\rho \wedge
   \mathit{d}\chi )\wedge d\text{Vol}(AdS_5)}{L}, \nn \\&
\end{align}
where $\tilde{\hat{F}}_7=-\star \tilde{\hat{F}}_3$.  It is interesting to point out that only an $F_3$ flux survives,
no $F_5$ is present.  This situation is rather unexpected in the
general framework of IIB solutions with an $AdS_5$ factor
in light of the results of \cite{Gauntlett:2005ww}. Since
this background breaks SUSY, we are not subject to those restrictions.

We proceed to compute the Page charge supergravity and central charge for the field theory.  Since $B_2=0$,
\beq
Q_{P, D5}=\frac{1}{2\kappa_{10}^2T_{D5}}\int_{\Sigma_3}F_3=N_{D5}.
\eeq
With the normalization we have used, the equations of motion imply $L^4=\frac{1}{2}N_{D5}\alpha'^2$, and therefore, the field theory central charge is:
\beq
\tilde{\hat{c}}=\frac{8\pi^5 L^8}{3\alpha'^4}=\frac{2\pi^5}{3}N_{D5}^2
\eeq
Here we can compare to the central charge after NATD and see that an Abelian T-duality on $\xi$ does not change the central charge.  We can then follow our previous comments and conclude that after
the final T-duality, the dual QFT has not changed.

Let us present another possible T-duality,
generating a different background.

\subsection{T-dual on $\theta$ for the NATD of $AdS_5\times S^5$}
In this section, we perform a T-duality on the $\theta$ angle
for the background
of Section \ref{NATD AdS5xS5} instead.
\begin{align}
\label{Eq:NATD-T-S5ontheta}
\tilde{\hat{ds}}^2=&ds^2(AdS_5)+4L^2d\alpha^2
+\frac{\alpha '^2}{L^2} (\frac{ \mathit{d}\theta ^2}{4\sin ^2\alpha }
  +  \frac{\mathit{d}\rho
   ^2}{\cos^2\alpha})+
   \frac{L^2 \rho ^2 \alpha '^2 \cos ^2\alpha (
   \mathit{d}\xi ^2 \sin ^2\chi +
   \mathit{d}\chi ^2)}{\rho ^2 \alpha '^2+L^4 \cos ^4\alpha }\nn \\[2mm]
\tilde{\hat{B}}=&\frac{\alpha'^3\rho^3\sin\chi\mathit{d}\xi\wedge\mathit{d}\chi}{\alpha'^2\rho^2+L^4\cos^4\alpha},\quad
e^{-2\tilde{\hat{\Phi}}}=\frac{4L^4\cos^2\alpha\sin^2\alpha(\alpha'^2\rho^2+L^4\cos^4\alpha)}{\alpha'^4}.
\end{align}
The RR fluxes are given by
\bea
\label{F3_equation on theta}
\tilde{\hat{F}}_1&=&\frac{8L^4\cos^3\alpha\sin\alpha\mathit{d}\alpha}{\alpha'^2},\nn \\[2mm]
\tilde{\hat{F}}_3&=& \frac{8L^4\alpha'\rho^3\cos^3\sin\alpha\sin\chi\mathit{d}\alpha\wedge\mathit{d}\xi\wedge\mathit{d}\chi}{\alpha'^2\rho^2+L^4\cos^4\alpha}, \nn \\[2mm]
\tilde{\hat{F}}_7&=&\frac{2\alpha'\rho}{L}\mathit{d}\rho\wedge d\text{Vol}AdS_5, \nn \\[2mm]
\tilde{\hat{F}}_9&=&\frac{2\alpha'^2\rho^2\cos^4\alpha\sin\chi d\text{Vol}AdS_5\wedge d\theta\wedge d\xi\wedge d\rho\wedge d\chi}{\alpha'^2\rho^2+L^4\cos^4\alpha}
\eea
We again compute the Page charges, this time we see only the Page charge associated with $D7$-branes survives,
\begin{align}
Q_{P, D7}=&\frac{1}{2\kappa_{10}^2T_{D7}}\int_{\Sigma_1}F_1=N_{D7}\nn \\[3mm]
Q_{P, D5}=&\frac{1}{2\kappa_{10}^2T_{D5}}\int_{\Sigma_3}F_3-B_2\wedge F_1=N_{D5}=0\nn \\[3mm]
Q_{P, D3}=&\frac{1}{2\kappa_{10}^2T_{D3}}\int_{\Sigma_5}F_5-B_2\wedge F_3=N_{D3}=0.
\end{align}
The $Q_{P,D7}$ imposes a relation $L^4=\frac{1}{2}N_{D7}\alpha'^2$.
Comparing these results to Eq.(\ref{NATDads5s5}), we see that $B_2$ is unchanged after the Abelian T-duality.  We can, thus perform a large gauge transformation in a similar way as in Section \ref{Sec:Page}.  Here we see that,
\be
\Delta Q_{P,D3}=nN_{D7},\quad \Delta Q_{P,D5}=0
\ee
Finally, we apply the $L^4\sim N$ relation above and find that the central charge is
\be
\tilde{\hat{c}}=\frac{8\pi^5 L^8}{3\alpha'^4}=\frac{2\pi^5}{3}N_{D7}^2.
\ee
Again, we see that the central charge is invariant under the Abelian T-duality.
Let us apply a similar logic for the case of $AdS_5\times T^{1,1}$.
\subsection{T-dual on $\xi$ for the NATD of $AdS_5\times T^{1,1}$}
In this section we perform an Abelian T-duality along
the $\xi$ direction of the background presented in Section \ref{Sec:NATD-T11}.

The resulting NS sector is given by
\bea
\tilde{\hat{ds}}^2&=&ds^2(AdS_5)+L^2\lambda_1^2(\mathit{d}\theta_1^2+\sin^2\theta_1\mathit{d}\phi_1^2)+\frac{1}{P}\lambda _2^2 L^2
   \alpha '^2 (\rho  \mathit{d}\chi  \sin \chi
   -\mathit{d}\rho  \cos \chi )^2
   \nn \\
   &~&+\frac{\alpha '^2}{L^2 QP} \left(\mathit{d}\rho
   \sin \chi  \left(\rho ^2 \alpha '^2+\lambda
   ^2 \lambda _2^2 L^4\right)+\lambda ^2 \lambda _2^2 L^4 \rho
    \mathit{d}\chi  \cos \chi \right){}^2\nn \\
   &~&+\frac{1}{4 \lambda
   ^2 \lambda _2^2 L^2 \rho ^2 Q\sin^2\chi} \left(\rho ^2 \alpha '^2
   \sin \chi  \left(\left(\lambda ^2-\lambda _2^2\right)
   \mathit{d}\rho  \sin 2 \chi +2 \rho  \mathit{d}\chi
   N\right)-2 Q \mathit{d}\xi \right){}^2,\nn  \\[2mm]
\tilde{\hat{B}}&=&\alpha'\cos\theta_1\mathit{d}\phi_1\wedge (\mathit{d}\xi+\cos\chi\mathit{d}\rho-\rho\sin\chi\mathit{d}\chi), \quad
e^{-2\tilde{\hat{\Phi}}}=\frac{L^4\lambda^2\lambda_2^2\rho^2\sin^2\chi}{\alpha'^2},
\eea
where, as in Eq.(\ref{Eq:QN}), we have $N=\lambda^2 \cos^2 \chi + \lambda_2^2 \sin^2 \chi$, $Q=(\alpha'^2\rho^2N+L^4\lambda^2\lambda_2^2)$ and $P=(\rho ^2 \alpha   '^2 \sin ^2\chi +\lambda ^2 \lambda _2^2   L^4)$.

The non-trivial RR fields resulting are,
\bea
\tilde{\hat{F}}_3&=&\frac{4L^4\lambda\lambda_1^2\lambda_2^2\sin\theta_1}{\alpha'}\mathit{d}\theta_1\wedge \mathit{d}\xi\wedge\mathit{d}\phi_1,\nn \\
\tilde{\hat{F}}_7&=&\frac{\alpha'\rho}{8L}(d\text{Vol}AdS_5\wedge(\mathit{d}\xi\wedge\mathit{d}\rho+\rho\sin\chi\mathit{d}\rho\wedge\mathit{d}\chi)).
\eea
The only non vanishing Page charge is given by
\beq
Q_{P, D5}=\frac{1}{2\kappa_{10}^2T_{D5}}\int_{\Sigma_3}F_3=N_{D5}.
\eeq
This implies a condition, $L^4=\frac{27}{2}N_{D5}\alpha'^2$. We can then compute the central charge,
\be
\tilde{\hat{c}}=\frac{\pi^5L^8}{162\alpha'^4}=\frac{9\pi^5}{8}N_{D5}^2
\ee
Similar to Section \ref{TonxiS5} above, we see that a T-duality on $\xi$ does not change the central charge of the dual QFT.

Despite its relative simplicity,  the results of Section \ref{Sec:Susyt11} show that this solution breaks SUSY. This is not surprising as it has been previously argued that the $\xi$ isometry plays the role of the R-symmetry in the NATD of $AdS_5\times T^{1,1}$ \cite{Itsios:2013wd,Gaillard:2013vsa}.
\subsection{T-dual on $\phi_1$ of NATD of $AdS_5\times T^{1,1}$}
The background corresponding to applying  T-duality on the $\phi_1$ direction  for the  NATD of $AdS_5\times T^{1,1}$ takes the form:
\bea
\tilde{\hat{ds}}^2&=&ds^2(AdS_5)+L^2\lambda_1^2\mathit{d}\theta_1^2\nn \\
&~&+\frac{\alpha '^2}{L^2} \bigg[\frac{\lambda _2^2 L^4}{P} (\mathit{d}\rho  \cos
   \chi -\rho  \mathit{d}\chi  \sin \chi )^2\nn \nn\\
   &~& +\frac{1}{QW}\left(-\lambda ^2 \mathit{d}\rho
   \cos \theta _1 \cos \chi  \left(\rho ^2 \alpha  '^2+\lambda _2^4 L^4\right)+\lambda ^2 \lambda _2^4 L^4 \rho
   \mathit{d}\chi  \cos \theta _1 \sin \chi +Q
   \mathit{d}\phi _1\right){}^2\nn \\
&~&+ \left.\frac{1}{P Q}\left(\mathit{d}\rho
    \sin \chi  \left(\rho ^2 \alpha '^2+\lambda ^2 \lambda _2^2
   L^4\right)+\lambda ^2 \lambda _2^2 L^4 \rho  \mathit{d}\chi  \cos \chi
   \right)^2 \right. \nn\\
   &~&+ \frac{1}{W^2}\lambda ^2 \lambda _1^2 \lambda _2^2 L^4 \rho ^2
   \mathit{d}\xi ^2 \sin ^4\theta _1 \sin ^2\chi \left(\lambda ^2
   \lambda _2^2 \rho ^2 \alpha '^2 \cot ^2\theta _1 \sin
   ^2\chi +\lambda _1^2 Q\right)\bigg],\nn \\[2mm]
W\tilde{\hat{B}}&=&\alpha'^3\lambda_1^2 \rho^2\sin\chi d\xi\wedge\bigg[\lambda^2 \sin\chi d\phi_1\nn\\
&~&+\sin^2\theta_1\big((\lambda^2-\lambda_1^2)\cos\chi\sin\chi d\rho+N\rho d\chi\big)+\lambda^2\cos^2\theta_1\sin\chi\big(\rho\sin\chi d\chi-\cos\chi d\rho\big)\bigg], \nn \\[2mm]
e^{-2\tilde{\hat{\Phi}}}&=&\frac{L^4}{\alpha'^4}W,
\eea
where
\beq
W=\alpha'^2\lambda^2\lambda_2^2\rho^2\cos^2\theta_1\sin^2\chi
+\lambda_1^2\sin^2\theta_1(\alpha'^2\rho^2N+L^4\lambda^2\lambda_2^4),
\eeq
where $P=(\rho ^2 \alpha   '^2 \sin ^2\chi +\lambda ^2 \lambda _2^2   L^4)$.
The RR sector has the following non-vanishing field strengths;
\bea
\tilde{\hat{F}}_1&=&\frac{4L^4\lambda\lambda_1^4 \sin\theta_1}{\alpha'^2}\mathit{d}\theta_1, \nn \\[3mm]
W\tilde{\hat{F}}_3&=&4L^4\alpha' \lambda \lambda_1^6 \rho^2 \sin\chi\sin\theta_1 d\theta_1\wedge d\xi\wedge\bigg[\lambda^2 \cos\theta_1\sin\chi d\phi_1\\
&~&+\sin^2\theta_1 \big((\lambda^2-\lambda_1^2)\cos\chi\sin\chi d\rho+N\rho d\chi\big)+\lambda^2\cos^2\theta_1 \sin\chi\big(\rho \sin\chi d\chi-\cos\chi d\rho\big)\bigg].\nn
\eea
For the sake of brevity, we only present the $F_1$ (if nontrivial) and $F_3$, and omit their corresponding Hodge duals for the rest of the results in this section.  The corresponding Page charges and central charge are
\begin{align}
Q_{P,D7}=&\frac{1}{2\kappa_{10}^2T_{D7}}\int_{\Sigma_1}F_1=N_{D7}\nn \\[3mm]
Q_{P, D5}=&\frac{1}{2\kappa_{10}^2T_{D5}}\int_{\Sigma_3}F_3-B_2\wedge F_1=N_{D5}=0\nn \\[3mm]
Q_{P, D3}=&\frac{1}{2\kappa_{10}^2T_{D3}}\int_{\Sigma_5}F_5-B_2\wedge F_3=N_{D3}=0,
\end{align}
and
\bea
\tilde{\hat{c}}=\frac{\pi^5L^8}{162\alpha'^4}=\frac{9\pi^5}{8}N_{D5}^2,
\eea
where, again we used $L^4=\frac{27}{2}N_{D5}\alpha'^2$.

Section \ref{Sec:Susyt11} shows that this solution preserves all supercharges  of  $AdS_5\times T^{1,1}$ in the form of an $SU(2)$-structure defined on the new 6d internal space.

\subsection{Shift $\phi_1\to \phi_1+\gamma \xi$, T-dual on $\xi$ of NATD of $AdS_5\times T^{1,1}$}
In this section we consider mixing the two $U(1)$'s symmetries via a shift. The resulting one-parameter solutions have a much smaller singularity loci, however as we once more dualise on the R-symmetry we break SUSY\footnote{The same is also true if we perform $\xi\to \xi+\gamma \phi_1$ and
T-dualise on $\partial_{\phi_1}$. This time because we introduce a $\phi_1$ dependence on the Killing spinor, which means the Kosmann derivative cannot vanish (see Section \ref{Sec:Susyt11}).}. Nonetheless $AdS_5$ solutions with parameters are uncommon so we believe this solution deserves some further study.

The NS sector is given by
\bea
\tilde{\hat{ds}}^2&=&ds^2(AdS_5)+L^2\lambda_1^2\mathit{d}\theta_1^2\nn \\
&~&+\frac{1}{P}\left(\lambda _2 L \alpha'   \mathit{d}\rho  \cos \chi -\lambda_2 L \rho  \alpha ' \mathit{d}\chi
   \sin \chi \right){}^2+\frac{1}{Y}\lambda ^2 \lambda _1^2
   \lambda _2^2 L^2 \rho ^2\alpha'^2 \mathit{d}\phi_1{}^2 \sin ^2\theta_1 \sin ^2\chi\nn \\
&~&+\frac{\alpha '^2}{L^2 QY}  \left(\rho  \mathit{d}\chi  \sin\chi  \left(\rho^2 \alpha '^2N -\gamma \lambda ^2 \lambda _2^4L^4 \cos\theta_1\right)\right.\nn\\
&~&+\frac{\alpha'^2}{L^2 QY}\left(\mathit{d}\rho \cos \chi  \left(\lambda ^2 \rho ^2 \alpha '^2 \left(\gamma \cos\theta _1+\sin^2\chi \right)-\lambda _2^2 \rho^2
   \alpha '^2 \sin ^2\chi+\gamma  \lambda ^2 \lambda_2^4 L^4 \cos \theta _1\right)-Q \mathit{d}\xi \right){}^2\nn \\
&~&+\frac{\alpha'^2 \left(\mathit{d}\rho
   \sin \chi  \left(\rho ^2 \alpha '^2+\lambda ^2
   \lambda _2^2 L^4\right)+\lambda^2 \lambda _2^2 L^4 \rho  \mathit{d}\chi \cos \chi \right){}^2}{L^2 Q P},\nn \\[2mm]
\frac{2Y}{\alpha '}\tilde{\hat{B}}&=& \gamma  \lambda _1^2 \sin ^2\theta _1 \bigg(\rho  \sin \chi
   \mathit{d}\phi _1\wedge \mathit{d}\chi  \left(-2 \rho ^2 \alpha '^2 N
   +\gamma  \lambda ^2 \lambda _2^4 L^4
   \cos \theta_1\right)\nn  \\
   &~&+\mathit{d}\rho \wedge \mathit{d}\phi_1
   \left(\lambda ^2 \rho ^2\alpha'^2 \left(\gamma  \cos
   \theta _1 \cos \chi   +\sin \chi  \sin 2 \chi\right) \right. \nn\\
&~&-2 \left. \lambda _2^2 \rho ^2 \alpha'^2 \sin ^2\chi  \cos \chi
   +\gamma  \lambda ^2 \lambda_2^4 L^4\cos \theta _1 \cos \chi
   - 2 Q \mathit{d}\xi \wedge   \mathit{d}\phi_1\right)\bigg) \nn\\
&~&-\frac{1}{Q} \lambda ^2
   \lambda _2^2 \rho ^2 \alpha
   '^2 \cos \theta _1
   \sin ^2\chi  \left(\gamma  \cos
   \theta _1+1\right)\bigg[\rho  \sin \chi
   \mathit{d}\phi _1\wedge
   \mathit{d}\chi  \left(2
   \rho ^2\alpha '^2 N
   +\lambda ^2 \lambda _2^4 L^4
   \left(1-\gamma  \cos\theta_1\right)\right)\nn \\
 &~&+\mathit{d}\rho \wedge \mathit{d}\phi
   _1 \bigg(\lambda ^2 \rho ^2
   \alpha '^2 \cos \chi
   \left(\cos 2 \chi -\gamma  \cos
  \theta _1\right)+\lambda
   _2^2 \rho ^2 \alpha '^2
   \sin \chi  \sin 2 \chi \nn \\
   &~&+\lambda ^2
   \lambda _2^4 L^4 \cos \chi
   \left(1-\gamma  \cos \theta
   _1\right)\bigg)2 Q
   \mathit{d}\xi \wedge \mathit{d}\phi_1\bigg], \nn \\[2mm]
e^{-2\tilde{\hat{\Phi}}}&=&\frac{L^4}{\alpha'^4} Y,
\eea
where
$Y=\alpha'^2 \lambda ^2 \lambda _2^2 \rho
   ^2 \sin ^2\chi  (\gamma  \cos
  \theta_1 +1)^2+\gamma ^2
   \lambda _1^2 \sin ^2\theta_1Q$. The above expression makes clear that a shift $\gamma$ has substantially reduced the singular locus. For example, taking into consideration that $Q$ is non-vanishing,  now to get a singular dilaton we need $\sin\chi=0$ simultaneously with $\sin\theta_1=0$; also $\rho=0$ and $\sin\theta_1=0$.

The RR sector contains:
\bea
\tilde{\hat{F}}_1&=&\frac{4 \gamma  \lambda  \lambda _1^2
   \lambda _2^2 L^4
   \sin
   \theta_1}{\alpha'^2} \mathit{d}\theta_1\nn \\
\tilde{\hat{F}}_3&=&2 \lambda  \lambda _1^2 \lambda _2^2 L^4
   \rho ^2 \alpha ' \sin \theta
   _1 \sin \chi
   \bigg[-\frac{1}{Q}\mathit{d}\rho \wedge
   \mathit{d}\theta _1\wedge
   \mathit{d}\phi _1
   \left(\frac{\lambda ^2 \sin 2 \chi
   \left(\rho ^2\alpha
   '^2+\lambda ^2 \lambda _2^2
   L^4\right)}{P}\nn \right.\\
   &~&-\frac{1}{Y}\lambda _2^2   \lambda ^2 \sin \chi  \left(\gamma
   \cos \theta _1+1\right)   \left(\lambda ^2 \rho ^2 \alpha
   '^2 \left(2 \gamma  \cos \theta _1 \cos \chi
   +\sin \chi  \sin 2 \chi \right)+2\gamma  \lambda ^2 \lambda _2^4 L^4
   \cos \theta _1 \cos\chi \nn \right. \\
&~&- 2\left.\left. \lambda _2^2 \rho ^2 \alpha'^2 \sin^2\chi  \cos\chi \right)-\frac{2}{P} \sin\chi  \cos \chi\right)\nn \\
&~&-\frac{2 }{Y} \gamma  \rho
   \mathit{d}\chi \wedge \mathit{d}\theta_1\wedge
   \mathit{d}\phi_1
   \left(\lambda ^2 \lambda _2^2 \cos\theta _1 \sin ^2\chi
   \left(\gamma  \cos \theta_1+1\right)+\gamma  \lambda_1^2 N \sin ^2\theta_1\right)\nn\\
   &~&-2\lambda^2 \lambda _2^2 \sin \chi
   \left(\gamma  \cos \theta_1+1\right) \mathit{d}\xi
   \wedge \mathit{d}\theta_1\wedge \mathit{d}\phi_1\bigg].
     \eea
\subsection{T-Dual on $\xi$ of NATD of $AdS_5\times Y^{p,q}$}
In this section we perform an Abelian T-duality along the $\xi$ direction on the background specified in Section \ref{NATD:ypq}.
\bea
\tilde{\hat{ds}}^2&=&ds^2(AdS_5)+L^2k^2\mathit{d}\alpha^2+\frac{L^2}{vw}\mathit{d}y^2\nn \\
&~&+\frac{1}{6
   L^2 \rho ^2 g m}(g \left(72 \alpha'^2 \rho ^2 \mathit{d}\xi  \cot
   \chi  (\mathit{d}\rho  \sin \chi +\rho  \mathit{d}\chi  \cos \chi )+36
   \alpha'^2 \rho ^2 (\mathit{d}\rho  \sin \chi +\rho
   \mathit{d}\chi  \cos \chi )^2\nn \right.\\
        &~&\left. +\mathit{d}\xi ^2 \left(36\alpha'^2 \rho ^2 \cot ^2\chi +L^4 m^2 \csc ^2\chi
   \right)\right)+6 \alpha'^2 \rho ^2 m (\mathit{d}\xi
   +\rho  \mathit{d}\chi  \sin \chi -\mathit{d}\rho  \cos \chi )^2)\nn \\[2mm]
        \tilde{\hat{B}}&=&-\frac{\alpha'h}{\sqrt{g}}\mathit{d}\alpha\wedge(\mathit{d}\xi-\cos\chi\mathit{d}\rho+\rho\sin\chi\mathit{d}\chi)\nn \\[2mm]
e^{-2\tilde{\hat{\Phi}}}&=&\frac{L^4gm\rho^2\sin^2\chi}{6\alpha'^2}.
\eea
Just as in the cases of dualizing along $\xi$ in the NATD of $AdS_5\times S^5$ and $AdS_5\times T^{1,1}$, we recover only an $F_3$ (and its Hodge dual),
\be
\tilde{\hat{F}}_3=\frac{2L^4m}{9\alpha'}\mathit{d}\alpha\wedge\mathit{d}y\wedge\mathit{d}\xi.
\ee
We compute the Page charge associated with the $N_{D5}$,
\beq
Q_{P, D5}=\frac{1}{2\kappa_{10}^2T_{D5}}\int_{\Sigma_3}F_3=N_{D5}.
\eeq
This implies a condition, $L^4=\frac{9}{l(y_2^2-y_1^2+2(y_1-y_2))}N_{D5}\alpha'^2$. We can then compute the central charge,
\bea
\tilde{\hat{c}}=\frac{3\pi^5}{4l(y_2^2-y_1^2+2(y_1-y_2))}N_{D5}^2.
\eea Notice that this is the same result we found after the NATD of $Y^{pq}$ with $N_{D6}$ being replaced by $N_{D5}$.

Like the equivalent NATD-T solution of  $AdS_5\times T^{1,1}$, this solution breaks SUSY. We propose that this is for the same reason as above.
Namely, the $\xi$ isometry once more plays the role of
the R-symmetry in the $SU(2)$ transformed $AdS_5 \times Y^{p,q}$ solution.
\subsection{T-Dual on $\alpha$ of NATD of $AdS_5\times Y^{p,q}$}
As a final solution, we consider performing a further Abelian T-duality along $\partial_{\alpha}$ of the NATD of $AdS_5\times Y^{p,q}$. As shown in section \ref{sec:GYpq}, such a solution preserves $\mathcal{N}=1$ SUSY in 4d via an $SU(2)$-structure defined on the 6-d internal space.

Performing the T-duality gives a NS sector of the form
\bea
d\tilde{\hat{s}}^2&=&ds^2(AdS_5)+\frac{L^2}{vw}dy^2+ \frac{\alpha'^2}{36\alpha'^2\rho^2\cos^2\chi+L^4 m^2}\bigg(6L^2m\big(\rho\cos\chi d\chi+\sin\chi d\rho\big)^2\nn\\
&~&+\frac{1}{L^2\Upsilon}\big((36\alpha'^2\rho^2+ L^4m^2)\cos\chi d\rho-L^4\rho m^2\sin\chi d\chi\big)^2\bigg)+\frac{L^2\alpha'^2\rho^2}{\Theta}k^2 gm\sin^2\chi d\xi^2\nn\\
&~&+\frac{\alpha'^2}{6 L^2 \Upsilon\Theta}\bigg(\Upsilon d\alpha+L^4\rho\sqrt{g}h m^2\sin\chi d\chi-\sqrt{g}h(36\alpha'^2\rho^2+L^4 m^2)\cos\chi d\rho\bigg)^2\nn,\\[3mm]
\tilde{\hat{B}}&=&\frac{\alpha'^3\rho^2}{\Theta}\sin\chi\bigg(6gk^2\cos\chi(\sin\chi d\xi\wedge d\rho+\rho\cos\chi d\xi\wedge d\chi)\nn\\
&~&+mw\sin\chi(-\cos\chi d\xi\wedge d\rho+\rho\sin\chi d\xi\wedge d\chi)-\sqrt{g}mh \sin\chi d\alpha\wedge d\xi\bigg),\nn\\[3mm]
e^{-2\tilde{\hat{\Phi}}}&=&\frac{L^4}{6\alpha'^4}\Theta,
\eea
where
$\Theta= \frac{1}{6}gk^2(36\alpha'^2\rho^2\cos^2\chi+L^4m^2)+\alpha'^2\rho^2 wm \sin^2\chi$, and \vskip5pt $\Upsilon=g(36\alpha'^2\rho^2\cos^2\chi+L^4m^2)+6\alpha'^2\rho^2m\sin^2\chi.$ \vskip5ptWhile the RR Sector has no trivial fluxes
\bea
F_1&=& \frac{2L^4}{9\alpha'^2}  m dy\\[2mm]
F_3&=& \frac{2\rho^2L^4 \alpha'm}{9\Theta}\sin\chi\bigg(6gk^2\cos\chi(\sin\chi d\xi\wedge d\rho+\rho\cos\chi d\xi\wedge d\chi)\nn\\
&~&+mw\sin\chi(-\cos\chi d\xi\wedge d\rho+\rho\sin\chi d\xi\wedge d\chi)-\sqrt{g}mh \sin\chi d\alpha\wedge d\xi\bigg)\wedge  dy\nn.
\eea
We compute the Page charge associated with the $N_{D7}$,
\beq
Q_{P, D7}=\frac{1}{2\kappa_{10}^2T_{D7}}\int_{\Sigma_3}F_1=N_{D7}.
\eeq
This implies a condition, $L^4=\frac{9}{l(y_2^2-y_1^2+2(y_1-y_2))}N_{D5}\alpha'^2$. We can then compute the central charge,
\bea
\tilde{\hat{c}}=\frac{3\pi^5}{4l(y_2^2-y_1^2+2(y_1-y_2))}N_{D5}^2.
\eea Notice that this is the same result we found after the NATD-T on $\xi$ of $Y^{pq}$ with $N_{D5}$ being replaced by $N_{D7}$.

In summary, we have presented a set of new solutions.
Some of them preserve minimal SUSY as will be proven in the next section.
These backgrounds escape the classification of
\cite{Gauntlett:2005ww}, in the sense that they present an
$AdS_5$ factor in IIB, but without
$F_5$ RR-field. More details will follow below.
\section{G-structures of the NATD-T Solutions}\label{Sec:Susy}
In this section we perform a SUSY analysis of some of
the new solutions generated in this paper. We
focus our attention on the NATD-T transformations of $AdS_5\times T^{1,1}$ and $AdS_5\times Y^{p,q}$ as these have relatively simple descriptions in terms of G-structures. As shown in Appendix \ref{sec: spinors} further $U(1)$ T-dualities performed on the NATD of $AdS_5\times S^5$ break all the remaining SUSY.

In \cite{Grana:2005sn} necessary and sufficient conditions for the
preservation of $\mathcal{N}=1$ SUSY where established in terms of
geometric quantities. For a solution with metric of the
form $\mathbb{R}^{1,3}\times \mathcal{M}^6$ and non trivial
RR sector these are,
\begin{enumerate}
\item The existence of an either an $SU(3)$ or $SU(2)$ structure defined on the internal manifold $\mathcal{M}^6$.
\item A NS 3-form $H_3$ that is closed.
\item A RR polyform F which obey certain differential relations in terms of the previously mentioned quantities.
\end{enumerate}
Let us briefly review what is required, we use the
notation of \cite{Andriot:2008va}---see also
the introduction of \cite{Andriot:2010sya} for a nice review.
We assume a metric of the form,
\beq
ds^2_{str}=e^{2A} dx_{1,3}^2+ ds^2(\mathcal{M}^6),
\eeq
where $A$ is an arbitrary function on the internal space.
The starting point is to introduce two Majorana-Weyl Killing spinors such that
\beq
\epsilon=\left(\begin{array}{l}
\epsilon_1\\
\epsilon_2\end{array}\right)
\eeq
These are then further split into 4-d $(\zeta)$ and 6-d $(\eta)$ components
\beq
\epsilon_1= e^{\frac{A}{2}}e^{i\frac{\theta_++\theta_-}{2}}(\zeta_{+}\otimes\eta^{1}_{+}+\zeta_{-}\otimes\eta^{1}_{-}),~~~\epsilon_1=e^{-i\frac{\theta_+-\theta_-}{2}}e^{\frac{A}{2}}( \zeta_{+}\otimes\eta^{2}_{\mp}+\zeta_{-}\otimes\eta^{2}_{\pm}),
\eeq
where $\pm$ labels chirality and so the upper/lower signs are taken in the above in type IIA/IIB respectively. The internal spinors also obey the relation $(\eta^{1,2}_+)*=\eta^{1,2}_-$. It is possible to define two bi-spinors on the internal space
\beq
\Psi_{\pm}= e^{A}e^{i\theta_{\pm}}\eta^{1}_+\otimes\eta^{2}_{\pm}
\eeq
where $\theta_{\pm}$ are arbitrary phases with support on $\mathcal{M}^6$. These bi-spinors may then be mapped to polyforms under the Clifford map at which point the conditions for $\mathcal{N}=1$ SUSY may be expressed as
\begin{align}
&\big(d-H\wedge \big)\big(e^{2A-\Phi})\Psi_{\pm}=0,\\[2mm]
&\big(d-H\wedge \big)\big(e^{2A-\Phi})\Psi_{\mp}=e^{2A-\Phi} dA\wedge \bar{\Psi}_{\mp}+\frac{ie^{3A}}{8}\tilde{F}\nn
\end{align}
where upper/lower signs are taken in type IIA/IIB, $\tilde{F}=
\iota_{(e^{tx^1x^2x^3})}(F)$ and $F$ is a sum over all the
RR fields in the democratic formalism.
The specific form of $\Psi_{\pm}$ is determined by the type of structure.
The two cases we will deal with
in this section are either $SU(3)$-structures which are characterised
by parallel internal spinors or orthogonal $SU(2)$ structures,
for which $\eta^{1\dag}_+\eta^2_+=0$.

In the case of  $SU(3)$-structures, the pure spinors are,
\beq
\Psi_+=- e^{i\theta_+}\frac{e^{A}}{8} e^{-i J},~~~\Psi_-=-ie^{i\theta_-}\frac{e^{A}}{8} \Omega_{hol}
\eeq
where $J$ is a (1,1)-form and $\Omega_{hol}$ is a holomorphic 3-form and they must satisfy,
\beq\label{SU(3)struct}
J\wedge\Omega_{hol}=0,~~~ J\wedge J\wedge J = \frac{4 i}{3}\Omega_{hol}\wedge \bar{\Omega}_{hol}.
\eeq
The components of these forms
may be calculated in terms of the 6-d gamma matrices $\gamma_a$
and the internal spinors via
\beq
J_{ab}=-i \eta^{1\dag}_+\gamma_{ab}\eta^1_+,~~~~\big(\Omega_{hol}\big)_{abc}=-i\eta^{1\dag}_+\gamma_{abc}\eta^{1}_{+}.
\eeq

For orthogonal $SU(2)$-structures, the pure spinors read,
\beq
\Psi_+=-i e^{i\theta_+}\frac{e^{A}}{8} e^{-v\wedge w}\wedge\omega,~~~\Psi_-=ie^{i\theta_-}\frac{e^{A}}{8} (v+i w)\wedge e^{-i j},
\eeq
where $j$ is real 2-form and $\omega$ is a holomorphic 2-form and $z=v+i w$ is a holomorphic 1-form. These must satisfy the $SU(2)$ structure conditions
\beq
j\wedge\omega=\o\wedge \o = \iota_{\bar{z}}(\omega)=\iota_{\bar{z}}(j)=0,~~~ j\wedge j=\frac{1}{2}\omega\wedge\bar{\omega}.
\eeq
The components of these forms may be calculated via,
\beq
\bar{z}_a=\eta^{1\dag}_-\gamma_a\eta^{2}_+,~~~j_{ab}
=-i\eta^{1\dag}\gamma_{ab}\eta^1_{+}+i \eta^{2\dag}_+\gamma_{ab}\eta^{2}_+,~~~\omega_{ab}=\eta^{1\dag}_-\gamma_{ab}\eta^{2}_{-}.
\eeq

Let us now just state some previously derived results we shall
be using. As reviewed at length in \cite{Itsios:2013wd},
a
Non-Abelian T-duality on an
$SU(2)$-isometry,
has a preferred basis of
vielbeins with respect to which the transformation of the RR sector
is given  by a simple bispinor transformation,
\beq\label{eq: RRtransformation}
e^{\Phi_{IIB}}F_{IIB}\Omega^{-1}= e^{\Phi_{IIA}}F_{IIA},
\eeq
where
\beq\label{eq su2trans}
\Omega_{SU(2)}= \Gamma^{(10)}\frac{1}{\sqrt{1+ \zeta_a\zeta^a}}\left(-\Gamma_{123}+\zeta_1\Gamma_1+\zeta_2\Gamma_2+\zeta_3\Gamma_3\right),
\eeq
and $F_{IIA/B}$ is the sum of all the democratic formalism RR fields\footnote{Strictly speaking one needs to map the RR-sector to bispinors under the Clifford map.}.
The $\zeta's$ are defined in terms of the
original vielbeins describing the $SU(2)$ isometry of the background
$e^{a}=e^{B_a}(\s_a+A_a)$ as
\beq
\zeta^a= v_ae^{-\sum_{b\neq a}B^b}.
\eeq
The $U(1)$ Omega matrix in the preferred frame of Abelian T-duality,
$e^{\theta}=e^{B}(d\theta+C)$ is simply
\beq
\Omega_{U(1)}=\Gamma^{(10)}\Gamma_{\theta}
\eeq
The action of T-duality on the MW Killing spinors is given by
\beq
\hat{\epsilon}_1=\epsilon_1,~~~\hat{\epsilon}_2=\Omega\epsilon_2,
\eeq
which was proven for $U(1)$ isometries in \cite{Hassan:1999bv} and $SU(2)$ isometries in \cite{Kelekci:2014ima}. The condition that SUSY is preserved is that the Kosmann derivative vanishes along the given isometry. The Kosmann derivative of the Killing spinor $\epsilon$  along a Killing vector $K$ is given by
\beq
\mathcal{L}_K\epsilon=K^a\nabla_a\epsilon+\frac{1}{8} \left(dK\right)_{ab}\Gamma^{ab}\epsilon.
\eeq
The vanishing of this object is equivalent to the independence of $\epsilon_{1,2}$ on the isometry directions in the appropriate preferred frame \cite{Kelekci:2014ima}.

Finally, it was established in
\cite{Barranco:2013fza,Macpherson:2013zba,Gaillard:2013vsa,Caceres:2014uoa}
 that the bi-spinors defined on a $d$-dimensional
internal space transform as
\beq
\hat{\Psi}_{\pm}= \Psi_{\mp}\Omega,
\eeq
at least up to conventionally dependent phases\footnote{Though it is yet
to be formally proven, it is
likely that this will hold when the Kosmann derivative along
the $SU(2)$ directions vanishes.}.

\subsection{G-structure for NATD-T of $AdS_5\times T^{1,1}$ }\label{Sec:Susyt11}
We start our analysis with the Klebanov-Witten solution
\cite{Klebanov:1998hh} which possesses a metric that may be
succinctly expressed in terms of the following
vielbein basis (notice that in this Section we rename $\theta_1\to\theta$ and $\phi_1\to\varphi$ and set $\alpha'=1$),
\begin{align}\label{eq:veil1}
&e^{x^{\mu}}= \frac{r}{L} x^{\mu},~~~ e^{r}= \frac{L}{r}dr,~~~ e^{\theta}= L\lambda_1 d\theta,~~~e^{\varphi}= L\lambda_1\sin\theta d\varphi,\nn\\[-2mm]
&~\\[-2mm]
&e^{1,2}= L \lambda_1 \s_{1,2},~~~ e^{3}= L \lambda (\s+\cos\theta d\varphi). \nn
\end{align}
With respect to Eq.(\ref{eq:veil1})
the projection conditions on the 10-d Majorana Killing spinor are
\beq
\Gamma_{r123}\epsilon=\epsilon,~~~\Gamma_{\theta\phi}\epsilon=\Gamma_{12}\epsilon.
\eeq
These define a canonical $SU(3)$-structure with
\beq
J= e^{r 3}+ e^{\varphi\theta}+e^{21},~~~\Omega_{hol}= (e^r+i e^3)\wedge(e^{\varphi}+i e^{\theta})\wedge(e^{2}+i e^{1}).
\eeq
From these and the warp factor on the Minkowski directions,
$e^{2A}= \frac{r^2}{L^2}$, it is possible to construct two
bi-spinors of the form in Eq.(\ref{SU(3)struct}),
 where $\theta_+=\frac{\pi}{2}$ and $\theta_-=0$.

We now turn our focus  to the $SU(2)$ T-dualised Klebanov-Witten
solution. As Eq.(\ref{eq:veil1}) is in the preferred frame and the associated killing spinor depends only on $r$ \cite{Arean:2006nc}, the result of \cite{Kelekci:2014ima} implies that all SUSY is preserved under the $SU(2)$ transformation. The transformed vielbeins are given by,
\begin{align}
\hat{e}^{1}&=-\lambda_1\frac{x_1(L^2\lambda_1 \cos\xi+x_2\sin\xi)(dx_2+L^2\lambda \hat{\sigma}_3)+(-L^2x_2\lambda^2\cos\xi+(x_1^2+L^4\lambda^2\lambda_1^2)\sin\xi)dx_1}{L\Delta},\nn\\[2mm]
\hat{e}^{2}&=\lambda_1\frac{x_1(L^2\lambda_1 \sin\xi-x_2\cos\xi)(dx_2+L^2\lambda \hat{\sigma}_3)-(L^2x_2\lambda^2\sin\xi+(x_1^2+L^4\lambda^2\lambda_1^2)\cos\xi)dx_1}{L\Delta},\nn\\[2mm]
\hat{e}^{3}&=- \lambda\frac{x_1x_2 dx_1+(x_2^2+L^4 \lambda_1^4)dx_2-L^2 x_1^2 \lambda_1^2 \hat{\sigma}_3}{L \Delta},
\end{align}
where
\beq
\Delta= L^4\l_1^4\l^2 +\l_1x_1^2+\l^2x_2^2,~~~ \hat{\s}_3= d\xi+ \cos\theta d\varphi.
\eeq
and we have introduced coordinates
\beq
x_1=\rho\sin\chi,~~~x_2=\rho\cos\chi.
\eeq
The matrix $\Omega_{SU(2)}$ is defined in terms of
\beq
\zeta_1= \frac{x_1}{L^2 \lambda\lambda_1} \sin\xi,~~~\zeta_2= \frac{x_1}{L^2 \lambda\lambda_1} \cos\xi,~~~\zeta_3= \frac{x_2}{L^2 \lambda_1},
\eeq
which implies that that the dual killing spinor depends on $\xi$. There are two isometries on which on can perform a further $U(1)$ T-duality, $\partial_\phi$ and $\partial_{\xi}$. A quick computation (for mathematica) gives
\beq
\mathcal{L}_{\partial_{\xi}}\hat{\epsilon}=\partial_{\xi}\hat{\epsilon},~~~~~~~\mathcal{L}_{\partial_{\phi_1}}\hat{\epsilon}=\partial_{\phi}\hat{\epsilon}
\eeq
which shows that SUSY will only be preserved when one performs a further T-duality on $\partial_{\phi}$. Let us now calculate the G-structure.

We first use the projections to simplify $\Omega$ and then rotate the preferred vielbein basis as $\tilde{e}=\mathcal{R}_1 \hat{e}$ where
\beq\label{eq:Rot1}
\mathcal{R}_1 = \frac{1}{\sqrt{1+\zeta_a\zeta^a}}\left(\begin{array}{cccccc}
1&0&0&\zeta_1&\zeta_2&\zeta_3\\
0&\sqrt{1+\zeta_a\zeta^a}&0&0&0&0\\
0&0&\sqrt{1+\zeta_a\zeta^a}&0&0&0\\
-\zeta_1&0&0&1&\zeta_3&-\zeta_2\\
-\zeta_2&0&0&-\zeta_3&1&\zeta_1\\
-\zeta_3&0&0&\zeta_2&-\zeta_1&1
\end{array}\right)
\eeq
and the ordering is $[r\theta\varphi 1 2 3]$.
Then, the $\Omega$ acting on $\epsilon_2$ is drastically simplified to
\beq
\Omega_{SU(2)}\epsilon_2 = \tilde{\Gamma}^r\epsilon_2.
\eeq
More remarkable is the fact the the projectors in this basis are also unchanged so that
\beq
\tilde{\Gamma}_{r123}\epsilon = \epsilon,~~~ \tilde{\Gamma}_{\theta\varphi}\epsilon=\tilde{\Gamma}_{12}\epsilon.
\eeq
Specifically the vielbeins $\tilde{e}$ are given by
\begin{align}\label{eq:veil2}
&\tilde{e}^r= \frac{L^4 \lambda \l_1^2 dr-r(x_1 dx_1+x_2 dx_2)}{L r\sqrt{\Delta}},\nn\\[2mm]
&\tilde{e}^{\theta}=L \l_1 d\theta_1,~~~\tilde{e}^{\varphi}=L \l_1 \sin\theta_1d\varphi,\nn\\[2mm]
&\tilde{e}^1 =-\frac{L \l_1\big(\sin\xi(x_1 dr+\l rdx_1)+\cos\xi r x_1 \l \hat{\s_3}\big)}{r \sqrt{\Delta}},\nn\\[2mm]
&\tilde{e}^2 =-\frac{L \l_1\big(\cos\xi(x_1 dr+\l rdx_1)-\sin\xi r x_1 \l \hat{\s_3}\big)}{r \sqrt{\Delta}},\nn\\[2mm]
&\tilde{e}^3 =- \frac{L}{r\sqrt{\Delta}}\big(\l x_2 dr+\l_1^2 r dx_2\big).
\end{align}
As  shown at length in \cite{Barranco:2013fza} it is now a rather simple
matter to derive the G-structure of the dual of Klebanov-Witten,
which turns out to be an orthogonal $SU(2)$-structure.
The relevant forms and phases are,
\begin{align}
z&=v+ i w= \tilde{e}^r+ i \tilde{e}^3,\nn\\[2mm]
 j&= \tilde{e}^{\varphi\theta} + \tilde{e}^{21},\\[2mm]
 \omega&= (\tilde{e}^2+i \tilde{e}^1)\wedge(\tilde{e}^{\varphi}+i \tilde{e}^{\theta}),\nn\\[2mm]
\tilde{\theta}_m&=\frac{\pi}{2},~~~\tilde{\theta}_m=0.\nn
\end{align}

The next step is to perform an Abelian T-duality on the
$\varphi$ direction. In order to do this it is convenient
to rotate to a frame in which $d\varphi$ appears in only one of the vielbein.
This can be achieved by rotating the basis $\tilde{e}$  by
\beq
\left(
\begin{array}{cccc}
 -\frac{\lambda  \lambda _1 x_1 \cos \theta_1}{\sqrt{\Xi }} & -\frac{\lambda  \lambda _1 \sin \theta_1
   \left(\lambda _1^2 L^2 \cos \xi +x_2 \sin \xi \right)}{\sqrt{\Xi }} & \frac{\lambda  \lambda _1 \sin \theta_1
   \left(\lambda _1^2 L^2 \sin \xi -x_2 \cos \xi \right)}{\sqrt{\Xi }} & \frac{\lambda _1^2 x_1 \sin (\text{$\theta
   $1})}{\sqrt{\Xi }} \\
 0 & \frac{x_2 \cos \xi -\lambda _1^2 L^2 \sin \xi }{\sqrt{\lambda _1^4 L^4+x_2^2}} & -\frac{\lambda _1^2 L^2 \cos \xi +x_2
   \sin \xi }{\sqrt{\lambda _1^4 L^4+x_2^2}} & 0 \\
 0 & -\frac{\lambda _1 x_1 \left(\lambda _1^2 L^2 \cos \xi +x_2 \sin \xi \right)}{\sqrt{\Delta } \sqrt{\lambda _1^4 L^4+x_2^2}}
   & \frac{\lambda _1 x_1 \left(\lambda _1^2 L^2 \sin \xi -x_2 \cos \xi \right)}{\sqrt{\Delta } \sqrt{\lambda _1^4 L^4+x_2^2}}
   & -\frac{\lambda  \sqrt{\lambda _1^4 L^4+x_2^2}}{\sqrt{\Delta }} \\
 \frac{\sqrt{\Delta } \lambda _1 \sin \theta_1}{\sqrt{\Xi }} & -\frac{\lambda ^2 \lambda _1 x_1 \cos \theta_1
   \left(\lambda _1^2 L^2 \cos \xi +x_2 \sin \xi \right)}{\sqrt{\Delta } \sqrt{\Xi }} & \frac{\lambda ^2 \lambda _1 x_1 \cos
   \theta_1 \left(\lambda _1^2 L^2 \sin \xi -x_2 \cos \xi \right)}{\sqrt{\Delta } \sqrt{\Xi }} & \frac{\lambda
   \lambda _1^2 x_1^2 \cos \theta_1}{\sqrt{\Delta } \sqrt{\Xi }}
\end{array}
\right),
\eeq
where this matrix acts on the flat directions $[\varphi123]$.
This gives a new vielbein basis
\begin{align}
&\tilde{e}^{a}\!\!~'=\tilde{e}^a= e^a,~~~~~a=x^{\mu},r,\theta\nn,\\[2mm]
&\tilde{e}^{\varphi}\!\!~'= \frac{L\l\l_1^2x_1 \sin\theta}{\sqrt{\Xi}}d\xi,~~~~~\tilde{e}^{1}\!\!~'=\frac{L \l_1}{\sqrt{x_2^2+L^4 \l_1^4}} dx_1,\nn\\[2mm]
&\tilde{e}^{2}\!\!~'=\frac{x_1x_2 dx_1+(x_2^2+ L^4 \l_1^4)dx_2}{L\sqrt{\Delta}\sqrt{x_2^2+ L^4 \l_1^4}}\\[2mm]
&\tilde{e}^{3}\!\!~'=\frac{L\l_1^2\big(\Xi d\varphi+\l^2\l_1^2x_1^2\cos\theta d\xi\big)}{\sqrt{\Delta}\sqrt{\Xi}}\nn,
\end{align}
where
\beq
\Xi= \Delta \l_1^2\sin\theta^2+ \l^2\l_1^2x_1^2\cos\theta^2.
\eeq
In this basis it is possible to follow the
standard Abelian T-duality rules in the presence of RR fields
\cite{Hassan:1999bv}.
In the frame Eq.(\ref{eq:veil2}),
the $U(1)$ $\Omega$ matrix is given by,
\beq
\Omega_{U(1)}=\Gamma^{(10)}\frac{\l_1\sqrt{\Delta}\sin\theta \tilde{\Gamma}^{\varphi}+\l x_1\cos\theta (-\cos\xi\tilde{\Gamma}^{1}+\sin\xi\tilde{\Gamma}^{2})}{\sqrt{\Xi}}.
\eeq
As one expects the MW Killing spinors  transform as,
\beq
\hat{\hat{\epsilon}}_1= \epsilon_1,~~~~\hat{\hat{\epsilon}}_2= \Omega_{U(1)}\Omega_{SU(2)}\epsilon_2
\eeq
following the logic of \cite{Barranco:2013fza} it would be advantageous to find a frame in which both $\Omega_{U(1)}$ and $\Omega_{SU(2)}$ are both
simple. Such a frame is provided by the Lorentz transformation $\tilde{\tilde{e}}=\mathcal{R}_2\tilde{e}$ where
\beq
\mathcal{R}_2 = \left(
\begin{array}{cccccc}
 1 & 0 & 0 & 0 & 0 & 0 \\
 0 & \frac{\sqrt{\Delta } \lambda _1 \sin \theta }{\sqrt{\Xi }} &
   0 & \frac{\lambda  \lambda _1x_1 \cos \theta  \sin
  \xi}{\sqrt{\Xi }} & \frac{\lambda  \lambda _1x_1 \cos
   \theta  \cos\xi}{\sqrt{\Xi }} & 0 \\
 0 & 0 & \frac{\sqrt{\Delta } \lambda _1 \sin \theta }{\sqrt{\Xi
   }} & -\frac{\lambda  \lambda _1x_1 \cos \theta  \cos
  \xi}{\sqrt{\Xi }} & \frac{\lambda  \lambda _1x_1 \cos
   \theta  \sin\xi}{\sqrt{\Xi }} & 0 \\
 0 & -\frac{\lambda  \lambda _1x_1 \cos \theta  \sin \xi
   }{\sqrt{\Xi }} & \frac{\lambda  \lambda _1x_1 \cos \theta
    \cos\xi}{\sqrt{\Xi }} & \frac{\sqrt{\Delta } \lambda _1 \sin
   \theta }{\sqrt{\Xi }} & 0 & 0 \\
 0 & -\frac{\lambda  \lambda _1x_1 \cos \theta  \cos \xi
   }{\sqrt{\Xi }} & -\frac{\lambda  \lambda _1x_1 \cos \theta \sin\xi}{\sqrt{\Xi }} & 0 & \frac{\sqrt{\Delta } \lambda _1 \sin
   \theta }{\sqrt{\Xi }} & 0 \\
 0 & 0 & 0 & 0 & 0 & -1
\end{array}
\right)
\eeq
One can check that this does indeed satisfy $\mathcal{R}_2\mathcal{R}_2^T=I$ and $\det\mathcal{R}=-1$. With respect to this basis we have
\beq
\Omega_{SU(2)}= -\Gamma^{(10)}\tilde{\tilde{\Gamma}}^r,~~~\Omega_{U(1)}= \Gamma^{(10)}\tilde{\tilde{\Gamma}}^{\varphi},
\eeq
and so
\beq
\hat{\hat{\epsilon}}_2= \tilde{\tilde{\Gamma}}^{r\varphi}\epsilon_2.
\eeq
The projections are only slightly modified to
\beq
\tilde{\tilde{\Gamma}}_{r123}\epsilon =- \epsilon,~~~ \tilde{\tilde{\Gamma}}_{\theta\varphi}\epsilon=\tilde{\tilde{\Gamma}}_{12}\epsilon.
\eeq
In this frame the Vielbein are,
\begin{align}\label{eq:veil3}
&\tilde{\tilde{e}}^r=\frac{\lambda  \lambda _1^2 L^4 dr-r (x_1
   dx_1+x_2 dx_2)}{\sqrt{\Delta } L r},\nn\\[2mm]
&\tilde{\tilde{e}}^{\theta}=\frac{\lambda _1^2 L (\Delta  r d\theta \sin \theta
   -\lambda  x_1 \cos \theta (\lambda  r
   dx_1+x_1 dr))}{\sqrt{\Delta } \sqrt{\Xi }
   r},\nn\\[2mm]
        &\tilde{\tilde{e}}^{\varphi} =\frac{\lambda ^2 \cos \theta \left(dx_2 \left(\lambda _1^4
   L^4+x_2^2\right)+x_1 x_2
   dx_1\right)-\Delta  d\varphi }{\sqrt{\Delta } L
   \sqrt{\Xi }},\nn\\[2mm]
&\tilde{\tilde{e}}^1 =-\frac{\lambda _1^2 L}{\sqrt{\Xi } r}\bigg(\cos\xi\lambda  r x_1\cos \theta d\xi +\sin\xi\big(\lambda  r (x_1 \cos \theta d\theta +\sin
   \theta dx_1)+x_1 \sin \theta dr\big)\bigg),\nn\\[2mm]
&\tilde{\tilde{e}}^2 =-\frac{\lambda _1^2 L}{\sqrt{\Xi } r}\bigg(-\sin\xi\lambda rx_1 \sin \theta d\xi +\cos\xi\big(\lambda  r (x_1\cos \theta d\theta +
   \sin \theta dx_1)+x_1 \sin \theta dr\big)\bigg),\nn\\[2mm]
&\tilde{\tilde{e}}^3 =\frac{L \left(\lambda _1^2 r dx_2+\lambda  x_2
   dr\right)}{\sqrt{\Delta } r}.
\end{align}
It is possible to show that these
solutions support an orthogonal $SU(2)$ structure where
\begin{align}
z&=v+i w= -\tilde{\tilde{e}}^1+i \tilde{\tilde{e}}^2,\nn\\[2mm]
j&=\tilde{\tilde{e}}^{3r}+\tilde{\tilde{e}}^{\varphi\theta},\\[2mm]
\omega&= (\tilde{\tilde{e}}^{\varphi}+i\tilde{\tilde{e}}^{\theta})\wedge(\tilde{\tilde{e}}^{3}+i\tilde{\tilde{e}}^{r})\nn.\\[2mm]
\tilde{\tilde{\theta}}_p&=\frac{\pi}{2},~~~\tilde{\tilde{\theta}}_m=0.\nn
\end{align}
Let us move to a similar study for the case of the NATD of
$AdS_5\times Y^{p,q}$.

\subsection{G-structures of $Y^{p,q}$ NATD-T}\label{sec:GYpq}
The vielbeins of $Y^{p,q}$ in the frame favoured by NATD are,
\begin{align}\label{eq: YPQNATDFrame}
&e^{x^{\mu}}= \frac{r}{L}dx^{\mu},~~~e^{r}=\frac{L}{r}dr,~~~,e^y=\frac{L}{\sqrt{vw}}dy~~~ e^{\alpha}=L k d\alpha,\\[2mm]
&e^{1,2}=\frac{L\sqrt{m}}{\sqrt{6}}\sigma_{1,2},~~~e^3= L(\sqrt{g} \sigma_3+h d\alpha).\nn
\end{align}
With respect to this basis the projection conditions that the
Majorana-Killing spinor $\epsilon$ obeys \cite{Canoura:2005uz},
can be succinctly expressed in terms the functions
\beq
\cos\kappa(y)=\frac{m}{3\sqrt{g}},~~~\sin\kappa(y)=-\frac{\sqrt{vw}}{6\sqrt{g}},
\eeq
as
\beq\label{Ypqproj}
\Gamma_{r \alpha}\epsilon=\Gamma_{y3}\epsilon,~~~\Gamma_{r123}\epsilon=(\cos \kappa+\sin \kappa \Gamma_{3\alpha})\epsilon.
\eeq
From these it is possible to define an $SU(3)$-structure,
however unlike in the case of the Klebanov-Witten background,
this will not be canonical in the NATD frame Eq.(\ref{eq: YPQNATDFrame}).
Instead it takes the form
\begin{align}
&J= e^r\wedge(\cos\kappa e^3+\sin \kappa e^{\alpha})+(-\sin \kappa e^3+\cos \kappa e^{\alpha})\wedge e^{y}+ e^{21},\\[2mm]
&\Omega_{hol}=(e^r+i (\cos \kappa e^3+\sin \kappa e^{\alpha}))\wedge((-\sin \kappa e^3+\cos \kappa e^{\alpha})+i e^{y})\wedge(e^2+i e^1)\nn.
\end{align}
Of course, this can
be put into canonical form by performing a
rotation in $e^{\alpha},e^{3}$ which is rather
more reminiscent of the wrapped D5 solution
(see for example \cite{Conde:2011rg}) rather than the Klebanov-Witten case.
The difference between NATD and canonical structure
frames makes the G-structure analysis of $Y^{p,q}$ more complicated
than the previous example. Indeed it was shown in
\cite{Gaillard:2013vsa} that a similar rotation in the wrapped
D5 solution leads to a dynamical $SU(2)$-structure in the NATD.
However we will learn shortly that this is not the case
for the NATD of $Y^{p,q}$. Indeed, as observed in
\cite{Sfetsos:2014tza} the structure is orthogonal.\\

The next step is to perform the NATD on the $\sigma_i$'s. To keep things compact we once more express the dual coordinates as
\beq
x_1=\rho\sin\chi,~~~~~x_2=\rho\cos\chi.
\eeq
The dual vielbeins are given by
\beq\label{eq:canframnatdypq}
\begin{array}{ll}
&\hat{e}^{a}=e^{a},~~~~a=x^{\mu},r,y,\alpha,\\
&\hat{e}^{1}=-\sqrt{m}\frac{x_1\big(dx_2+L^2(gd\xi+\sqrt{g}h d\alpha)\big)\big(L^2m\cos\xi+gx_2 \sin\xi)\big)+\big(gx_1^2\sin\xi+L^2g(-6x_2\cos\xi+L^2m \sin\xi)\big)dx_1}{6\sqrt{6}\Pi}\\[2mm]
&\hat{e}^{2}=-\sqrt{m}\frac{x_1\big(dx_2+L^2(gd\xi+\sqrt{g}h d\alpha)\big)\big(L^2m\sin\xi-gx_2 \cos\xi)\big)-\big(gx_1^2\cos\xi+L^2g(6x_2\sin\xi+L^2m \cos\xi)\big)dx_1}{6\sqrt{6}\Pi}\\[2mm]
&\hat{e}^3=-\frac{36x_2(x_1dx_1+x_2dx_2)\sqrt{g}-6 L^2 x_1^2(\sqrt{g}d\xi+hd\alpha)m+L^4\sqrt{g}m^2 dx_2}{36 L \Pi}.
\end{array}
\eeq
The action on the spinor MW killing spinors is once more
\beq
\hat{\epsilon}_1=\epsilon_1,~~~\hat{\epsilon}_2=\Omega_{SU(2)}\epsilon_2,
\eeq
where in the frame of Eq.(\ref{eq:canframnatdypq})
$\Omega_{SU(2)}$ is given by Eq.(\ref{eq su2trans}) with $\zeta^a$
\beq\label{eqzetaypq}
\zeta_1=\frac{\sqrt{6}}{L^2 \sqrt{gm}}x_1\sin\xi,~~~\zeta_2=\frac{\sqrt{6}}{L^2 \sqrt{gm}}x_1\cos\xi,~~~\zeta_3=\frac{6}{L^2 m}x_2.
\eeq
One can once more calculate the Kosmann derivative at this stage and find that the only isometry preserving SUSY under a further Abelian T-duality is $\partial_{\alpha}$. Using the projections in Eq.(\ref{Ypqproj}) one arrives at,
\beq
\hat{\epsilon}_2=(\cos\kappa \hat{\Gamma}^r+\sin\kappa\hat{\Gamma}^y+\zeta_a\hat\Gamma^a)\epsilon_2,
\eeq
which  suggests performing a rotation such that,
\begin{align}
&\hat{e}^r~\!\!'=\cos\kappa \hat{e}^r+\sin\kappa\hat{e}^y,\nn\\[2mm]
&\hat{e}^y~\!\!'=\cos\kappa \hat{e}^y-\sin\kappa\hat{e}^r,\nn\\[2mm]
& \hat{e}^a~\!\!'=\hat{e}^a,~~a\neq r,y
\end{align}
so that $\hat{\epsilon}_2$ takes the same form as it
did for the Klebanov-Witten NATD.
Indeed if we then rotate $\hat{e}^r~\!\!'$ with the matrix $\mathcal{R}_1$
in Eq.(\ref{eq:Rot1}) but with $\zeta_a$ now
defined by Eq.(\ref{eqzetaypq}) we find the orthogonal $SU(2)$ structure,
\begin{align}
&z=v+i w=\tilde{e}^r+i \tilde{e}^3,\nn\\[2mm]
&j=\tilde{e}^{\alpha y}+\tilde{e}^{21},\\[2mm]
&\omega=(\tilde{e}^{\alpha}+i \tilde{e}^y)\wedge(\tilde{e}^2+i\tilde{e}^1)\nn\\[2mm]
&\tilde{\theta}_p=0,~~~\tilde{\theta}_m=\frac{\pi}{2}.
\end{align}
The new vielbeins are given by
\begin{align}
&\tilde{e}^{r}=\frac{L^4\sqrt{g} m(\cos\kappa d\log r+\sin\kappa\frac{dy}{\sqrt{vw}})-6(x_1dx_1+x_2dx_2)}{6 L \sqrt{\Pi}}\nn\\[2mm]
&\tilde{e}^{y}=L\cos\kappa d\log r-L\sin\kappa \frac{dy}{\sqrt{vw}},~~~ \tilde{e}^{\alpha}= L k d\alpha,\nn\\[2mm]
&\tilde{e}^{1}=-\frac{L \sqrt{m}}{\sqrt{6}\sqrt{\Pi}}\bigg[x_1\bigg(\sin\xi\big(\cos\kappa d\log r+\sin\kappa\frac{dy}{\sqrt{vw}}\big)+\cos\xi h d\alpha\bigg)+\sqrt{g}\big(\sin\xi dx_1+x_1 \cos\xi d\xi\big)\bigg]\nn\\[2mm]
&\tilde{e}^{2}=-\frac{L \sqrt{m}}{\sqrt{6}\sqrt{\Pi}}\bigg[x_1\bigg(-\cos\xi\big(\cos\kappa d\log r+\sin\kappa\frac{dy}{\sqrt{vw}}\big)+\sin\xi h d\alpha\bigg)+\sqrt{g}\big(-\cos\xi dx_1+x_1 \sin\xi d\xi\big)\bigg]\nn\\[2mm]
&\tilde{e}^3=-\frac{L}{6\sqrt{\Pi}}\big(6x_2\sqrt{g}(\cos\kappa d\log r+\sin\kappa\frac{dy}{\sqrt{vw}})+m dx_2\bigg),
\end{align}
where
\beq
\Pi=\frac{1}{36}\big(6x_1^2+g(36 x_2^2+L^4m^2)\big).
\eeq

Finally we turn consider the structure of the $Y^{p,q}$ NATD-T on $\alpha$ solution, we omit the details of the the derivation and just present the result.
The geometry
also presents an orthogonal $SU(2)$ structure with forms given by
\begin{align}
&z=v+i w=\tilde{\tilde{e}}^{\alpha}-i \tilde{\tilde{e}}^r,\nn\\[2mm]
&j=\tilde{\tilde{e}}^{y3}+\tilde{\tilde{e}}^{21},\\[2mm]
&\omega=(\tilde{\tilde{e}}^{y}+i \tilde{\tilde{e}}^3)\wedge(\tilde{\tilde{e}}^2+i\tilde{\tilde{e}}^1)\nn,\\
&\tilde{\tilde{\theta}}_p=\frac{\pi}{2}+\arctan \frac{6x_2}{L^2 m},~~~\tilde{\tilde{\theta}}_m=\xi.
\end{align}
where here the vielbein basis is
\begin{align}
&\tilde{\tilde{e}}^r=\frac{L \sqrt{m}}{\sqrt{\Theta}}\bigg[x_1\sqrt{h^2+k^2}\big(\sin\beta d\log r+\cos\beta\frac{dy}{\sqrt{vw}}\big)+\sqrt{g}k dx_1\bigg],\nn\\[2mm]
&\tilde{\tilde{e}}^y=L\big(-\cos\beta d\log r+\sin\beta \frac{dy}{\sqrt{vw}}\big),~~~
\tilde{\tilde{e}}^{\alpha}=\frac{Ldx_1 k \sqrt{gm}}{\sqrt{\Theta}}d\alpha,\nn\\[2mm]
&\tilde{\tilde{e}}^{1}=\frac{L}{\sqrt{6}\sqrt{\Theta}\sqrt{36x_2^2+L^4 m^2}}\bigg[\sqrt{g}k\big(36x_2^2+L^4m^2\big)\big(\sin\beta d\log r+\cos\beta \frac{dy}{\sqrt{vw}}\big)-6m\sqrt{h^2+k^2}x_1dx_1\bigg],\nn\\[2mm]
&\tilde{\tilde{e}}^{2}=\frac{\big(h^2+k^2\big)\big(36x_1x_2dx_1+(36x_2^2+L^4m^2)dx_2\big)-\sqrt{g}h\big(36x_2^2+L^4m^2\big)d\alpha}{\sqrt{6}L\sqrt{\Theta}\sqrt{h^2+k^2}\sqrt{36 x_2^2+L^4 m^2}},\nn\\[2mm]
& \tilde{\tilde{e}}^{3}=\frac{d\alpha}{L\sqrt{h^2+k^2}},
\end{align}
where we have defined the following functions for concision
\beq
\cos\beta(y)=\frac{k\cos\kappa-h\sin\kappa}{\sqrt{h^2+k^2}},~~~\sin\beta(y)=\frac{h\cos\kappa+k\sin\kappa}{\sqrt{h^2+k^2}},
\eeq
and
\beq
\Theta=6k^2\Pi+h^2m x_1^2.
\eeq
In summary, we have shown in this Section that two of
the new backgrounds we obtained by a successive application of
a NATD and a T-duality on $\alpha$ are SUSY preserving; both of them support an orthogonal
$SU(2)$-structure.
\subsection{Comments on supersymmetry and relation to other works}
It has been established that NATD-T of the $T^{1,1}$ and $Y^{p,q}$ backgrounds are supersymmetric solutions when the final $U(1)$ transformation is performed on $\phi$ or $\alpha$ respectively. Similar arguments run into trouble when the final T-duality is applied along the $\xi$ direction. Indeed, this plays the role of the $U(1)$ R-symmetry in the dual solutions which is inherited from the original backgrounds.  SUSY preservation is fairly intuitive in the frame preferred by T-duality on either $U(1)$ or $SU(2)$ isometries. It
merely requires the Killing spinor to be independent of the isometry directions \cite{Hassan:1999bv,Kelekci:2014ima}. The frame dependence of this statement is akin to saying that a metric is stationary because it does not depend explicitly on time. The general statement in the case of a stationary metric is the existence of a time-like Killing vector. Here too, a general frame independent statement can be made  in terms of the Kosmann derivative \cite{Kelekci:2014ima}. When the Kosmann derivative vanishes along the isometry then SUSY is preserved under a T-duality transformation.


It is worth commenting on related work which sought supersymmetric backgrounds with AdS${}_5$ factors. The most relevant work in this context is the classification of \cite{Gauntlett:2005ww} where a large class of supersymmetric solutions of IIB supergravity with AdS${}_5$ were classified. A natural question we need to address is the place of the solutions generated in this manuscript in the above classification. There are two properties of the solutions classified in \cite{Gauntlett:2005ww} that our solutions do not satisfy. Namely, in order to make further progress, \cite{Gauntlett:2005ww} considered solutions with non-vanishing $F_5$ form and with trivial axion. As can be seen clearly in the explicit expressions for our backgrounds we have  {\it (i) $F_5=0$, (ii) Nontrivial axion, that is,  $F_1 \neq 0$. This class of solutions was not considered in  \cite{Gauntlett:2005ww}}\footnote{We thank D. Martelli for insightful comments and clarifications on this point.}.

There is also work along the lines of a systematic classification of solutions with AdS${}_5$ which is relevant but somehow more restrictive since they demand the existence of an $S^2$ factor inside the $M_5$ submanifold \cite{Colgain:2011hb}. In particular, \cite{Colgain:2011hb} demands that the $S^2$ not be fibered over the  $M_3$ manifold that complements it inside $M_5$. This direct product requirement is motivated by the goal of having $\mathcal{N}=2$ SCFT on the field theory side whereby this $SU(2)$ would  be dual to $SU(2)_R$; without this direct product it is not possible to get doublets under $SU(2)$ which are required by SUSY. Clearly, our solutions evade this classifications as the $S^2$ part is always fibered over the remaining $M_3$ base.

Let us finally comment on a general property of the classification approach at large. In the  papers under comparison, \cite{Gauntlett:2005ww} and   \cite{Colgain:2011hb}, but in the generic situation in IIB, it is assumed that supersymmetry imposes a holomorphic condition on the  axion-dilaton \cite{Grana:2001xn}. Then, using the statement that in a compact manifold the only regular harmonic  functions are constant functions one arrives at a constant axion-dilaton.  By looking at the form of the dilaton in some of the various solutions presented, it does not look anything like a holomorphic  function. In fact, in some cases it seems to depend on three coordinates, for example,   $(\chi,\rho, \theta_1)$.  Therefore, the supersymmetry mechanism underlying our solutions seems to be of a quite different nature and deserves to be scrutinized further.

\section{Comments on the dual field theory:
A generalization of toric duality?}\label{Sec:FT}

In this section, we will try to put together different comments
on the field theories dual to our different backgrounds.

An important source of information comes from
the role of Page charges. These are particularly useful in
backgrounds with regular as well as fractional D-branes.
The prototypical example was worked out in this language
explicitly in \cite{Benini:2007gx}. It was shown  there,
that the transformation of the Page charges matches precisely
the transformation of the rank of gauge groups under Seiberg duality.
Namely, under large gauge transformation in the
Klebanov-Strassler background, the Page charges transform exactly as
the ranks of the gauge groups of the dual field theory.
Given the transformation of the Page charges in the background
we discussed in Section \ref{Sec:Page}, we believe it
is plausible that there is a version of Seiberg duality at work.
This duality does not involved the energy scale as is the case
in the Klebanov-Strassler background \cite{Strassler:2005qs}.
Hence, we propose that this Seiberg-like duality,
represented by large gauge transformations of the $B_2$ NS field,
 is a transformation between conformal theories.
This interpretation was first made in \cite{Lozano:2014ata}
where the NATD of $AdS_4\times \mathbb{CP}^3$ was studied,
the solutions we consider here also
exhibit such behaviour.

As mentioned above, we can always `counter' a motion in the
$\rho$-coordinate with a large gauge transformation of the $B$-field.
this mixing of geometry and NS-fields points to some relation with
non-geometric backgrounds.

The transformations in the $\rho$-coordinate and the
large gauge transformations of the $B$-field that
 we discussed in
Section \ref{Sec:Page}, implied a change in the Page charges of
D6 and D4 branes given by
\beq
\Delta Q_{P, D6}=0,\;\;\; \Delta Q_{P, D4}= - n N_{D6};
\eeq
that reads exactly like the changes in Page charges for D5 and D3 branes in
the cascade of dualities for the Klebanov-Strassler system.

Moreover,
in the  KS-case, it was shown
that motions in the radial direction (an `Energy' direction labelled
$\tau$ in the KS background) implied changes in
the quantity $b_0$ \cite{Benini:2007gx}. In order
to keep the quantity $b_0$ bounded, a Seiberg duality was applied
when flowing down
or up in the radial coordinate $\tau$.
In the cases considered in this paper, and as first observed in \cite{Lozano:2014ata}, motion in the $\rho$-direction (that is
{\it not} a motion in energies in the dual field theory) also implies the need
for a change in description to keep $b_0$ bounded;
identifying $\rho\sim \rho+\pi$ and changing
description of the QFT every time we cross $\rho=n\pi$ with ($n=1,2,3,4...$).
This change in description can be seen, by analogy, as a Seiberg duality.

Notice that the coordinate $\rho$ is {\it not} periodic---in the same way that the radial
$\tau$-coordinate is certainly not periodic in the KS-system.
However, as suggested in \cite{Lozano:2014ata}, there is a `minimal cell'
of length $\pi$ in the $\rho$-coordinate. This is somewhat
analogous to what happens
for an $H_2$
manifold (or any other negatively curved Riemann manifold),
that can be locally written in terms of coordinates $(x,y)$ with metric
$ds^2\sim y^{-2}(dx^2+dy^2)$. In principle $(x,y)$ are unbounded, but
the $H_2$ is described in terms of a minimal cell, with finite volume.

Notice also that while the Klebanov-Witten field theory is self-dual
under Seiberg duality, the quiver dual to the NATD or our new backgrounds need not be.
Indeed, examples of this sort have been studied in the
context of toric duality.
A toric duality on the gravity (geometry) side
and a Seiberg duality on the field theory side have
been shown to be equivalent
\cite{Beasley:2001zp}, \cite{Feng:2001bn}, \cite{Franco:2002mu}.
The origin of the toric duality is the
inherent ambiguity in the definition of a toric diagram that
arises from unimodular transformations on the lattice defining the diagram.
For example, in the case of
Calabi-Yau toric three folds defined on $\mathbb{Z}_3$,
the set of $SL(3,\mathbb{C})$ transformation leaving the endpoints
of the vector defining the toric diagram invariant gives the
same toric variety.
In those cases, toric duality has nothing to do with the energy scale
of the dual field theory, as the CFT is conformal and all the gauge
groups of the quiver have the same rank.

When we apply these ideas to our context,
we argued by analogy that there is certain ambiguity in the range
of the coordinate $\rho\in [n\pi, \pi (n+1)]$; which whenever a
`boundary' is crossed, changes
the field theory description, which leads to Seiberg dual versions
(different toric manifolds)
of the same theory. Either that or that we 'undo' the crossing
with a large gauge transformation of the $B$-field, that would also
change the vacuum where the QFT is defined.

An obvious comment we need to make regarding this analogy:
the cone over the resulting
compact 5-manifold, ${\cal C}(W_5)$,
is not necessarily toric; it is certainly not Calabi-Yau.
Therefore, the situation we are describing must be related to
a non-toric and not Calabi-Yau extension of the
standard  argument for D3 branes place at a toric singularity.

Another quantity that provided important information is the central charge.
We have found, along our different dualities that it is always possible to
write the central charge in terms of the Page charge $N_c$ characterizing the
background in the form $c\sim N_c^2$. In the examples before
the NATD $N_c=N_{D3}$ is the Page charge of D3 branes. After the NATD we
found $N_c=N_{D6}$ is the number of D6 branes. In all cases, this
is a characteristically `gauge theoretic' behavior
and different to what happens in
Gaiotto-like theories (either in the extended or minimally SUSY cases).
We conclude that our theories are different from those proposed by
Gaiotto, in spite of the dual metrics in M-theory being quite similar
(an important difference is that the Gaiotto-like duals contain
an hyperbolic plane, whilst ours contain a two sphere).
It is also of interest the comparison with the backgrounds of
\cite{ReidEdwards:2010qs}, \cite{Aharony:2012tz}.

In the same line of central charges, we observe in all of our
examples given in Section \ref{Sec:Page}, that the quotient of the
central charges, before and after the NATD--- is given by
\beq
\frac{c_{\text{before}}}{c_{\text{after}}}= 3\frac{N_{D3}^2}{N_{D6}^2}.
\eeq
This gives an interesting hint about the complete
description of these new field theories, for which we
are finding a dual description.

Let us move to comment something about gauge couplings.
Let us focus on the Klebanov-Witten system and its NATD.
In the KW case, it was shown that the gauge couplings
of the two groups are given by (we take, like above, $g_s=1$),
\beq
\frac{4\pi^2}{g_1^2}+\frac{4\pi^2}{g_2^2}=\pi e^{-\Phi},\;\;\;
\frac{4\pi^2}{g_1^2}-\frac{4\pi^2}{g_2^2}=\pi e^{-\Phi}(\frac{1}{2\pi^2\alpha'}
\int_{\Sigma_2}B_2 -1).\;\;\;
\eeq
Using these definitions,
we obtain the expressions for two individual
couplings $\frac{8\pi^2}{g_{1,2}^2}$
\beq
\frac{8\pi^2}{g_1^2}=2\pi e^{-\Phi}b_0,\;\;\;
\frac{8\pi^2}{g_2^2}=2\pi e^{-\Phi}(1-b_0),
\label{g1g2}\eeq
where as above $4\pi^2\alpha' b_0=\int_{\Sigma_2} B_2$.
It is from expressions like these in
Eq.(\ref{g1g2}), that the condition that $b_0$
is bounded in the interval $(0,1)$ is imposed.
The coupling $g_1^2$ can be calculated using a
D5 brane that wraps a two cycle inside
the conifold with an electric field in its worldvolume.
Conversely, one can equate the inverse coupling with the
BI part of the Action of an Euclidean D1 wrapping
the same cycle.

In our NATD geometries written in Section \ref{Sec:NATD-T11},
we can define a configuration
representing an instanton in two ways.
First, by wrapping an Euclidean D2 brane on the cycle parameterized by
$\Sigma_3=[\theta_1,\phi_1,\xi]$, with $\chi=\frac{\pi}{2}$ and
at constant $\rho$.
We can also consider an Euclidean D0 brane, that extends along the direction
$\phi_1$, with $\chi=\frac{\pi}{2}$ and constant $\rho$
\footnote{Calculating the gauge coupling and theta angle,
with a D6 brane extended in
$R^{1,3}\times \Sigma_3$ and switching on an electric field
on the Minkowski directions
would give a similar result. }.
Let us discuss in detail the calculation with an Euclidean D0 brane.
The induced metric and gauge field on the cycle $\Sigma_1=[\phi_1]$ (
restricted such that $2\chi=\pi$ and $\theta_1=0$) are,
\bea
& & ds_{D0}^2= \frac{\alpha'^2 L^2}{Q}\lambda^2 \lambda_2^2 \rho^2 d\phi_1^2,
\nonumber\\
& & C_1=4\frac{\lambda \lambda_1^2\lambda_2^2 L^4}{\alpha'^{3/2}} d\phi_1.
\eea
We calculate the Born-Infeld and the Wess-Zumino parts of the Action
for this D0,
\bea
& & S_{BI}= - T_{D0}\int d\phi_1 e^{-\Phi}\sqrt{\det[g_{ind}]}=\frac{L^2}{\alpha'}\lambda\lambda_2 2\pi \rho=\sqrt{N_{D6}}\pi \rho\sim b_0.\nonumber\\
& & S_{WZ}=T_{D0}\int C_1= N_{D6}\pi.
\eea
We have used the explicit values of $\lambda,\lambda_1, \lambda_2$
and the relation $2L^4=27 \alpha'^2 N_{D6} $, discussed above.
We then equate this to the Action of an instanton
$S_{inst}=\frac{8\pi^2}{g^2} + i \Theta$, obtaining expressions
\beq
\frac{8\pi^2}{g^2}\sim b_0, \;\;\; \Theta= \pi N_{D6}.
\eeq
In the case of the calculation with
the Euclidean D2 brane, things are quite analogous.
There, we see that the induced metric and $B$-field
on the three manifold are,
\bea
ds_{ind}^2= L^2\lambda_1^2 (d\theta_1^2+\sin^2\theta_1d\phi_1^2)+
\frac{\alpha'^2}{Q}L^2 \lambda^2\lambda_2^2 \rho^2
(d\xi+\cos\theta_1d\phi_1)^2; \;\;\;B_2=0.
\eea
We then calculate the BI-Action and get
\bea
T_{D2}\int d\theta_1 d\phi_1 d\xi e^{-\Phi}\sqrt{\det[g_{ind}]}=
\frac{2 \pi L^4\lambda_1^2\lambda_2 \lambda}{\alpha'^2} \frac{\rho}{\pi}.
\eea
Equating the BI action with the Action for an instanton
$S_{BI}\sim \frac{1}{g^2}$, we obtain that the coupling defined in this way is
\beq
\frac{1}{g^2}\sim b_0.
\eeq
These results make contact with the one summarised in Eq.(\ref{g1g2})
for the
Klebanov-Witten background and reinforces the point that the quantity
$b_0$ should be bounded, as we imposed above.
It would be interesting to attempt a
similar calculation for the cases of $Y^{p,q}$ and $S^5$.

Let us finally comment on interesting future problems and
close this section with a general comment about our procedure.
It would be interesting to use the variables we introduced and
apply our methodology to the full duality cascade
represented by the Baryonic Branch of the Klebanov-Strassler theory
(choosing a mesonic branch is also possible, but more complicated technically).
Working out this problem will produce as anticipated
in \cite{Gaillard:2013vsa} a configuration in Massive IIA.
It would be interesting to study the interplay (if any) between the
usual Seiberg duality in the 'Energy' direction $r$ and the one described in
this section.
Finally, we should point out that our smooth backgrounds and procedure
can be thought of, as a way of defining a field theory.
Indeed, any of our non-singular backgrounds defines the large $N_c$ strong
coupled regime of a (presumably new!) QFT.
Obtaining properties of these field theories
using the supergravity backgrounds is a way of learning about new QFTs.
The set of variables and techniques developed in this work make this
task clearer.

\section{Conclusions}\label{Sec:Conclusions}

In this paper we have presented several genuinely
new supergravity backgrounds.
Whilst these new solutions are singular,  it
is worth pointing out that in some cases the singularity structure is
very mild and one can hope for a smoothing  mechanism.

Our solutions ``evade'' previous classification efforts \cite{Gauntlett:2005ww} due
to the fact that they, generically do not contain an  $F_5$  and
have non-trivial and non-holomorphic axion-dilaton.
More recent classification efforts have focused on solutions with a round
$S^2$ which is not fibered over the rest of the manifold \cite{Colgain:2011hb};
our solutions certainly do not contain such $S^2$ factor.

We tried to clarify the extent to which
field theory data (central charge, entanglement entropy)
were invariant under NATD. Our analysis was limited to the concrete cases
we tackled. We found an intriguing relation for the quotient
between the central charges before and after the NATD, that seems to be universal.
It would be interesting to prove this universality (if correct), in full generality,
that is, including situations with generic B-field.
We host the hope that such approach could shed some light on the extension to which the
Ryu-Takayanagi formula goes beyond simple supergravity backgrounds, embodying
deeper string-theoretic principles.
We also presented a proposal for a Seiberg-like duality acting on the field
theories dual to our backgrounds. This proposal was based on the study of supergravity
quantities. By the same study, a relation with
non-geometric backgrounds is suggested by the interplay between the motions in
the $\rho$-coordinate and large gauge transformations of the $B_2$-field.

There are a few interesting venues that we believe are worth exploring.
One interesting generalization, would be to attempt to generate more general
solutions exploiting `spinor rotations' like in   \cite{Maldacena:2009mw},
\cite{Gaillard:2010qg}, \cite{Caceres:2011zn}.
Indeed, given that we know the $SU(2)$-structure
of some of our new solutions, one
could speculate with `rotating' the structure to obtain new solutions.
This approach should lead to interesting solutions
with exciting gravity duals.
It would also be interesting to have a
better understanding of the central charge,
when calculated using the M-theory backgrounds.
Using our supergravity backgrounds,
we can calculate different observables
in the initial and final CFTs to compare them.
This is probably a fruitful line of work, that will give information
on the structure of the new CFTs. The relation with non-geometric
backgrounds is of obvious interest.
It would also be interesting to place the backgrounds studied here
within the formalism recently developed in \cite{Gutowski:2014ova},
\cite{Beck:2014zda}.

Another  interesting direction would be to pursue some of the guidelines of the analysis of Lunin and Maldacena in
\cite{Lunin:2005jy}
where it was clarified that the gravity transformation corresponding to T-s-T applies to any field
theory with a global $U(1)\times U(1)$ symmetry. We can similarly
argue that our transformation applies to any field theory with a gravity dual and a $SU(2)\times U(1)$
global symmetry; it is important to remark that this global symmetry should be different from
the R-symmetry so as to preserve supersymmetry.
Ultimately, similar to how \cite{Lunin:2005jy} presented the
transformation as a symmetry of string theory compactified on a torus
where there is a natural action on the torus complex parameter by $SL(2,\mathbb{R})$,
it should be possible to compactify on the appropriate manifold and formulate our NATD-T
as a symmetry of the lower dimensional theory.
This approach might shed some light on the structure of NATD as well.
Finally, perhaps the most interesting question is related to the elusive $h$-deformation of ${\cal N}=4$ SYM.
Given the reduced number of symmetries of the supergravity background dual to the h-deformed ${\cal N}=4$ SYM,
the best chances for finding the background should rely on solution generating techniques.
Na\"ively, however, NATD in its present form is not an invertible transformation and
thus prevents an approach mimicking that of Lunin and Maldacena in \cite{Lunin:2005jy}.
But probably an approach along the lines discussed in this paper and
 \cite{Kelekci:2014ima}  may help
to `invert' the NATD.
We hope to return to some of these questions in the future.

\section*{Acknowledgments}
We are particularly grateful to Elena
C\'aceres for collaboration in related topics
and to I. Bah and S. Cremonesi
for extensive discussions on the field theory
interpretation of some of the solutions.
We are thankful to A. Ardehali, I. Bah, Y. Bea Besada,
E. C\'aceres,  S. Cremonesi, J. de Boer, N. Halmagyi, U. Kol, K. Kooner, G. Itsios,
Y. Lozano,  D. Martelli, E. O'Colgain,
K. Sfetsos, A. Sierra, D. Thompson,
for useful comments and suggestions that helped
to improve the presentation of this paper. Carlos Nunez thanks IHES (Paris) for hospitality while this work was being completed.
Carlos Nunez is a Wolfson Fellow of the Royal Society.
This research was supported in part by the National Science Foundation under Grant Nos. 1067889
(University of Iowa) and the Department of Energy under grant DE-SC0007859 (University of Michigan).
\appendix
\section{Review of Non-Abelian and Abelian T-duality rules}\label{App:Rules}
\subsection{Non-Abelian T-Duality}
We follow \cite{Itsios:2013wd} in the generalized 3-step B{\"u}scher procedure and consider only backgrounds with an SU(2) isometry such that the metric can be written in the form
\begin{equation}
ds^2=G_{\mu\nu}(x)dx^{\mu}dx^{\nu}+2G_{\mu i}(x)d
x^{\mu}L^i+g_{ij}(x)L^iL^j
\end{equation}
where $\mu,\nu = 1,...7$  and $i,j=1,2,3$.  The $L^i$'s are the SU(2) Maurer-Cartan forms. ($L_{\pm}^i=-i\text{Tr}(t^ig^{-1}\partial_{\pm}g)$).  We also consider a similar decomposition of the antisymmetric 2-form,
\begin{equation}
B=\frac{1}{2}B_{\mu\nu}(x)dx^{\mu}\wedge dx^{\nu}+B_{\mu i}(x)dx^{\mu}\wedge L^i+\frac{1}{2}b_{ij}L^i\wedge L^j
\end{equation}
The Lagrangian density for the NS sector fields is given below, where we omit the dilaton contribution.  (The transformation of the dilaton is given below in \ref{dualdilaton}.)
\begin{equation}
\label{L0short}
\mathcal{L}_0=Q_{AB}\partial_{+} X^A\partial_{-}X^B
\end{equation}
where $A,B=1,...,10$ and
\begin{equation}
Q_{AB}=
\left(
\begin{array}{c|c}
\\ \ Q_{\mu\nu}\quad & Q_{\mu i} \\ \\  \hline \
 Q_{i\mu} \quad  & E_{ij} \\
\end{array}
\right),
\quad  \text{and}\quad  \partial_{\pm}X^A=\left(\partial_{\pm}X^{\mu},\ L_{\pm}^i\right)
\end{equation}
with \begin{equation}
Q_{\mu\nu}=G_{\mu\nu}+B_{\mu\nu},\quad Q_{\mu i}=G_{\mu i}+B_{\mu i},\quad Q_{i\mu}=G_{i \mu}+B_{i\mu},\quad E_{ij}=g_{ij}+b_{ij}
\end{equation}
We then gauge the SU(2) isometry by changing derivatives to covariant derivatives according to, $\partial_{\pm}g\to D_{\pm}g=\partial_{\pm}g-A_{\pm}g$. The next step is to add a Lagrange multiplier term to \ref{L0short} to ensure the gauge fields, $A_{\pm}$ are non-dynamical.
\begin{equation}
-i\text{Tr}(\alpha'vF_{\pm}),\quad F_{\pm}=\partial_+A_--\partial_-A_+-[A_+,A_-]
\end{equation}
We must now eliminate three of the variables by making a gauge fixing choice, described in detail in \ref{App:GFix} below. A natural choice is $g=\mathbb{I}$, so that all 3 of the Lagrange multipliers, $v_i$, become dual coordinates.
The last step is to integrate out the gauge fields to obtain the dual Lagrangian density,
\begin{equation}
\hat{\mathcal{L}}=\hat{Q}_{AB}\partial_{+} \hat{X}^A\partial_{-}\hat{X}^B \label{dualL}
\end{equation}
where we can read off the dual components of $\hat{Q}_{AB}$ from,
\begin{equation}
\hat{Q}_{AB}=
\left(
\begin{array}{c|c}
\\ \ Q_{\mu\nu}-Q_{\mu i}M_{ij}^{-1}Q_{j\nu}\quad & Q_{\mu j}M_{ji}^{-1} \\ \\  \hline \
 -M_{ij}^{-1}Q_{j\mu} \quad  & M_{ij}^{-1} \\
\end{array}
\right),
\quad  \text{and}\quad  \partial_{\pm}\hat{X}^A=\left(\partial_{\pm}X^{\mu},\ \partial_{\pm}v^i\right)
\end{equation}
where we have defined,
\begin{equation}
M_{ij}=E_{ij}+f_{ij},\quad \text{with}\quad f_{ij}=\alpha'\e_{ij}^{\ k}v_k
\end{equation}
(Note that If we wish to carry through the correct factors of $\alpha'$, we must include one factor of $\alpha'$ in front of the $v_i$'s.)  The $v_i$'s originating in the Lagrange multiplier term may now take on the role of dual coordinates, depending on the gauge fixing choice.  We can identify the dual metric and $\hat{B}_2$ field as the symmetric and antisymmetric components of $\hat{Q_{AB}}$, respectively. The transformation of the dilation is given by
\begin{equation}
\label{dualdilaton}
\hat{\Phi}=\Phi-\frac{1}{2}\text{ln}(\frac{\text{det}M}{\alpha'^3})
\end{equation}

\subsubsection{RR Flux Transformation}
In order to transform the RR Fluxes, one must construct a bispinor out of the RR forms and their Hodge duals, (in Type IIB)
\be
P=\frac{e^{\Phi}}{2}\sum^5_{n=0}\slashed{F}_{2n}
\ee
where $\slashed{F}_p=\frac{1}{p!}\Gamma_{\mu_1...\mu_p}F_p^{\ \mu_1...\mu_p}$. Then, the dual fluxes arise from inverting $\Omega$
\be
\hat{P}=P\cdot \Omega^{-1}
\ee
where
$\Omega=(A_0\Gamma^1\Gamma^2\Gamma^3+A_a\Gamma^a)\Gamma_{11}/\sqrt{\alpha'^3}$ and
\be
A_0=\frac{1}{\sqrt{1+\zeta^2}},\quad \quad A^a=\frac{\zeta^a}{\sqrt{1+\zeta^2}}
\ee
where $\zeta^a=\kappa^a_{\ i}z^i$ with $\kappa^a_{\ i}\kappa^a_{\ j}=g_{ij}$ and $z^i=\frac{y^i}{\det \kappa}$, $y_i=b_i+v_i$.

\subsection{Abelian T-Duality}
Our conventions for the B{\"u}scher rules of Abelian T-duality  \cite{Hassan:1999bv} for the NS sector, including the appropriate factors of $\alpha'$, are
\bea
& &\tilde{G}_{99}=\frac{\alpha'}{G_{99}},\quad
\tilde{G}_{9i}=-\frac{\alpha'B_{9i}}{G_{99}},\quad
\tilde{G}_{ij}=G_{ij}-\frac{\alpha'}{G_{99}}(G_{9i}G_{9j}-B_{9i}B_{9j})\nn \\ & &
\tilde{B}_{9i}=-\frac{G_{9i}}{G_{99}},\quad \tilde{B}_{ij}=B_{ij}-\frac{1}{G_{99}}(G_{9i}B_{9j}-B_{9i}G_{9j})
\eea
where $x_9$ is the direction with the U(1) isometry we wish to dualize along.  One can construct the form $\tilde{B}_2$ using,
\be
\tilde{B}_2=\tilde{B}_{ij}\mathit{d}x_i\wedge\mathit{d}x_j+\alpha'\tilde{B}_{9i} \mathit{d}x_9\wedge\mathit{d}x_i.
\ee
The dilaton transformation is
\be
\tilde{\Phi}=\Phi-\frac{1}{2}\ln \frac{G_{99}}{\alpha'}.
\ee

\subsubsection{RR Flux Transformation}

Similar to the NATD RR Flux transformation, one must construct a bispinor out of the RR forms and their Hodge duals, (in Type IIA)
\be
P=\frac{e^{\Phi}}{2}\sum^4_{n=0}\slashed{F}_{2n+1}
\ee
The dual fluxes arise from inverting $\Omega$
\be
\hat{P}=P\cdot \Omega^{-1}
\ee
where
\be
\Omega=\frac{\alpha'}{\sqrt{G_{99}}}\Gamma_{11}\Gamma_9
\ee

\section{Multiparametric Families of Solutions in Type IIA and Type IIB}
\subsection{NATD-s-T on $\xi$ of $AdS_5\times S^5$}
Motivated by the work of Lunin-Maldacena \cite{Lunin:2005jy} we also consider constructing a one-parameter family of solutions. Namely, inspired by the TsT transformation that lead \cite{Lunin:2005jy} to the construction of a large class of gravity solutions with interesting field theory duals,
we perform a shift with parameter $\gamma$ such that
\be
\theta\to\theta+\gamma \xi,
\ee
where $\xi$ is a U(1) angle in the $S^2$ leftover from NATD and $\theta$ is the U(1) angle originating in the $S^2$ (see the first line in Eq.(\ref{Eq:NATD-T-S5}) for our non-standard notation of the angles) that was unaffected by the NATD.
If we then T-dualize along $\xi$, we obtain,
\bea
\tilde{\hat{ds}}^2=&&4ds^2(AdS_5)+4L^2d\alpha^2+\frac{\alpha'^2\mathit{d}\rho^2}{L^2\cos^2\alpha}+\frac{1}{W}\bigg(L^2\alpha'^2\rho^2\sin^22\alpha \sin^2\chi d\theta^2\nn \\&&\left.+\frac{\alpha'^2}{L^2}(\rho^2(L^4\gamma^2\sin^22\alpha+\alpha'^2\rho^2\sin^2\chi) \mathit{d}\chi^2 -\alpha'^22\rho^3\sin\chi \mathit{d}\chi \mathit{d}\xi+(\alpha'^2\rho^2+L^4\cos^4\alpha)\mathit{d}\xi^2)\right),\nn \\
\tilde{\hat{B}}&=&\frac{4\alpha'\gamma\sin^2\alpha \mathit{d}\theta\wedge((\alpha'^2\rho^2+L^4\cos^4\alpha)\mathit{d}\xi-\alpha'^2\rho^3\sin\chi  \mathit{d}\chi)}{W}\nn \\
e^{-2\tilde{\hat{\Phi}}}&=&\frac{L^4}{\alpha'^4}\cos^2\alpha W
\eea
where
\be
W=4\gamma^2(\alpha'^2\rho^2+L^4\cos^4\alpha)\sin^2\alpha+\alpha'^2\rho^2\cos^2\alpha\sin^2\chi.
\ee
As in the previous cases $\alpha=\pi/2$ is a singularity of the dilaton. We verified that this singularity is indeed a curvature singularity by direct computation of the Ricci scalar.  The dual RR fluxes are,
\bea
\tilde{\hat{F}}_1&=&\frac{8L^4}{\alpha'^2}\gamma\cos^3\alpha\sin\alpha d\alpha,  \\
W\tilde{\hat{F}}_3&=&-8L^4\alpha'\rho^2\cos^3\alpha\sin\alpha\sin\chi\mathit{d}\alpha\wedge \mathit{d}\theta\wedge\left(4\rho\gamma^2\sin^2\alpha \mathit{d}\chi+\cos^2\alpha\sin\chi  \mathit{d}\xi \right).\nn
\eea
All of the Type IIB equations have been satisfied for this case.

\subsection{Generic gauge fixing in NATD}\label{App:GFix}
In this section we discuss gauge ambiguities and consider new solutions that can be generated by exploiting these intrinsic ambiguities in the NATD procedure. As explained in detail in \cite{Itsios:2013wd} and \cite{Sfetsos:2010uq}, the NATD procedure requires gauge fixing leading to potentially dim$(G)$ degrees of freedom.  For  our case of $SU(2)$, we have up to three parameters that can be exploited.
Given a specific $SU(2)$ matrix $D^{ij}$, we can perform an orthogonal transformation on the Lagrange multipliers via
\be
\hat{v}_i=D_{ij}v^j \label{vhat}
\ee
where $D^{ij}=Tr(t^igt^jg^{-1})$, and
which explicitly can be written as\footnote{Shorthand notation has been used such that $C_x=\cos x$ and $S_x=\sin x$}
\be
D_{ij}=\left(
\begin{array}{ccc}
 C_{\beta _0} C_{\psi _0} C_{\phi _0}-S_{\psi _0} S_{\phi _0} & C_{\beta
   _0} C_{\phi _0} S_{\psi _0}+C_{\psi _0} S_{\phi _0} & -C_{\phi _0}
   S_{\beta _0} \\
 -C_{\beta _0} C_{\psi _0} S_{\phi _0}-C_{\phi _0} S_{\psi _0} & C_{\psi
   _0} C_{\phi _0}-C_{\beta _0} S_{\psi _0} S_{\phi _0} & S_{\beta _0}
   S_{\phi _0} \\
 C_{\psi _0} S_{\beta _0} & S_{\beta _0} S_{\psi _0} & C_{\beta _0} \\
\end{array}
\right).
\ee
We will define $v=\alpha'(x_1, x_2, x_3)$ and henceforth,  set $\alpha'=1$ and $L=1$. Note that numerical coefficients in front of the $x$'s may be used for convenience.

From Eq(\ref{vhat}) one can see that $\hat{v}$ is covariant with respect to an SO(3) transformation. We would like to investigate how ambiguities might arise through the process of identifying the dual metric coordinates.   In order to see how the parameters $(\b_0,\psi_0,\phi_0)$ and $(x_1,x_2,x_3)$ responds to a generic $SO(3)$ transformation, say $O(\b,\psi,\phi)$, we write $v'=O v$ and $D'=O D O^T $.  Here $(\b,\psi,\phi)$ are generic angles for the  $SO(3)(SU(2))$ transformation matrix.  One may collect  the coordinates  and the parameters of  $D^{ij}$  in the six-tuple $\Phi=(x_1,x_2,x_3,\b_0,\psi_0,\phi_0)$. Then, under an infinitesimal
transformation, where  $(\b\approx \e_1,\psi\approx \e_2,\phi\approx \e_3),$ we find $\delta\Phi=(\d x_1, \d x_2,\d x_3,\d \b_0,\d \psi_0, \d \phi_0)$, where
\bea
\delta x_1&=&x_2\Lambda_1-x_3\Lambda_2\nonumber\\
\delta x_2&=&-x_1 \Lambda_1 \nonumber\\
\d x_3 &=&x_1 \Lambda_2 \\ \label{trans1}
\d \b_0&=&(-\L_1+ \L_2 \csc(\b_0) \sin(\phi_0) +
\L_2 \cot(\b_0)\sin(\psi_0),\psi_0,\phi_0)\nonumber\\
\d \psi_0 &=& \L_2 (\cos(\phi_0)-\cos(\psi_0))\nonumber\\
\d \phi_0 &=& \L_1- \L_2 \cot(\b_0) \sin(\phi_0) -\L_2 \csc(\b_0) \sin(\psi_0).\nonumber
\eea
where $\L_1 \equiv \e_1+\e_2,$ and $\L_2\equiv \e3$.  $\Lambda_1$ displays an isotropy which reduces the access of gauge fixing conditions to only $SU(2)/U(1)$. This suggests that in  this case there will always be gauge redundancy in the choice of dual coordinates \cite{Giveon:1993ai}.

  We define the dual coordinates by pulling back these six coordinates onto the three manifold that will serve as the dual volume. Let $\xi=(\xi_1,\xi_2,\xi_3)$ denote the  dual coordinates and denote $\cal X$ as the pullback function that maps $\Phi$ into $\xi$.   Then the pullback functions  becomes, $$
{\cal X}:\Phi\rightarrow(x_1(\xi),x_2(\xi),x_3(\xi),\b_0(\xi),\psi_0(\xi),\phi_0(\xi))\equiv \Phi(\xi).$$ The dual metric is then constructed from frame fields given by,$$e^a_j = \frac{\partial \hat{v}^a}{\partial \xi^j} = \frac{\partial \hat{v}^a}{\partial \Phi^L}\frac{\partial \Phi(\xi)^L}{\partial \xi^j},$$ where $L=1\cdots6$.   We can then determine the differential volume density of the internal  dual metric via, $g_{ij} =e^a_ie^b_j \d_{ab}$.  If the volume, $\det(e)$, is zero, the pullback is tantamount to a ``poor gauge fixing'' choice as the coordinates have dependence on each other. Three degrees of freedom have been used to specify the dual frame and three remaining degrees of freedom serve as parameters that can span a family of dual volumes.  One may ask how the dual space metric transforms under Eq.(\ref{trans1}) to see if there is residual symmetry.  One strategy would be to compute  the matrix, $$c_{a b } = \frac{\partial^2\det(e')}{\partial \L_a \partial \L_b}|_{\L=0},\,\,\,\, a,b=1,2,$$ where here ${e'}^a_j$ is the response of the dual frame fields to the infinitesimal transformations in Eq.(\ref{trans1}). When $\det(c_{a b})$ vanishes, this  suggests that the two remaining parameters are not independent and residual symmetry  exists.

As an example, consider the pullback $\cal{X}$ such that ${\cal X}:\Phi -> (\xi_1, \xi_2, \xi_3, \frac{1}{2}\pi, 0, 0)$.  This gives the frame fields
\be e^a_j= \begin{pmatrix}0 & 0 & -1 \\
0 & 1 & 0 \\
1 & 0 & 0 \\
\end{pmatrix}, \ee
and the $\det(e)=1$.  Had we chosen our pullback to be ${\cal X}:\Phi -> (\xi_1, \xi_2, \xi_3, \o_1,\o_2, \o_2)$, where $\o_1,\o_2,$ and $ \o_2$ are constants. The frame fields become
\be e^a_j(\o)= \left(
\begin{array}{ccc}
 C_{\o_1} C_{\o_2} C_{\o_3}-S_{\o_2} S_{\o_3} & C_{\o_1} C_{\o_3} S_{\o_2}+C_{\o_2} S_{\o_3} & -C_{\o_3}
   S_{\o_1} \\
 -C_{\o_1} C_{\o_2} S_{\o_3}-C_{\o_3} S_{\o_2} & C_{\o_2} C_{\o_3}-C_{\o_1} S_{\o_2} S_{\o_3} & S_{\o_1}
   S_{\o_3} \\
 C_{\o_2} S_{\o_1} & S_{\o_1} S_{\o_2} & C_{\o_1} \\
\end{array}
\right),  \ee which is precisely an $SO(3)$ matrix and therefore $\det(e)=1$.

Below are the most obvious pullback choices for the dual volume. Extra parameters are labeled with a ${}_0$ subscript.
\bea
\left(
\begin{array}{cc}
\text{Dual Coordinates, $ \xi$} & \text{det (e)} \\\hline
 \left(x_1,x_2,x_3\right) & 1 \\
 \left(x_1,x_2,\beta \right) & \left(x_2 \sin\psi _0+x_1 \cos \psi _0\right) \\
  \left(x_1,x_2,\psi \right) & 0 \\
 \left(x_1,x_2,\phi \right) & \sin \beta _0 \left(x_2 \cos\psi _0-x_1 \sin \psi
   _0\right) \\
  \left(x_1,x_3,\beta \right) & x_3 \sin \psi _0\\
 \left(x_1,x_3,\psi \right) & x_1 \\
  \left(x_1,x_3,\phi \right) & \left(x_1 \cos \beta _0-x_3 \sin \beta _0 \cos \psi
   _0\right)\\
  \left(x_1,\beta ,\psi \right) & x_1 \left(x_{20} \sin \psi +x_1 \cos \psi \right) \\
  \left(x_1,\beta ,\phi \right) & x_1 \left(x_{30} \sin \beta -\cos \beta  \left(x_{20} \sin \psi _0+x_1
   \cos \psi _0\right)\right) \\
  \left(x_1,\psi ,\phi \right) & x_1 \sin \beta _0 \left(x_{20} \cos \psi -x_1 \sin \psi \right) \\
 \left(x_2,x_3,\beta \right) & x_3\cos \psi _0 \\
  \left(x_2,x_3,\psi \right) & x_2 \\
  \left(x_2,x_3,\phi \right) & \left(x_2 \cos \beta _0-x_3 \sin \beta _0 \sin \psi
   _0\right) \\
  \left(x_2,\beta ,\psi \right)& x_2 \left(x_2 \sin \psi +x_{10} \cos \psi \right) \\
  \left(x_2,\beta ,\phi \right) & x_2 \left(x_{30} \sin \beta -\cos \beta  \left(x_2 \sin \psi _0+x_{10}
   \cos\psi _0\right)\right) \\
  \left(x_2,\psi ,\phi \right) & x_2 \sin \beta _0 \left(x_2 \cos \psi -x_{10} \sin \psi \right) \\
  \left(x_3,\beta ,\psi \right) & x_3 \left(x_{20} \sin \psi +x_{10} \cos \psi \right) \\
  \left(x_3,\beta ,\phi \right) & x_3 \left(x_3 \sin \beta -\cos \beta  \left(x_{20} \sin \psi _0+x_{10}
   \cos \psi _0\right)\right)\\
 \left(x_3,\psi ,\phi \right) & x_3 \sin \beta _0 \left(x_{20} \cos \psi -x_{10} \sin \psi \right)
   \\
 \left(\beta ,\psi ,\phi \right) & 0 \\
\end{array}
\right)
\eea
Note that in some of the cases the parameters support the internal manifold's volume.  For example, when ($x_1,\ x_2,\ \beta$) are coordinates, the parameter $\psi_0$ cannot be set to zero, or else the metric will have zero volume.  Therefore, certain ``poor gauge fixing" choices can be remedied by introducing these parameters.
\subsection{Multiparametric solutions of NATD of $AdS_5\times T^{1,1}$}
In the remaining sections we present a few examples of new Type IIA solutions with extra parameters, generated by exploiting the gauge fixing ambiguities of NATD discussed above.  In all of the following examples, we have checked explicitly that all of the Type IIA equations are satisfied.
Here we present an example using $AdS_5\times T^{1,1}$.
\begin{enumerate}

\item $x_1,\ x_3,\ \psi$ are coordinates and $x_{20}$ is an extra parameter ($\beta_0=0$, $\phi_0=0$)

\bea
& &\hat{ds}^2=ds^2(AdS_5)+\lambda _1^2 \left(\mathit{d}\phi
   _1{}^2 \sin ^2\theta
   _1+\mathit{d}\theta
   _1{}^2\right)\nn \\& &+\frac{1}{\Delta _0} \left((x_1^2+\lambda^2\lambda_2^2)\mathit{d}x_1^2+(x_3^2+\lambda_2^4)\mathit{d}x_3^2+(x_1^2+x_{20}^2)\lambda^2\lambda_2^2(\cos\theta_1\mathit{d}\phi_1+\mathit{d}\psi)^2\right.\nn \\& & \left.+2\mathit{d}x_1(x_1x_3\mathit{d}x_3-x_{20}\lambda^2\lambda_2^2(\cos\theta_1\mathit{d}\phi_1+\mathit{d}\psi))\right)\nn \\& &
   \hat{B}= \frac{1}{\Delta _0}\left(\lambda_2^2x_{20}\mathit{d}x_1\wedge\mathit{d}x_3+(\lambda^2x_1x_3\mathit{d}x_1+\lambda_2^2(x_1^2+x_{20}^2)\mathit{d}x_3)\wedge\mathit{d}\psi\right.\nn \\& &\left.-\lambda ^2 \cos \theta_1((x_3^2+\lambda_2^4)\mathit{d}x_3+x_1x_3\mathit{d}x_1)\wedge \mathit{d}\phi_1\right) \nn\\& &
   e^{-2\hat{\Phi}}=\Delta_0,\quad \Delta_0=\lambda ^2 \left(\lambda
   _2^4+x_3^2\right)+\lambda _2^2
   \left(x_1^2+x_{20}^2\right)
   \eea

   \bea
   & &\hat{F}_2=4 \lambda  \lambda _1^2 \lambda _2^2
   \sin\theta_1
   \mathit{d}\theta_1\wedge
   \mathit{d}\phi_1 \\& &
   \hat{F}_4=\frac{1}{\Delta
   _0}(4\lambda\lambda_1^2\lambda_2^2\sin\theta_1(\lambda_2^2x_{20}\mathit{d}x_1\wedge\mathit{d}x_3\wedge\mathit{d}\theta_1\wedge\mathit{d}\phi_1\nn \\& &+(\lambda_2^2(x_1^2+x_{20}^2)\mathit{d}x_3-\lambda^2x_1x_3\mathit{d}x_1)\wedge \mathit{d}\theta_1\wedge\mathit{d}\phi_1\wedge\mathit{d}\psi))
   \nn
   \eea

\end{enumerate}

\subsection{Multiparametric solutions of NATD of $AdS_5\times S^5$}
Here we present additional examples using $AdS_5\times S^5$.


\begin{enumerate}
\item$x_1$, $x_2$, $x_3$ are coordinates, $\beta_0,\ \phi_0\ \psi_0$ are parameters
\bea
&\hat{ds}^2=4ds^2(AdS_5)+4d\alpha^2+4\sin^2\alpha d\theta^2\nn \\&
+\frac{1}{\cos^2\alpha\Delta_1}((x_1^2+\cos^4\alpha)dx_1^2+(x_2^2+\cos^4\alpha)dx_2^2+(x_3^2+\cos^4\alpha)dx_3^2\nn \\&+2x_1dx_1(x_2dx_2+x_3dx_3)+2x_2x_3dx_2dx_3)\nn \\&
\hat{B}=-\frac{1}{\Delta_1}(x_3dx_1\wedge dx_2-x_2dx_1\wedge dx_3+x_1dx_2\wedge dx_3)\nn \\&
e^{\hat{-2\Phi}}=\cos^2\alpha\Delta_1
\eea
where $\Delta_1=((x_1^2+x_2^2+x_3^2)+\cos^4\alpha)$
\bea
&\hat{F}_2=8L^4\cos^3\alpha\sin\alpha d\alpha\wedge d\theta,\nn \\&
\hat{F}_4=-\frac{8\cos^3\alpha\sin\alpha}{\Delta_1}(x_3dx_1\wedge dx_2 -x_2dx_1\wedge dx_3 +x_1dx_2\wedge dx_3) \wedge d\alpha\wedge d\theta\nn
\eea
This is precisely the answer we would have obtained if we had chosen a general gauge fixing (i.e. ($x_1$, $x_2$, $x_3$, $\beta=0,\ \phi=0\ \text{and}\ \psi=0$) with no extra parameters.

\item $x_1$, $x_3$, $\psi$ are coordinates, $x_{20}$, $\beta_0$, $\phi_0$ are parameters (though only $x_{20}$ appears)
\bea
& &\hat{ds}^2=4ds^2(AdS_5)+4d\alpha^2+4\sin^2\alpha d\theta^2\nn \\& &
+\frac{\sec^2\alpha}{\Delta_2}((x_1^2+\cos^4\alpha)dx_1^2+(x_3^2+\cos^4\alpha)dx_3^2+(x_1^2+x_{20}^2)\cos^4\alpha d\psi^2\nn \\& &+2dx_1(x_1x_3dx_3+x_{20}\cos^4\alpha d\psi))\nn \\& &
\hat{B}=-\frac{1}{\Delta_2}(x_{20}dx_1\wedge dx_3+x_1x_3dx_1\wedge d\psi-(x_1^2+x_{20}^2)dx_3\wedge d\psi)\nn \\& &
e^{\hat{-2\Phi}}=\cos^2\alpha\Delta_2,\quad \Delta_2=(x_1^2+x_{20}^2+x_3^2+\cos^4\alpha)
\eea
\bea
& &\hat{F}_2=-4 \cos^3\alpha\sin\alpha d\alpha\wedge d\theta,\nn \\& &
\hat{F}_4=-\frac{4\cos^3\alpha}{\Delta_2}\left(x_{20}\sin\alpha dx_1\wedge dx_3\wedge d\alpha\wedge d\theta+x_1x_3\sin\alpha dx_1\wedge d\alpha\wedge d\theta\wedge d\psi\right. \nn \\& & \left.
-(x_1^2+x_{20}^2)\sin\alpha dx_3\wedge d\alpha\wedge d\theta\wedge d\psi\right)
\eea

\item $x_3$, $\psi$,  $\beta$ are coordinates, $x_{10}$, $x_{20}$, $\phi_0$ are parameters (though only $x_{10}$ and $x_{20}$ appear)
\bea
& &\hat{ds}^2=4ds^2(AdS_5)+4d\alpha^2+4\sin^2\alpha d\theta^2 \nn \\& &
+\frac{\cos^2\alpha}{\Delta_3}\left( (1+x_3^2\sec^4\alpha)dx_3^2+(x_{10}^2+x_{20}^2)d\psi^2+2(x_{10}\cos\psi +x_{20} \sin\psi)dx_3d\beta \right. \nn \\& &
\left. - 2x_3(x_{20}\cos\psi-x_{10}\sin\psi)d\psi d\beta +\nn\right. \\& &\left. (\frac{1}{2}(x_{10}^2+x_{20}^2+2 x_3^2+(x_{10}^2-x_{20}^2)\cos2\psi)+2x_{10}x_{20}\sin2\psi)d\beta^2 \right)\nn \\& &
\hat{B}=\frac{1}{\Delta_3}((x_{10}^2+x_{20}^2)dx_3\wedge d\psi+x_3(-x_{20}\cos\psi+x_{10}\sin\psi)dx_3\wedge d\beta \nn \\& &
+(x_{10}^2+x_{20}^2+x_3^2)(x_{10}\cos\psi+x_{20}\sin\psi)d\beta\wedge d\psi)\nn  \\& &
e^{\hat{-2\Phi}}=4\cos^2\alpha\Delta_3,\quad \Delta_3=(x_{10}^2+x_{20}^2+x_3^2+\cos^4\alpha)
\eea
\bea
& &\hat{F}_2=-4 \cos^3\alpha\sin\alpha d\alpha\wedge d\theta,\nn \\& &
\hat{F}_4=-\frac{4\cos^3\alpha\sin\alpha}{\Delta_3}(-(x_{10}^2+x_{20}^2) dx_3\wedge d\psi\wedge d\alpha\wedge d\theta \nn \\& &
+x_3(x_{20}\cos\psi-x_{10}\sin\psi) dx_3\wedge d\beta\wedge d\alpha\wedge d\theta\nn \\& &+(x_{10}^2+x_{20}^2+x_3^2)(x_{10}\cos\psi+x_{20}\sin\psi) d\psi\wedge d\beta\wedge d\alpha\wedge d\theta))
\eea

\end{enumerate}

\section{Killing Spinor on $AdS_5\times S^5$}\label{sec: spinors}
In this appendix we derive a Killing spinor for $AdS_5\times S^5$ that is independent of the $SU(2)$ directions on which the NATD is performed.

To start we choose the vielbein basis
\begin{align}
e^{x^\mu}&=\frac{2r}{L}dx^{\mu},~~~ e^{r}=\frac{2L}{r}dr,~~~ e^{i}=L \cos\alpha \sigma_{i},\\[2mm]
e^4&=2L d\alpha,~~~ e^5=2L \sin\alpha d\theta,
\end{align}
where $i=1,2,3$.
With respect to this basis the non zero components of the spin connection are
\beq
\omega^{x^{\mu}r}= \frac{1}{2L} e^{x^{\mu}},~~~\omega^{45}= -\frac{1}{2L}\cot\alpha e^{5},~~~ \omega^{i5}= -\frac{1}{2L}\tan\alpha e^{i},~~~\omega^{ij}=\frac{1}{2L}\sec\alpha \epsilon_{ijk} e^k.
\eeq
which clearly indicates a 5+5 split, so the gravitino variation along the $AdS_5$ and $S^5$ directions can be treated independently\footnote{since the dilaton is constant and only the 5-form flux is non trivial the dilatino variation is automatically satisfied}.  As $F_5$ is given by
\beq
F_5=\frac{2}{L}\bigg(e^{tx^1x^2x^3r}-e^{12345}\bigg)
\eeq
and we choose
\beq
\Gamma^{tx^1x^2x^3 r12345}\epsilon= \Gamma_{AdS_5}\Gamma_{S^5}\epsilon=-\epsilon,
\eeq
the $AdS_5$ part leads to \footnote{Here $\epsilon=\epsilon_1+i \epsilon_2$, where $\epsilon_i$ are the MW Killing spinors in 10-d}
\beq
(\nabla_{\mu} + \frac{i}{2L}\Gamma_{AdS_5}\Gamma_{\mu})\epsilon=0
\eeq
where $\mu=t,x^1,x^2,x^3,r$, which is a standard Killing spinor equation on $AdS_5$ and so for our purposes it is sufficient to solve the gravitino variation on the $S^5$ directions. This is given by
\beq
(\nabla_{a} - \frac{i}{2L}\Gamma_{S^5}\Gamma_{a})\epsilon=0
\eeq
where $a=1,2,3,4,5$. If we make the assumption that $\epsilon$ is independent of the $SU(2)$ directions this gives the following set of coupled differential and algebraic equations
\begin{align}
&\big(2\partial_{\alpha}+i\Gamma_{1235})\epsilon=0,\nn\\[2mm]
&\big(2\partial_{\theta}- \cos\alpha\Gamma_{45}-i \sin\alpha \Gamma_{1234}\big)\epsilon=0,\nn\\[2mm]
&\big(\Gamma_{23}-\sin\alpha\Gamma_{14}-i \cos\alpha\Gamma_{2345}\big)\epsilon=0,\nn\\[2mm]
&\big(\Gamma_{13}+\sin\alpha\Gamma_{24}-i \cos\alpha\Gamma_{1345}\big)\epsilon=0,\nn\\[2mm]
&\big(\Gamma_{12}-\sin\alpha\Gamma_{34}-i \cos\alpha\Gamma_{1245}\big)\epsilon=0.
\end{align}
These reduce to a projection
\beq
\Gamma_{45}\epsilon=(\cos\alpha+i\sin\alpha\Gamma_{1235})\epsilon
\eeq
and two differential equations
\begin{align}
&\big(2\partial_{\alpha}+i\Gamma_{1235})\epsilon=0,\nn\\[2mm]
&\big(2\partial_{\theta}+i \big)\epsilon=0.
\end{align}
The whole Killing spinor then takes the form
\beq\label{S5killingspinor}
\epsilon= \mathcal{M}(AdS_5)e^{-\frac{i}{2}\theta}e^{-\frac{i\alpha}{2}\Gamma_{1235}}\eta,
\eeq
where $\eta$ is a constant spinor obeying
\beq
\Gamma_{45}\eta=\eta,
\eeq
and $\mathcal{M}(AdS_5)$ is a matrix which commutes with the projection and depends on the $AdS_5$ directions.
There for a total of 16 real supercharges are preserved.

As we have found a Killing spinor preserving $\mathcal{N}=2$ SUSY in 4-d which is independent of the $SU(2)$ direction \cite{Kelekci:2014ima} tells us that this is the SUSY preserve by the NATD solution, confirming the result of \cite{Sfetsos:2010uq}.

The NATD solution of $AdS_5\times S^5$ contains two $U(1)$ isometries, $\partial_{\theta}$ and $\partial_{\xi}$. We have checked that in the preferred frame of NATD the the Kosmann derivative in each case reduces to
\beq
\mathcal{L}_{\partial_{\theta}}\hat\epsilon=\partial_{\theta}\hat\epsilon,~~~\mathcal{L}_{\partial_{\xi}}\hat\epsilon=\partial_{\xi}\hat\epsilon.
\eeq
The dual MW Killing spinors are given in this frame by
\beq
\hat{\epsilon}_1=\epsilon_1,~~ \hat{\epsilon}_2=\Omega \epsilon_2.
\eeq
where
\beq
\Omega=\frac{-L^2 \Gamma_{123}+\rho\big(\sin\chi\cos\xi\Gamma_1+\sin\chi \sin\xi\Gamma_2+\cos\chi \Gamma_3\big)}{\sqrt{\rho^2+L^4\cos^4\alpha}}.
\eeq
Thus it is easy to see that a further Abelian T-duality along $\theta$ or $\xi$ will break SUSY completely because the angular dependence in Eq.(\ref{S5killingspinor}) ensures that neither Kosmann derivative can vanish  \cite{Kelekci:2014ima}.
\bibliographystyle{JHEP}
\bibliography{Ypqbib}

\providecommand{\href}[2]{#2}\begingroup\raggedright\begin{thebibliography}{10}

\bibitem{Maldacena:1997re}
J.~M. Maldacena, {\it {The large N limit of superconformal field theories and
  supergravity}},  {\em Adv. Theor. Math. Phys.} {\bf 2} (1998) 231--252,
  [\href{http://xxx.lanl.gov/abs/hep-th/9711200}{{\tt hep-th/9711200}}].

\bibitem{Witten:1998qj}
E.~Witten, {\it {Anti-de Sitter space and holography}},  {\em Adv. Theor. Math.
  Phys.} {\bf 2} (1998) 253--291,
  [\href{http://xxx.lanl.gov/abs/hep-th/9802150}{{\tt hep-th/9802150}}].

\bibitem{Gubser:1998bc}
S.~Gubser, I.~R. Klebanov, and A.~M. Polyakov, {\it {Gauge theory correlators
  from noncritical string theory}},  {\em Phys.Lett.} {\bf B428} (1998)
  105--114, [\href{http://xxx.lanl.gov/abs/hep-th/9802109}{{\tt
  hep-th/9802109}}].

\bibitem{Aharony:1999ti}
O.~Aharony, S.~S. Gubser, J.~M. Maldacena, H.~Ooguri, and Y.~Oz, {\it {Large N
  field theories, string theory and gravity}},  {\em Phys. Rept.} {\bf 323}
  (2000) 183--386, [\href{http://xxx.lanl.gov/abs/hep-th/9905111}{{\tt
  hep-th/9905111}}].

\bibitem{Gauntlett:2004zh}
J.~P. Gauntlett, D.~Martelli, J.~Sparks, and D.~Waldram, {\it {Supersymmetric
  AdS(5) solutions of M theory}},  {\em Class.Quant.Grav.} {\bf 21} (2004)
  4335--4366, [\href{http://xxx.lanl.gov/abs/hep-th/0402153}{{\tt
  hep-th/0402153}}].

\bibitem{Gauntlett:2005ww}
J.~P. Gauntlett, D.~Martelli, J.~Sparks, and D.~Waldram, {\it {Supersymmetric
  AdS(5) solutions of type IIB supergravity}},  {\em Class.Quant.Grav.} {\bf
  23} (2006) 4693--4718, [\href{http://xxx.lanl.gov/abs/hep-th/0510125}{{\tt
  hep-th/0510125}}].

\bibitem{Colgain:2011hb}
E.~O~Colgain and J.~Stefanski, Bogdan, {\it {A search for AdS5 X S2 IIB
  supergravity solutions dual to N = 2 SCFTs}},  {\em JHEP} {\bf 1110} (2011)
  061, [\href{http://xxx.lanl.gov/abs/1107.5763}{{\tt arXiv:1107.5763}}].

\bibitem{Stephani:624239}
H.~Stephani, D.~Krämer, M.~MacCallum, C.~Hoenselaers, and E.~Herlt, {\em {Exact
  solutions of Einstein's field equations; 2nd ed.}}
\newblock Cambridge Univ. Press, Cambridge, 2003.

\bibitem{Peet:2000hn}
A.~W. Peet, {\it {TASI lectures on black holes in string theory}},
  \href{http://xxx.lanl.gov/abs/hep-th/0008241}{{\tt hep-th/0008241}}.

\bibitem{Lunin:2005jy}
O.~Lunin and J.~M. Maldacena, {\it {Deforming field theories with U(1) x U(1)
  global symmetry and their gravity duals}},  {\em JHEP} {\bf 0505} (2005) 033,
  [\href{http://xxx.lanl.gov/abs/hep-th/0502086}{{\tt hep-th/0502086}}].

\bibitem{Ossa:1992vc}
X.~C. de~la Ossa and F.~Quevedo, {\it {Duality symmetries from nonAbelian
  isometries in string theory}},  {\em Nucl.Phys.} {\bf B403} (1993) 377--394,
  [\href{http://xxx.lanl.gov/abs/hep-th/9210021}{{\tt hep-th/9210021}}].

\bibitem{Fridling:1983ha}
B.~Fridling and A.~Jevicki, {\it {Dual Representations and Ultraviolet
  Divergences in Nonlinear $\sigma$ Models}},  {\em Phys.Lett.} {\bf B134}
  (1984) 70.

\bibitem{Fradkin:1984ai}
E.~Fradkin and A.~A. Tseytlin, {\it {Quantum Equivalence of Dual Field
  Theories}},  {\em Annals Phys.} {\bf 162} (1985) 31.

\bibitem{Giveon:1993ai}
A.~Giveon and M.~Rocek, {\it {On nonAbelian duality}},  {\em Nucl.Phys.} {\bf
  B421} (1994) 173--190, [\href{http://xxx.lanl.gov/abs/hep-th/9308154}{{\tt
  hep-th/9308154}}].

\bibitem{Gasperini:1993nz}
M.~Gasperini, R.~Ricci, and G.~Veneziano, {\it {A Problem with nonAbelian
  duality?}},  {\em Phys.Lett.} {\bf B319} (1993) 438--444,
  [\href{http://xxx.lanl.gov/abs/hep-th/9308112}{{\tt hep-th/9308112}}].

\bibitem{Alvarez:1994np}
E.~Alvarez, L.~Alvarez-Gaume, and Y.~Lozano, {\it {On nonAbelian duality}},
  {\em Nucl.Phys.} {\bf B424} (1994) 155--183,
  [\href{http://xxx.lanl.gov/abs/hep-th/9403155}{{\tt hep-th/9403155}}].

\bibitem{Elitzur:1994ri}
S.~Elitzur, A.~Giveon, E.~Rabinovici, A.~Schwimmer, and G.~Veneziano, {\it
  {Remarks on nonAbelian duality}},  {\em Nucl.Phys.} {\bf B435} (1995)
  147--171, [\href{http://xxx.lanl.gov/abs/hep-th/9409011}{{\tt
  hep-th/9409011}}].

\bibitem{Sfetsos:1994vz}
K.~Sfetsos, {\it {Gauged WZW models and nonAbelian duality}},  {\em Phys.Rev.}
  {\bf D50} (1994) 2784--2798,
  [\href{http://xxx.lanl.gov/abs/hep-th/9402031}{{\tt hep-th/9402031}}].

\bibitem{Curtright:1994be}
T.~Curtright and C.~K. Zachos, {\it {Currents, charges, and canonical structure
  of pseudodual chiral models}},  {\em Phys.Rev.} {\bf D49} (1994) 5408--5421,
  [\href{http://xxx.lanl.gov/abs/hep-th/9401006}{{\tt hep-th/9401006}}].

\bibitem{Alvarez:1995uc}
O.~Alvarez and C.-H. Liu, {\it {Target space duality between simple compact Lie
  groups and Lie algebras under the Hamiltonian formalism: 1. Remnants of
  duality at the classical level}},  {\em Commun.Math.Phys.} {\bf 179} (1996)
  185--214, [\href{http://xxx.lanl.gov/abs/hep-th/9503226}{{\tt
  hep-th/9503226}}].

\bibitem{Alvarez:1994dn}
E.~Alvarez, L.~Alvarez-Gaume, and Y.~Lozano, {\it {An Introduction to T duality
  in string theory}},  {\em Nucl.Phys.Proc.Suppl.} {\bf 41} (1995) 1--20,
  [\href{http://xxx.lanl.gov/abs/hep-th/9410237}{{\tt hep-th/9410237}}].

\bibitem{Giveon:1994fu}
A.~Giveon, M.~Porrati, and E.~Rabinovici, {\it {Target space duality in string
  theory}},  {\em Phys.Rept.} {\bf 244} (1994) 77--202,
  [\href{http://xxx.lanl.gov/abs/hep-th/9401139}{{\tt hep-th/9401139}}].

\bibitem{Sfetsos:2010uq}
K.~Sfetsos and D.~C. Thompson, {\it {On non-abelian T-dual geometries with
  Ramond fluxes}},  {\em Nucl.Phys.} {\bf B846} (2011) 21--42,
  [\href{http://xxx.lanl.gov/abs/1012.1320}{{\tt arXiv:1012.1320}}].

\bibitem{Lozano:2011kb}
Y.~Lozano, E.~O~Colgain, K.~Sfetsos, and D.~C. Thompson, {\it {Non-abelian
  T-duality, Ramond Fields and Coset Geometries}},  {\em JHEP} {\bf 1106}
  (2011) 106, [\href{http://xxx.lanl.gov/abs/1104.5196}{{\tt
  arXiv:1104.5196}}].

\bibitem{Itsios:2013wd}
G.~Itsios, C.~Nunez, K.~Sfetsos, and D.~C. Thompson, {\it {Non-Abelian
  T-duality and the AdS/CFT correspondence:new N=1 backgrounds}},  {\em
  Nucl.Phys.} {\bf B873} (2013) 1--64,
  [\href{http://xxx.lanl.gov/abs/1301.6755}{{\tt arXiv:1301.6755}}].

\bibitem{Caceres:2014uoa}
E.~Caceres, N.~T. Macpherson, and C.~Nunez, {\it {New Type IIB Backgrounds and
  Aspects of Their Field Theory Duals}},
  \href{http://xxx.lanl.gov/abs/1402.3294}{{\tt arXiv:1402.3294}}.

\bibitem{Lozano:2012au}
Y.~Lozano, E.~O~Colgain, D.~Rodriguez-Gomez, and K.~Sfetsos, {\it
  {Supersymmetric $AdS_6$ via T Duality}},  {\em Phys.Rev.Lett.} {\bf 110}
  (2013), no.~23 231601, [\href{http://xxx.lanl.gov/abs/1212.1043}{{\tt
  arXiv:1212.1043}}].

\bibitem{Itsios:2012zv}
G.~Itsios, C.~Nunez, K.~Sfetsos, and D.~C. Thompson, {\it {On Non-Abelian
  T-Duality and new N=1 backgrounds}},  {\em Phys.Lett.} {\bf B721} (2013)
  342--346, [\href{http://xxx.lanl.gov/abs/1212.4840}{{\tt arXiv:1212.4840}}].

\bibitem{Macpherson:2013lja}
N.~T. Macpherson, {\it {Non-abelian T-duality, generalised geometry and
  holography}},  {\em J.Phys.Conf.Ser.} {\bf 490} (2014) 012122,
  [\href{http://xxx.lanl.gov/abs/1309.1358}{{\tt arXiv:1309.1358}}].

\bibitem{Gevorgyan:2013xka}
E.~Gevorgyan and G.~Sarkissian, {\it {Defects, Non-abelian T-duality, and the
  Fourier-Mukai transform of the Ramond-Ramond fields}},  {\em JHEP} {\bf 1403}
  (2014) 035, [\href{http://xxx.lanl.gov/abs/1310.1264}{{\tt
  arXiv:1310.1264}}].

\bibitem{Lozano:2013oma}
Y.~Lozano, E.~O. Colgain, and D.~Rodriguez-Gomez, {\it {Hints of 5d Fixed Point
  Theories from Non-Abelian T-duality}},
  \href{http://xxx.lanl.gov/abs/1311.4842}{{\tt arXiv:1311.4842}}.

\bibitem{Gaillard:2013vsa}
J.~Gaillard, N.~T. Macpherson, C.~Nunez, and D.~C. Thompson, {\it {Dualising
  the Baryonic Branch: Dynamic SU(2) and confining backgrounds in IIA}},  {\em
  Nucl.Phys.} {\bf B884} (2014) 696--740,
  [\href{http://xxx.lanl.gov/abs/1312.4945}{{\tt arXiv:1312.4945}}].

\bibitem{Elander:2013jqa}
D.~Elander, A.~F. Faedo, C.~Hoyos, D.~Mateos, and M.~Piai, {\it {Multiscale
  confining dynamics from holographic RG flows}},  {\em JHEP} {\bf 1405} (2014)
  003, [\href{http://xxx.lanl.gov/abs/1312.7160}{{\tt arXiv:1312.7160}}].

\bibitem{Zacarias:2014wta}
S.~Zacarias, {\it {Semiclassical strings and Non-Abelian T-duality}},
  \href{http://xxx.lanl.gov/abs/1401.7618}{{\tt arXiv:1401.7618}}.

\bibitem{Pradhan:2014zqa}
P.~M. Pradhan, {\it {Oscillating Strings and Non-Abelian T-dual Klebanov-Witten
  Background}},  {\em Phys.Rev.} {\bf D90} (2014) 046003,
  [\href{http://xxx.lanl.gov/abs/1406.2152}{{\tt arXiv:1406.2152}}].

\bibitem{Lozano:2014ata}
Y.~Lozano and N.~T. Macpherson, {\it {A new $AdS_4/CFT_3$ Dual with Extended
  SUSY and a Spectral Flow}},  \href{http://xxx.lanl.gov/abs/1408.0912}{{\tt
  arXiv:1408.0912}}.

\bibitem{Barranco:2013fza}
A.~Barranco, J.~Gaillard, N.~T. Macpherson, C.~Nunez, and D.~C. Thompson, {\it
  {G-structures and Flavouring non-Abelian T-duality}},  {\em JHEP} {\bf 1308}
  (2013) 018, [\href{http://xxx.lanl.gov/abs/1305.7229}{{\tt
  arXiv:1305.7229}}].

\bibitem{Sfetsos:2014tza}
K.~Sfetsos and D.~C. Thompson, {\it {New ${\cal N} = 1$ supersymmetric $AdS_5$
  backgrounds in Type IIA supergravity}},
  \href{http://xxx.lanl.gov/abs/1408.6545}{{\tt arXiv:1408.6545}}.

\bibitem{Klebanov:1998hh}
I.~R. Klebanov and E.~Witten, {\it {Superconformal field theory on threebranes
  at a Calabi-Yau singularity}},  {\em Nucl. Phys.} {\bf B536} (1998) 199--218,
  [\href{http://xxx.lanl.gov/abs/hep-th/9807080}{{\tt hep-th/9807080}}].

\bibitem{Herzog:2001xk}
C.~P. Herzog, I.~R. Klebanov, and P.~Ouyang, {\it {Remarks on the warped
  deformed conifold}},  \href{http://xxx.lanl.gov/abs/hep-th/0108101}{{\tt
  hep-th/0108101}}.

\bibitem{Gauntlett:2004yd}
J.~P. Gauntlett, D.~Martelli, J.~Sparks, and D.~Waldram, {\it {Sasaki-Einstein
  metrics on S(2) x S(3)}},  {\em Adv. Theor. Math. Phys.} {\bf 8} (2004)
  711--734, [\href{http://xxx.lanl.gov/abs/hep-th/0403002}{{\tt
  hep-th/0403002}}].

\bibitem{Gauntlett:2004hh}
J.~P. Gauntlett, D.~Martelli, J.~F. Sparks, and D.~Waldram, {\it {A new
  infinite class of Sasaki-Einstein manifolds}},  {\em Adv. Theor. Math. Phys.}
  {\bf 8} (2006) 987--1000, [\href{http://xxx.lanl.gov/abs/hep-th/0403038}{{\tt
  hep-th/0403038}}].

\bibitem{Itsios:2012dc}
G.~Itsios, Y.~Lozano, E.~O~Colgain, and K.~Sfetsos, {\it {Non-Abelian T-duality
  and consistent truncations in type-II supergravity}},  {\em JHEP} {\bf 1208}
  (2012) 132, [\href{http://xxx.lanl.gov/abs/1205.2274}{{\tt
  arXiv:1205.2274}}].

\bibitem{Gaiotto:2009gz}
D.~Gaiotto and J.~Maldacena, {\it {The Gravity duals of N=2 superconformal
  field theories}},  {\em JHEP} {\bf 1210} (2012) 189,
  [\href{http://xxx.lanl.gov/abs/0904.4466}{{\tt arXiv:0904.4466}}].

\bibitem{ReidEdwards:2010qs}
R.~Reid-Edwards and j.~Stefanski, B., {\it {On Type IIA geometries dual to N =
  2 SCFTs}},  {\em Nucl.Phys.} {\bf B849} (2011) 549--572,
  [\href{http://xxx.lanl.gov/abs/1011.0216}{{\tt arXiv:1011.0216}}].

\bibitem{Aharony:2012tz}
O.~Aharony, L.~Berdichevsky, and M.~Berkooz, {\it {4d N=2 superconformal linear
  quivers with type IIA duals}},  {\em JHEP} {\bf 1208} (2012) 131,
  [\href{http://xxx.lanl.gov/abs/1206.5916}{{\tt arXiv:1206.5916}}].

\bibitem{Cvetic:1999zs}
M.~Cvetic, H.~Lu, C.~Pope, and K.~Stelle, {\it {T duality in the Green-Schwarz
  formalism, and the massless / massive IIA duality map}},  {\em Nucl.Phys.}
  {\bf B573} (2000) 149--176,
  [\href{http://xxx.lanl.gov/abs/hep-th/9907202}{{\tt hep-th/9907202}}].

\bibitem{Maldacena:2000mw}
J.~M. Maldacena and C.~Nunez, {\it {Supergravity description of field theories
  on curved manifolds and a no go theorem}},  {\em Int.J.Mod.Phys.} {\bf A16}
  (2001) 822--855, [\href{http://xxx.lanl.gov/abs/hep-th/0007018}{{\tt
  hep-th/0007018}}].

\bibitem{Gauntlett:2006ux}
J.~P. Gauntlett, O.~A. Mac~Conamhna, T.~Mateos, and D.~Waldram, {\it {AdS
  spacetimes from wrapped M5 branes}},  {\em JHEP} {\bf 0611} (2006) 053,
  [\href{http://xxx.lanl.gov/abs/hep-th/0605146}{{\tt hep-th/0605146}}].

\bibitem{Bah:2012dg}
I.~Bah, C.~Beem, N.~Bobev, and B.~Wecht, {\it {Four-Dimensional SCFTs from
  M5-Branes}},  {\em JHEP} {\bf 1206} (2012) 005,
  [\href{http://xxx.lanl.gov/abs/1203.0303}{{\tt arXiv:1203.0303}}].

\bibitem{Bah:2013qya}
I.~Bah, {\it {Quarter-BPS $AdS_{5}$ solutions in M-theory with a $T^{2}$ bundle
  over a Riemann surface}},  {\em JHEP} {\bf 1308} (2013) 137,
  [\href{http://xxx.lanl.gov/abs/1304.4954}{{\tt arXiv:1304.4954}}].

\bibitem{Cucu:2003bm}
S.~Cucu, H.~Lu, and J.~F. Vazquez-Poritz, {\it {A Supersymmetric and smooth
  compactification of M theory to AdS(5)}},  {\em Phys.Lett.} {\bf B568} (2003)
  261--269, [\href{http://xxx.lanl.gov/abs/hep-th/0303211}{{\tt
  hep-th/0303211}}].

\bibitem{Fayyazuddin:1999zu}
A.~Fayyazuddin and D.~J. Smith, {\it {Localized intersections of M5-branes and
  four-dimensional superconformal field theories}},  {\em JHEP} {\bf 9904}
  (1999) 030, [\href{http://xxx.lanl.gov/abs/hep-th/9902210}{{\tt
  hep-th/9902210}}].

\bibitem{Alvarez:1993qi}
E.~Alvarez, L.~Alvarez-Gaume, J.~Barbon, and Y.~Lozano, {\it {Some global
  aspects of duality in string theory}},  {\em Nucl.Phys.} {\bf B415} (1994)
  71--100, [\href{http://xxx.lanl.gov/abs/hep-th/9309039}{{\tt
  hep-th/9309039}}].

\bibitem{Klebanov:2000nc}
I.~R. Klebanov and A.~A. Tseytlin, {\it {Gravity Duals of Supersymmetric SU(N)
  x SU(N+M) Gauge Theories}},  {\em Nucl. Phys.} {\bf B578} (2000) 123--138,
  [\href{http://xxx.lanl.gov/abs/hep-th/0002159}{{\tt hep-th/0002159}}].

\bibitem{Klebanov:2000hb}
I.~R. Klebanov and M.~J. Strassler, {\it {Supergravity and a confining gauge
  theory: Duality cascades and chiSB-resolution of naked singularities}},  {\em
  JHEP} {\bf 08} (2000) 052,
  [\href{http://xxx.lanl.gov/abs/hep-th/0007191}{{\tt hep-th/0007191}}].

\bibitem{Strassler:2005qs}
M.~J. Strassler, {\it {The Duality cascade}},
  \href{http://xxx.lanl.gov/abs/hep-th/0505153}{{\tt hep-th/0505153}}.

\bibitem{Benini:2007gx}
F.~Benini, F.~Canoura, S.~Cremonesi, C.~Nunez, and A.~V. Ramallo, {\it
  {Backreacting flavors in the Klebanov-Strassler background}},  {\em JHEP}
  {\bf 0709} (2007) 109, [\href{http://xxx.lanl.gov/abs/0706.1238}{{\tt
  arXiv:0706.1238}}].

\bibitem{Sfetsos:2014cea}
K.~Sfetsos and D.~C. Thompson, {\it {Spacetimes for $\lambda$-deformations}},
  \href{http://xxx.lanl.gov/abs/1410.1886}{{\tt arXiv:1410.1886}}.

\bibitem{Berman:2013eva}
D.~S. Berman and D.~C. Thompson, {\it {Duality Symmetric String and M-Theory}},
   \href{http://xxx.lanl.gov/abs/1306.2643}{{\tt arXiv:1306.2643}}.

\bibitem{Henningson:1998gx}
M.~Henningson and K.~Skenderis, {\it {The Holographic Weyl anomaly}},  {\em
  JHEP} {\bf 9807} (1998) 023,
  [\href{http://xxx.lanl.gov/abs/hep-th/9806087}{{\tt hep-th/9806087}}].

\bibitem{Gabella:2009ni}
M.~Gabella, J.~P. Gauntlett, E.~Palti, J.~Sparks, and D.~Waldram, {\it {The
  Central charge of supersymmetric AdS(5) solutions of type IIB supergravity}},
   {\em Phys.Rev.Lett.} {\bf 103} (2009) 051601,
  [\href{http://xxx.lanl.gov/abs/0906.3686}{{\tt arXiv:0906.3686}}].

\bibitem{Ramallo:2013bua}
A.~V. Ramallo, {\it {Introduction to the AdS/CFT correspondence}},
  \href{http://xxx.lanl.gov/abs/1310.4319}{{\tt arXiv:1310.4319}}.

\bibitem{Klebanov:2007ws}
I.~R. Klebanov, D.~Kutasov, and A.~Murugan, {\it {Entanglement as a probe of
  confinement}},  {\em Nucl.Phys.} {\bf B796} (2008) 274--293,
  [\href{http://xxx.lanl.gov/abs/0709.2140}{{\tt arXiv:0709.2140}}].

\bibitem{Kol:2014nqa}
U.~Kol, C.~Nunez, D.~Schofield, J.~Sonnenschein, and M.~Warschawski, {\it
  {Confinement, Phase Transitions and non-Locality in the Entanglement
  Entropy}},  {\em JHEP} {\bf 1406} (2014) 005,
  [\href{http://xxx.lanl.gov/abs/1403.2721}{{\tt arXiv:1403.2721}}].

\bibitem{Horowitz:1993wt}
G.~T. Horowitz and D.~L. Welch, {\it {Duality invariance of the Hawking
  temperature and entropy}},  {\em Phys.Rev.} {\bf D49} (1994) 590--594,
  [\href{http://xxx.lanl.gov/abs/hep-th/9308077}{{\tt hep-th/9308077}}].

\bibitem{Grana:2005sn}
M.~Grana, R.~Minasian, M.~Petrini, and A.~Tomasiello, {\it {Generalized
  structures of N=1 vacua}},  {\em JHEP} {\bf 0511} (2005) 020,
  [\href{http://xxx.lanl.gov/abs/hep-th/0505212}{{\tt hep-th/0505212}}].

\bibitem{Andriot:2008va}
D.~Andriot, {\it {New supersymmetric flux vacua with intermediate SU(2)
  structure}},  {\em JHEP} {\bf 0808} (2008) 096,
  [\href{http://xxx.lanl.gov/abs/0804.1769}{{\tt arXiv:0804.1769}}].

\bibitem{Andriot:2010sya}
D.~Andriot, {\it {String theory flux vacua on twisted tori and Generalized
  Complex Geometry}}, .

\bibitem{Hassan:1999bv}
S.~Hassan, {\it {T duality, space-time spinors and RR fields in curved
  backgrounds}},  {\em Nucl.Phys.} {\bf B568} (2000) 145--161,
  [\href{http://xxx.lanl.gov/abs/hep-th/9907152}{{\tt hep-th/9907152}}].

\bibitem{Kelekci:2014ima}
O.~Kelekci, Y.~Lozano, N.~T. Macpherson, and E.~O. Colgain, {\it {Supersymmetry
  and non-Abelian T-duality in type II supergravity}},
  \href{http://xxx.lanl.gov/abs/1409.7406}{{\tt arXiv:1409.7406}}.

\bibitem{Macpherson:2013zba}
N.~T. Macpherson, {\it {Non-Abelian T-duality, $G_2$-structure rotation and
  holographic duals of $N=1$ Chern-Simons theories}},  {\em JHEP} {\bf 1311}
  (2013) 137, [\href{http://xxx.lanl.gov/abs/1310.1609}{{\tt
  arXiv:1310.1609}}].

\bibitem{Arean:2006nc}
D.~Arean, {\it {Killing spinors of some supergravity solutions}},
  \href{http://xxx.lanl.gov/abs/hep-th/0605286}{{\tt hep-th/0605286}}.

\bibitem{Canoura:2005uz}
F.~Canoura, J.~D. Edelstein, L.~A. Pando~Zayas, A.~V. Ramallo, and D.~Vaman,
  {\it {Supersymmetric branes on AdS(5) x Y**p,q and their field theory
  duals}},  {\em JHEP} {\bf 0603} (2006) 101,
  [\href{http://xxx.lanl.gov/abs/hep-th/0512087}{{\tt hep-th/0512087}}].

\bibitem{Conde:2011rg}
E.~Conde, J.~Gaillard, and A.~V. Ramallo, {\it {On the holographic dual of
  $N=1$ SQCD with massive flavors}},  {\em JHEP} {\bf 1110} (2011) 023,
  [\href{http://xxx.lanl.gov/abs/1107.3803}{{\tt arXiv:1107.3803}}].

\bibitem{Grana:2001xn}
M.~Grana and J.~Polchinski, {\it {Gauge / gravity duals with holomorphic
  dilaton}},  {\em Phys.Rev.} {\bf D65} (2002) 126005,
  [\href{http://xxx.lanl.gov/abs/hep-th/0106014}{{\tt hep-th/0106014}}].

\bibitem{Beasley:2001zp}
C.~E. Beasley and M.~R. Plesser, {\it {Toric duality is Seiberg duality}},
  {\em JHEP} {\bf 0112} (2001) 001,
  [\href{http://xxx.lanl.gov/abs/hep-th/0109053}{{\tt hep-th/0109053}}].

\bibitem{Feng:2001bn}
B.~Feng, A.~Hanany, Y.-H. He, and A.~M. Uranga, {\it {Toric duality as Seiberg
  duality and brane diamonds}},  {\em JHEP} {\bf 0112} (2001) 035,
  [\href{http://xxx.lanl.gov/abs/hep-th/0109063}{{\tt hep-th/0109063}}].

\bibitem{Franco:2002mu}
S.~Franco and A.~Hanany, {\it {Toric duality, Seiberg duality and
  Picard-Lefschetz transformations}},  {\em Fortsch.Phys.} {\bf 51} (2003)
  738--744, [\href{http://xxx.lanl.gov/abs/hep-th/0212299}{{\tt
  hep-th/0212299}}].

\bibitem{Maldacena:2009mw}
J.~Maldacena and D.~Martelli, {\it {The Unwarped, resolved, deformed conifold:
  Fivebranes and the baryonic branch of the Klebanov-Strassler theory}},  {\em
  JHEP} {\bf 1001} (2010) 104, [\href{http://xxx.lanl.gov/abs/0906.0591}{{\tt
  arXiv:0906.0591}}].

\bibitem{Gaillard:2010qg}
J.~Gaillard, D.~Martelli, C.~Nunez, and I.~Papadimitriou, {\it {The warped,
  resolved, deformed conifold gets flavoured}},
  \href{http://xxx.lanl.gov/abs/1004.4638}{{\tt arXiv:1004.4638}}.

\bibitem{Caceres:2011zn}
E.~Caceres, C.~Nunez, and L.~A. Pando~Zayas, {\it {Heating up the Baryonic
  Branch with U-duality: A Unified picture of conifold black holes}},  {\em
  JHEP} {\bf 1103} (2011) 054, [\href{http://xxx.lanl.gov/abs/1101.4123}{{\tt
  arXiv:1101.4123}}].

\bibitem{Gutowski:2014ova}
J.~Gutowski and G.~Papadopoulos, {\it {Supersymmetry of AdS and flat
  backgrounds in M-theory}},  \href{http://xxx.lanl.gov/abs/1407.5652}{{\tt
  arXiv:1407.5652}}.

\bibitem{Beck:2014zda}
S.~Beck, J.~Gutowski, and G.~Papadopoulos, {\it {Supersymmetry of AdS and flat
  IIB backgrounds}},  \href{http://xxx.lanl.gov/abs/1410.3431}{{\tt
  arXiv:1410.3431}}.

\end{thebibliography}\endgroup

\end{document}